%% file: main.tex
\documentclass[a4paper]{JHEP3} 

\usepackage{amsmath}
\usepackage{amsfonts}
 
\usepackage{feynmp}
 
\usepackage[dvips]{graphicx}

\newcommand{\bea}{\begin{eqnarray}}
\newcommand{\eea}{\end{eqnarray}}
\newcommand{\bi}{\begin{itemize}}
\newcommand{\ei}{\end{itemize}}
\newcommand{\g}{\gamma}
\newcommand{\G}{\Gamma}
\newcommand{\eq}{=}
\newcommand{\GF}{G_{\text{F}} }
\newcommand{\ud}{\text{d}}
\newcommand{\bve}[1]{\big({#1}\big)}
\newcommand{\ds}[1]{\dash{#1}}
\newcommand{\ph}[1]{\phantom{#1}}
\newcommand{\pl}{P_{\text{L}}}
\newcommand{\rightproj}{P_{\text{R}}}
\newcommand{\half}{\frac{1}{2}}

\newcommand{\bra}{\langle}
\newcommand{\ket}{\rangle}

\newcommand{\dash}[1]{{#1}\!\!\!/}

\title{A phenomenological model for radiative corrections in exclusive semileptonic $B$-meson decays to (pseudo)scalar final state mesons}
\author{Florian Urs Bernlochner \\  
              Department of Physics, Humboldt University of Berlin, Newtonstr. 15, 12489 Berlin, Germany\\ 
              E-mail: \email{florian@slac.stanford.edu} }
\author{Heiko Lacker\\
              Department of Physics, Humboldt University of Berlin, Newtonstr. 15, 12489 Berlin, Germany\\ 
              E-mail: \email{lacker@physik.hu-berlin.de}   }

\abstract{Next-to-leading order corrections are an important aspect in the extraction of the Cabibbo- Kobayashi-Maskawa matrix elements $|V_{\text{cb}}|$ and $|V_{\text{ub}}|$ at $B$-factory experiments: virtual and real photons couple to all charged particles of the decay, alter the resulting decay dynamics and enhance the total decay rate. We present a phenomenologically motivated model and a Monte-Carlo implementation to study electromagnetic radiative corrections to differential and total decay rates for semileptonic $B$-meson decays into exclusive pseudoscalar and scalar final states and apply it to $B \to D \, l \, \nu$ and $B \to \pi \, l \, \nu$, and $B \to D_0^* \, l \, \nu$ decays, respectively. We study such corrections with a phenomenological picture of point-like mesons (with some structure-dependent corrections), which is valid in the low-energy region of the photons and whose results we extrapolate over the complete phase space. The largest quantifiable uncertainty is due to the approximative matching to the Standard Model. In addition unknown structure-dependent contributions and model dependencies might have an impact on our findings. 
 }

\preprint{HU-EP-10/09\\ arXiv:1003.1620v4 }

\begin{document} 

 \begin{fmffile}{graphs}

  \input{introduction} \newpage
  \input{treelevel}

  \input{nlo_formalism}

  \input{nlo}
  \input{numerics} 
  \input{results}

 \clearpage


\begin{appendix}
  \input{formfactors} \newpage
  \input{loopintegrals} \newpage
\end{appendix}

\end{fmffile}

\bibliographystyle{JHEP}
\bibliography{journal}

\end{document}

%% file: introduction.tex
\vspace{-9ex}

\section{Introduction}
\vspace{-2ex}
In the Standard Model, the Cabibbo-Kobayashi-Maskawa (CKM) matrix \cite{PhysRevLett.10.531} governs the weak transitions between the up- and down-type quark generations. The precision determination of its matrix elements and CP violating phase is the focus of intense research over the past decade. The combination of such measurements to test unitarity of the CKM matrix as a whole is considered a strong instrument for the search of physics beyond the Standard Model \cite{Charles:2004jd,Bona:2005eu}.

In this paper, we present a calculation and a simulation tool for the electromagnetic one-loop corrections in exclusive semileptonic $B$-meson decays based on a phenomenological model, which is applied to $B \to D \, l \, \nu$, $B \to \pi \, l \, \nu$, and $B \to D_0^* \, l \, \nu$ decays. Next-to-leading order corrections to such decays are an important aspect in the extraction of the CKM matrix elements $|V_{\text{cb}}|$ and $|V_{\text{ub}}|$ at $B$-factory experiments: virtual and real photons couple to all charged particles of the decay, alter the resulting decay dynamics and enhance the weak decay rate. To correct for the altered decay dynamics, experimentalists use approximative all-purpose next-to-leading order algorithms. These exploit the fact that such corrections become universal in the soft photon energy limit \cite{PhysRev.140.B516, Yennie:1961ad}. In addition, the total leading order decay rate of semileptonic decays are corrected by the known leading logarithm of the virtual corrections at parton level \cite{Sirlin:1977sv,Sirlin:1981ie}. As implied by \cite{Becirevic:2009fy} such a treatment might not be sufficient: corrections beyond the very soft limit alter the leading-order decay dynamics and can potentially alter acceptance studies of experimental results - and therefore measured values for $|V_{\text{cb}}|$ and $|V_{\text{ub}}|$.

Experience from exclusive semileptonic $K$-meson decays illustrate the importance of having a good understanding of radiative effects: until 2004 the global average of the extracted value of $|V_{\text{us}}|$ from $K_{l3}^+$ and $K_{l3}^0$ decays implied the violation of CKM unitarity by two standard deviations \cite{Eidelman:2004wy}. Further measurements proved dissonant with these findings \cite{Sher:2003fb, Alexopoulos:2004sw, Alexopoulos:2004sx, Alexopoulos:2004sy}, indicating that the achieved experimental precision needed an understanding of electromagnetic corrections at the percent level. Since for many decays next-to-leading order calculations do not exist, experiments often use the approximative all-purpose algorithm \texttt{PHOTOS} \cite{Barberio:1990ms,Barberio:1993qi} to study the reconstruction efficiency and acceptance. The accuracy of this approach was tested by the KTeV collaboration, using the measured photon spectra from radiative $K_{l3}^0$ decays: the angular distribution of the simulated photons did not agree well with the predicted spectrum \cite{Andre:2004fs}. This lead to the development of the next-to-leading order Monte Carlo generator \texttt{KLOR} (see \cite{Andre:2004fs}), whose next-to-leading order calculation is based on a phenomenological model. Its predicted angular photon distribution agreed satisfactorily with the measured spectra. Although using next-to-leading order calculations are more preferable than approximative all-purpose algorithms, they are often complicated to adopt for an experiment: electromagnetic next-to-leading order calculations only exist for a few decay modes, sometimes only valid in a limited region of phase-space. 

Over the last 10 years, an increasing amount of data and a better understanding of detector effects leads to a very accurate picture of physics at both $B$-factory experiments BaBar and Belle. This increased precision lead to the demand of knowledge of next-to-leading order electromagnetic effects beyond the precision of approximative all-purpose algorithms. Our paper was written to help closing this important gap and improving the status quo in a twofold way: first by providing a Monte Carlo generator that produces differential decay rates based on a next-to-leading order calculation valid in the low-energy limit of phase space with respect to the photon; and second by predicting the unknown isospin breaking contributions to the next-to-leading order enhancement of the total decay rate. The former can be used to conduct acceptance studies to improve the understanding of the role of such radiative events in the extraction of CKM matrix elements. The latter result give small corrections to the known parton level result, and therefore to the extracted values of $|V_{\text{cb}}|$ and $|V_{\text{ub}}|$ from semileptonic decays. The uncertainties due to corrections beyond the soft limit for the latter are hard to estimate, and we restrict ourselves to conduct studies within the phenomenological model itself. The error budget to the number should therefore be used with care. 

Our model is based on the considerations of \cite{Andre:2004fs} and makes use of some of the formal ideas of \cite{Gasser:2005uq}: The latter calculates the real next-to-leading order effects of semileptonic $K_{l3}$ decays. We use the formalism therein to split the radiative hadronic current in contributions that depend on the tree-level hadronic current and further corrections. The virtual one-loop diagrams are determined in a phenomeonological model are matched to the parton level correction. Our paper proceeds as follows: Section \ref{treelevel} briefly reviews exclusive $B \to X \, l \nu$ decays at tree-level. Section \ref{nloform} thereafter develops the  next-to-leading order formalism and introduces the phenomenological model. Section \ref{nlo} present the next-to-leading order matrix elements and the next-to-leading order differential decay rate. Section \ref{numerics} describes the numerical methods used to integrate the studied next-to-leading order matrix elements. Section \ref{results} presents the results for the next-to-leading order total and differential decay rate predictions, and Section \ref{conclusion} summarizes our result and presents our conclusions\footnote{Charge-conjugated modes are implied throughout this paper and that natural units, $c = 1$ and $\hbar = 1$, are used.}.

%% file: treelevel.tex
\section{Phenomenological tree-level decay revisited}\label{treelevel}
 \begin{figure}[h!]\begin{center}  
\vspace{0.5cm}
 \unitlength = 1mm
 \begin{fmfgraph*}(40,20)
 	\fmfleft{on1,on2,is,on3,on4}
	\fmfright{of1,of2,fs,of3,of4}
	\fmf{plain,tension=2/3}{is,v1,fs}
	\fmf{plain,tension=1/3}{of4,v2,v1}
	\fmf{plain,tension=1/3}{of1,v3,v1}
	\fmf{fermion,tension=0}{of4,v1,of1}
         \fmfblob{0.1w}{v1}     
	\fmflabel{$p_{B}$}{is}  
	\fmflabel{$p$}{fs}  
	\fmflabel{$p_{l}$}{of4} 
	\fmflabel{$p_{\nu}$}{of1}  
	 \fmffreeze
\end{fmfgraph*}
\hspace{2cm}
 \begin{fmfgraph*}(40,25)
	\fmfleft{ll1,ll2,ql1,ql2}
	\fmfright{lr1,lr2,qr1,qr2}
	\fmf{plain,tension=5}{ql1,v0,v1,v2,v3,v4,v5,v6,qr2}
	\fmf{fermion,tension=5}{ql1,v3,qr2}
         \fmf{plain,tension=1}{lr1,w0,w1,w2,v3,w3,w4,w5,lr2}
         \fmf{fermion,tension=1}{lr1,v3,lr2}
         \fmflabel{$q$}{ql1}  
         	\fmflabel{$q'$}{qr2}  
         	\fmflabel{$p_l$}{lr1}  
         	\fmflabel{$p_\nu$}{lr2}  
         \fmfblob{0.1w}{v3}         
       \fmffreeze
\end{fmfgraph*} 
\end{center}
\caption{ The tree-level weak $B \to X l \nu$ decay using Fermi theory is shown: the left hand side shows the phenomenological picture of mesons; the right hand side the decay at parton level ignoring the light spectator quark. The shaded circle in the transition represents the effective Fermi vertex.}   \label{treelevelgraphs}
 \end{figure}
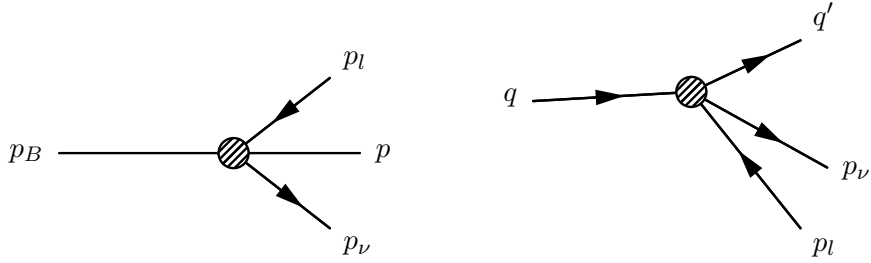  

In the Standard Model the tree-level weak $b \to x l \nu$ decay in Feynman gauge is described by the transition matrix element\footnote{In the following, $\mathcal{M}^n_m$ ($\G^n_m$) denotes the matrix element (total decay rate) at $\mathcal{O}(\alpha^n)$ with \emph{m} photons in the final state, the tree-level matrix element (decay rate) consequently is indicated as $\mathcal{M}^0_0$ ($\G^0_0$) .} 
\bea\label{treelevelmatrixelSM}
 \mathcal{M}^0_{0,\, \text{SM}} & \eq & - i \, \frac{\GF}{\sqrt{2}} \, V_{\text{xb}}  \, \frac{m_W^2}{t - m_W^2} \left[ \bar u (p_\nu)\, \g^\mu \, \pl \,  v(p_l) \right] \, h_\mu  , \nonumber \\
  & \eq &  - i \, \frac{\GF}{\sqrt{2}} \, V_{\text{xb}} \left[ \bar u (p_\nu) \, \g^\mu \, \pl \,  v(p_l) \right] \,  h_\mu + \mathcal{O}\left( \frac{t}{m_W^2} \right) \, ,
\eea
with the left-hand projection operator $\pl = 1 - \g_5$ and the quark current $h_\mu = \left[ \bar x (q') \, \g_\mu \, \pl \, b(q) \right]$. Since the momentum transfer $\sqrt{t}$ through the $W$ propagator is very small compared to the $W$-boson mass $m_W$, terms of $\mathcal{O}\left( t / m_W^2 \right)$ can safely be neglected and the full amplitude can be approximated by the first term in Eqn. (\ref{treelevelmatrixelSM}). This term can also be obtained by an effective Hamiltonian 
\bea\label{partonlevelweakhamiltonian}
 \mathcal{H}_{\text{eff.}} & \eq & \frac{\GF}{\sqrt{2}} \,  V_{\text{xb}}  \, \left[ \psi_\nu \,  \g^\mu \, \pl \, \psi_l \right] \, \hat h_\mu \, ,
\eea
with the quark current $\hat h_\mu$ in position space. At leading order in the $W$ propagator expansion the effective Hamiltonian Eqn. (\ref{partonlevelweakhamiltonian}) and the Standard Model yield identical expressions for the $b \to x \, l \, \nu$ transition amplitude. 
In order to describe the weak decay of the bound state of a $B$-meson, the quark current in Eqn. (\ref{partonlevelweakhamiltonian}) needs to be replaced with a hadronic current $\hat H_\mu$ describing the hadronic matrix element. The resulting matrix element can be described by means of an effective Hamiltonian containing the phenomenological meson fields for the initial $B$-meson and the final state $X$-meson. For the $B \to X \, l \, \nu$ decay into a (pseudo)scalar final state the effective Hamiltonian can be written as
\bea\label{phenoweakhamiltonian}
 \mathcal{H}_{\text{eff.}} & \eq & \frac{\GF}{\sqrt{2}} \, V_{\text{xb}} \, \bar\psi_\nu\, \g_\mu \, \pl \, \psi_l \big[ \bve{\hat f_+ + \hat f_-} \phi \, \partial^\mu \phi_B + \bve{\hat f_+ - \hat f_-} \phi_B \, \partial^\mu \phi \big] \, + \dots ,
\eea
where the ellipses denote further higher order operators. 
The relevant hadronic degrees of freedom are the initial state $B$-meson and the final state $X$-meson and can be described by two scalar fields $\phi_B$ and $\phi$, respectively. 
The resulting amplitude for a $B \to X l \nu$ decay from the leading term in Eqn. (\ref{phenoweakhamiltonian}) is
\bea\label{treelevelmatrixel}
 \mathcal{M}^0_{0} & \eq &  - i \, \frac{\GF}{\sqrt{2}} \, V_{\text{xb}} \left[ \bar u (p_\nu) \, \g^\mu \, \pl \,  v(p_l) \right] \,  H_\mu 
\eea
with 
\bea \label{hadcur}
H_\mu  & = &  \bra X(p) | \, h_\mu \,  | B(p_B) \ket  \, \eq \,   \bve{p_B + p}_\mu \, f_+ + \bve{p_B - p}_\mu \, f_- \, .
 \eea
 The effective Hamiltonian Eqn. (\ref{phenoweakhamiltonian}) reproduces the correct tree-level matrix element. In the hadronic current Eqn. (\ref{hadcur}) the form factors $f_{\pm} = f_{\pm}(t)$ are functions of the four-momentum transfer squared of the $B$-meson to the $X$-meson system, e.g.
\bea\label{squaredmomentumtrans}
 t & \eq & \bve{p_B - p}^2 \, ,
\eea
and describe the influence of the weak decay on the bound state quarks caused by the strong interaction. This dependence can be made explicit in  Eqn. (\ref{phenoweakhamiltonian}) by adding appropriate higher order operators. As a consequence the effective Hamiltonian Eqn. (\ref{phenoweakhamiltonian}) cannot teach us anything about the consitutents of these mesons, but it proofs useful to study low-energy electromagnetic corrections that do not resolve the bound state structure of the involved mesons.The tree-level differential decay rate in the $B$-meson rest frame is given by \bea\label{treeleveldifferentialrate}
\ud \Gamma^0_0 & \eq & \frac{1}{64 \, \pi^3 m_B} \big| \mathcal{M}^0_0 \big|^2 \, \ud E \, \ud E_l \, ,
\eea
with $E = p^0$ and $E_l = p_l^0$, and the explicit expressions of $f_{\pm}$ as a function of the momentum transfer squared for $B \to D \, l \, \nu$, $B \to D_0^* \, l \, \nu$, and   $B \to \pi \, l \, \nu$ decays can be found in App. \ref{appa}.

%% file: nlo_formalism.tex
\section{Next-to-leading order formalism and phenomenological model}\label{nloform}

The aim of this section is to develop a formalism to calculate the corrections at $\mathcal{O}( \alpha \, \GF)$ for $B \to X \, l \, \nu$ decays and introduce a phenomenological model to perform the actual calculation in an approximative manner. First the formalism used for virtual corrections and its connection to renormalization is discussed. Thereafter the splitting of the electromagnetic coupling into parts that can be described by minimal coupling to the leading term in the effective Hamiltonian Eqn. (\ref{phenoweakhamiltonian}), and terms that arise from further (known and unknown) higher order operators, is discussed. In the end of this section our strategy to obtain the approximative corrections at  $\mathcal{O}( \alpha \, \GF)$ is summarized. 

\subsection{Formalism for virtual corrections: short- and long-distance contributions}

Electromagnetic loop corrections to the $B \to X \, l \, \nu$ decay amplitude modify the Hamiltonian Eqn. (\ref{phenoweakhamiltonian}) by adding further terms of $\mathcal{O}(\alpha \, \GF)$. Such corrections to the Hamiltonian are calculated by comparing amplitudes in the full theory with the $W$-boson present as a dynamical degree of freedom to amplitudes in the effective theory with the $W$-boson removed as such. The arising corrections come from regions of loop momenta of order $\backsim m_W$, since the effective Hamiltonian has been constructed in a way to correctly reproduce the full theory for momenta much smaller than $m_W$. The standard approach to proceed involves calculating the one-loop graphs for the $B \to X \, l \, \nu$ decay in the effective theory with counter terms and compare it to the renormalized Standard Model result. Fixing the counter-terms results in the desired matching of both results. The effective theory itself is nonrenormalizable, but the Standard Model can be renormalized to measured quantities, e.g.  Fermi's constant, in order to produce finite predictions. The arising infrared divergencies in such one-loop graphs are canceled with divergencies that show up in the infrared limit of of the $B \to X \, l \, \nu \, \g$ decay. Such a matching procedure was carried out in much detail by \cite{DescotesGenon:2005pw} for semileptonic Kaon decays. For semileptonic $B$-meson decays usually the result of \cite{Sirlin:1977sv,Sirlin:1981ie} is used to correct in a similar manner for virtual corrections of $\mathcal{O}(\alpha \, \GF)$ of the quark current $h_\mu$. Corrections that do not resolve the involved hadrons to parton level are small, but in order to study the full real and virtual corrections at $\mathcal{O}(\alpha \, \GF)$ consistently they need to be included. 

In order to proceed we assume that further (unknown) higher-order operators in Eqn. (\ref{phenoweakhamiltonian}) can be constructed in such a way, such that the full Hamiltonian reproduces the virtual electromagnetic corrections at all energies -- explicitly that its high energy limit reproduces the parton level correction to Eqn. (\ref{partonlevelweakhamiltonian}), i.e. the findings of \cite{Sirlin:1977sv,Sirlin:1981ie}. The ellipses in Eqn. (\ref{phenoweakhamiltonian}) then contain two classes of higher order operators: derivatives of fields that derive the tree-level $t$ dependence, and operators that describe the corrections at  $\mathcal{O}(\alpha \, \GF)$ and maintain the transition from the phenomenological degrees of freedom of hadrons to the partonic degrees of freedom of quarks at high-energies, i.e., the Standard Model picture. 

The large separation of scales of the masses of the involved decay products in contrast to the $W$-boson mass, explicit $m_X, m_B \ll m_W$, allows for an alternative prescription to match such an effective field theory 
to the Standard Model and to obtain a meaningful result with respect to renormalization.
A general divergent $N$-point tensor integral of order $p$ in the effective theory has the form of
\bea
 T^{\mu_1 \dots \mu_p} (p_1, \dots, p_{N-1}) & \propto & \int \ud k^2 \frac{ k^{\mu_1} \dots k^{\mu_p} }{d_0 \dots d_{N-1}} \, ,
\eea
with denominators $d_0 = k^2$ and $d_i = \left( (p_i - k)^2 - m_i^2 \right)$. This integral can be split by an Euclidian regulator $ \mu_0 \backsim m_X,m_B \ll m_W$ into two regions
\bea\label{npointsep}
 T^{\mu_1 \dots \mu_p} (p_1, \dots, p_{N-1}) & \propto & \int_{\lambda^2}^{\mu_0^2} \ud k^2 \frac{ k^{\mu_1} \dots k^{\mu_p} }{d_0 \dots d_{N-1}} +  \int_{\mu_0^2}^{\infty} \ud k^2 \frac{ k^{\mu_1} \dots k^{\mu_p} }{d_0 \dots d_{N-1}} \,,
\eea
where we introduced an infrared regulator $\lambda$. Choosing the Euclidian regulator near $m_X$ and $m_B$ ensures that in the region $k^2 < \mu_0^2$ the relevant degrees of freedom are the bound-state mesons, i.e., the arising dynamic can be described by minimal coupling to the leading term of Eqn. (\ref{phenoweakhamiltonian}) and the tree-level operators.
Large photon momenta resolve the bound state quarks and the corrections from operators that restore the partonic degrees of freedom will no longer be negligible. This qualitative argument is true up to intermediate resonances, that can show up in the energy regime of $k^2 < \mu_0^2$, and breaks down at loop momenta $\backsim \mu_0^2$, when the virtual photon starts to resolve the bound state meson. Note in particular that the first term of Eqn. (\ref{npointsep}) contains only infrared divergencies, whereas the second integral is ultraviolet divergent only.
Using this splitting the virtual next-to-leading order matrix element decomposes as
\bea
\mathcal{M}_0^1 & \eq & \mathcal{M}^1_{0, \, \text{ld}}(\mu_0)+ \mathcal{M}^1_{0, \, \text{sd}} (\mu_0) \, ,
\eea
where $ \mathcal{M}^1_{0, \, \text{ld}}(\mu_0)$ and $ \mathcal{M}^1_{0, \, \text{sd}}(\mu_0)$ denote the contributions from the first and second integral in Eqn. (\ref{npointsep}), respectively. Introducing an Euclidian regulator will in general destroy gauge invariance of the electromagnetic interaction. A splitting that leaves gauge invariance intact is given by
\bea\label{npointseppv}
 T^{\mu_1 \dots \mu_p} (p_1, \dots, p_{N-1}) & \propto & \int_{\lambda^2}^{\infty} \ud k^2 \,  \left( \frac{ k^{\mu_1} \dots k^{\mu_p} }{ d_0 \dots d_{N-1}} -  \frac{  k^{\mu_1} \dots k^{\mu_p} }{\left(k^2 -\mu_0^2 \right) d_1 \dots d_{N-1}} \right) \nonumber \\
  & & +  \int_{\lambda^2}^{\infty} \ud k^2 \,  \frac{  k^{\mu_1} \dots k^{\mu_p} }{ \left(k^2 -\mu_0^2 \right) d_1 \dots d_{N-1}} \, . 
\eea
The second term in the first integral of in Eqn. (\ref{npointseppv})  acts as an additional convergence factor that leaves the low-energy behavior intact, but suppresses  the exchange of virtual photons of momentum larger than $\mu_0$, what introduces an unphysical photon-like vector field with opposite norm, as proposed by Pauli and Villars \cite{Pauli:1949zm}. The photon mass $\mu_0$  leaves the high-energy behavior of the second integral of Eqn. (\ref{npointseppv}) intact, but cuts off the exchange of particles with momentum lower than $\mu_0$. This again renders the first integral of Eqn. (\ref{npointseppv}) ultraviolet finite, and the second integral of Eqn. (\ref{npointseppv}) infrared finite.
One interesting feature of the splitting of Eqn. (\ref{npointseppv}) lies in its properties under renormalization: The term $\mathcal{M}^1_{0, \, \text{sd}}(\mu_0)$ carries the full ultraviolet behavior of $\mathcal{M}_0^1$ and thus can be used to renormalize all parameters. Consequently, because the splitting is exact, all parameters in $\mathcal{M}^1_{0, \, \text{ld}}(\mu_0)$ are renormalized automatically. In the following we will refer to contributions like the first integral of Eqn. (\ref{npointseppv}) as \emph{long-distance} (ld) contribution. If the matching scale is chosen such that $\mu_0 \backsim m_B$ or $m_X$, the electromagnetic interaction at this energy cannot resolve the underlying structure of the involved mesons, and only feels the scalar long-distance coupling -- up to corrections from excited intermediate states and the apparent break down at loop momenta close to $\mu_0$. Contributions like the second integral will be denoted as \emph{short-distance} (sd) contribution to the amplitude, that can resolve the partonic structure of the involved mesons.

\subsection{Formalism for real corrections: inner bremsstrahlung and structure-dependent terms}

In this section we will clarify the idea of long- and short-distance interactions in a more formal way in order to make it applicable for real emissions by splitting the electromagnetic current of the hadronic system into two components: \emph{inner-bremsstrahlung} (IB) contributions, which account for photon radiation from the external charged particles (i.e. described by scalar QED) and are completely determined by the non-radiative process, e.g. the leading operator of Eqn. (\ref{phenoweakhamiltonian}) and further tree-level operators; and \emph{structure-dependent} (SD) contributions, which describe intermediate hadronic states and represent new information with respect to the IB contributions, e.g. from additional higher order operators of  Eqn. (\ref{phenoweakhamiltonian}). This approach will allow us to split the real corrections correspondingly into IB and SD contributions. 

Coupling the electromagnetic current to the semileptonic $B$-meson decay amplitude yields
\bea\label{eleccoupl}
    i\,  e \, \frac{\GF}{\sqrt{2}} V_{\text{xb}} \bar u(p_\nu) \, P_{\text{R}} \, \gamma^\mu  \, \Bigg( -  \frac{ H_{\mu}}{2 p_l \cdot k}   \left( \g^\rho \ds{k}  + 2 p_l^\rho \right)  + V_{\mu\nu} - A_{\mu\nu}  \Bigg) v(p_l) \,  \, ,
\eea
with the hadronic current $H_\mu$, as introduced in Eqn. (\ref{hadcur}). The hadronic vector and axial form factors of the $B / X \, \g$ coupling are given by the unknown non-local operator
\bea\label{nonlocalop}
  V_{\mu\nu} - A_{\mu\nu} & \eq & \int \ud^4 x \, e^{i \, k \cdot x} \, \bra X(p') | T \, [   \hat h_{\mu} (0) \, J^{\text{em}}_\nu(x) ]  \, | B(p_B') \ket \, .
\eea
It is $  J^{\text{em}}_\nu$ the electromagnetic current and $\hat h_{\mu}$ the quark-current in position-space. Eqn. (\ref{nonlocalop}) obeys the electromagnetic Ward-identity \cite{PhysRev.78.182}:
\bea \label{wardident}
 k^\nu \, V_{\mu\nu} & \eq & H_\mu \, , \nonumber \\
 k^\nu \, A_{\mu\nu} & \eq & 0 \, .
\eea
Gauge invariance of the resulting amplitudes reveals a connection between the tree-level and the next-to-leading form factors: Low's theorem \cite{PhysRev.110.974,PhysRevLett.20.86} states that the leading contributions to the next-to-leading order amplitude in powers of the photon four-momentum $k$, explicit $k^{-1}$ and $k^0$, are completely determined by the on-shell form factors of the tree-level decay. The work of \cite{Gasser:2005uq} presents a set of arguments to include corrections beyond $\mathcal{O}(k^0)$, by separating the non-local operator Eqn. (\ref{nonlocalop}) into SD and IB contributions, as depicted in Fig. \ref{sdibdepicted}.
 We briefly restate them:
\begin{itemize}
 \item  In order to describe two different physical mechanisms, the IB and SD amplitudes must be separately gauge invariant;
 \item The SD amplitude contains terms of order $k$ and higher.
\end{itemize}
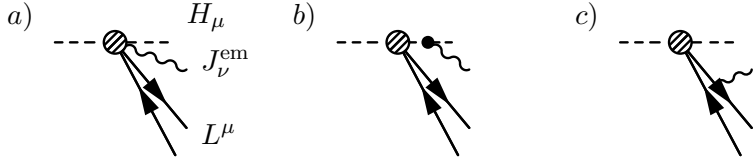
\begin{figure}[h!]\begin{center}  
\vspace{1cm}
 \unitlength = 1mm
 \begin{fmfgraph*}(20,15)
 	\fmfleft{on1,on2,is,on3,on4}
	\fmfright{of1,of2,fs,of3,of4}	
	\fmf{dashes,tension=2/3}{on4,v0,v1,v2,of4}
	\fmffreeze
	\fmf{plain,tension=1/3}{v1,v3,of2}
	\fmf{plain,tension=1/3}{v1,v4,of1}
	\fmf{fermion,tension=0}{of1,v1,of2}
	\fmf{photon,tension=0}{v1,of3}
         \fmfblob{0.15w}{v1}     
	\fmflabel{$a)$$\,\,$}{on4}  
	\fmflabel{$H_\mu$}{of4}  
	\fmflabel{$L^\mu$}{of2} 
	\fmflabel{$J^{\text{em}}_\nu$}{of3}
\end{fmfgraph*} \hspace{1.5cm}
 \begin{fmfgraph*}(20,15)
 	\fmfleft{on1,on2,is,on3,on4}
	\fmfright{of1,of2,fs,of3,of4}	
	\fmf{dashes,tension=2/3}{on4,v0,v1,v2,of4}
	\fmffreeze
	\fmf{plain,tension=1/3}{v1,v3,of2}
	\fmf{plain,tension=1/3}{v1,v4,of1}
	\fmf{fermion,tension=0}{of1,v1,of2}
	\fmf{photon,tension=0}{v2,of3}
         \fmfblob{0.15w}{v1}     
	\fmflabel{$b)$$\,\,$}{on4}  
        \fmfdot{v2}     
\end{fmfgraph*} \hspace{1.5cm}
 \begin{fmfgraph*}(20,15)
 	\fmfleft{on1,on2,is,on3,on4}
	\fmfright{of1,of2,fs,of3,of4}	
	\fmf{dashes,tension=2/3}{on4,v0,v1,v2,of4}
	\fmffreeze
	\fmf{plain,tension=1/3}{v1,v3,of2}
	\fmf{plain,tension=1/3}{v1,v4,of1}
	\fmf{fermion,tension=0}{of1,v1,of2}
	\fmf{photon,tension=0}{v3,of3}
         \fmfblob{0.15w}{v1}     
	\fmflabel{$c)$$\,\,$}{on4}  
\end{fmfgraph*} \hspace{1.5cm}
\vspace{0.25cm}
\end{center}
\caption{The coupling of the electromagnetic current to the hadronic and leptonic currents is shown: \emph{a)} structure-dependent , \emph{b)} scalar coupling to a intermediate particle with a $B$- or $X$-meson pole, and \emph{c)} the leptonic coupling. The shaded circle represents the one-particle irreducible graph of the weak decay.}   \label{sdibdepicted}
 \end{figure}  
The second condition does not prevent the IB amplitude from containing terms of order $k$ and higher. Splitting the amplitude under this restriction allows us to collect more terms in the IB part, by still using only the knowledge of the non-radiative matrix element. This offers the advantage to obtain a more precise predictions for the decay process, without formulating the (mostly unknown) SD contributions. The splitting of the transition matrix element requires a corresponding splitting of the non-local operator Eqn. (\ref{nonlocalop}) into SD and IB parts. 

We first consider the axial contributions: They are strictly zero for photon emissions producing a (pseudo)scalar intermediate $B$- or $X$-meson state \emph{b)} of Fig. \ref{sdibdepicted}, and therefore can be considered purely SD. They can be written in the form \cite{Bijnens:1992en} 
\bea
 A_{\mu\nu} & \eq & A_{\mu\nu}^{\text{SD}} \,\, \eq \,\, - i \, \varepsilon_{\mu\nu\rho\sigma} \left( A_1 \, p^\rho \, k^\sigma  + A_2 \, k^\rho \bve{p_l + p_\nu}^\sigma  \right) \nonumber \\
   & & \ph{ A_{\mu\nu}^{\text{SD}} \,\, \eq \,\,- } - i \, \varepsilon_{\nu\lambda\rho\sigma} \, p^\lambda \, k^\rho \, \bve{p_l + p_\nu}^\sigma  \left( A_3 \, \bve{p_l + p_\nu}_\mu + A_4 \, p_\mu \right) \, .
\eea
Note that the Lorentz-invariant scalars $A_{i}$ are non-singular in the vanishing photon-energy limit by construction and are functions of the three independent scalar variables that can be constructed out of $p_B$, $p$, and $k$. 

The decomposition of the vector current is
\bea\label{veccursplit}
 V_{\mu\nu} & \eq &  V_{\mu\nu}^{\text{IB}} + V_{\mu\nu}^{\text{SD}} \, , 
\eea
where the IB piece is chosen in such a way, that within Eqn. (\ref{wardident}) 
\bea \label{IBWard}
 k^\nu   V_{\mu\nu}^{\text{IB}}  & \eq & H_{\mu} \,,
\eea
and therefore for the SD contributions it is imperative that
\bea
 k^\nu   V_{\mu\nu}^{\text{SD}}  & \eq & 0 \,.
\eea
The decay amplitude separates now as
\bea \label{sdibsplit}
&&    i\,  e \, \frac{\GF}{\sqrt{2}} V_{\text{xb}} \bar u(p_\nu) \, P_{\text{R}} \, \gamma^\mu  \, \Bigg( -  \frac{ H_{\mu}}{2 p_l \cdot k}   \left( \g^\rho \ds{k}  + 2 p_l^\rho \right)  + V^{\text{IB}}_{\mu\nu}  \Bigg) v(p_l) \,  \nonumber \\
 && \ph{- } +  i\,  e \, \frac{\GF}{\sqrt{2}} V_{\text{xb}} \bar u(p_\nu) \, P_{\text{R}} \, \gamma^\mu  \, \Bigg( V^{\text{SD}}_{\mu\nu} - A^{\text{SD}}_{\mu\nu}  \Bigg) v(p_l) \,  \, .
\eea
It remains to construct the explicit form of $ V_{\mu\nu}^{\text{IB}}$, what is done in Sec. \ref{realIB} for the real emission process. For the $B \to B \g$ IB contributions, one finds 
\bea\label{IBBBg}
 \g^\mu \, V_{\mu\nu}^{\text{IB}} & \eq & - \frac{ H_{\mu} }{p_B \cdot k} \, \g^\mu \, \bve{p_B}_\nu   + \frac{ f_3 }{p_B \cdot k} \ds{k}\, \bve{p_B}_\nu \nonumber \\
& &\ph{ - }   - \frac{ Z_{f_{+}} }{p_B \cdot k} \left( 2 \ds{p} + m_l \right)  \bve{p_B}_\nu - \frac{ Z_{f_{-}} }{p_B \cdot k} \left( m_l \right)  \bve{p_B}_\nu   - f_3 \, \g_\nu \nonumber \\ 
 & & \ph{ - }   + \left(2 \ds{p} + m_l \right) \, Z_{f_+ \, \nu}      + \left(m_l \right) \, Z_{f_- \, \nu} \, , 
 \eea
 with 
 \bea
 Z_{f_\pm \, \nu} & \eq &  \frac{\bve{p_B - p}_\nu}{ k \cdot \bve{p_B - p}} \left( f_\pm(t') - f_\pm(t) \right) \, , \quad
 Z_{f_\pm} \,\, \eq \,\, k^\nu \, Z_{f_\pm \, \nu} \, , \nonumber \\
 f_3 &\eq &  \left(f_+(t) + f_-(t)\right) \, ,
\eea
Furthermore $t' \eq \bve{p_B - p - k}^2$ is the radiative four-momentum transfer squared. Similarly the $X \to X \g$ IB contributions are 
\bea\label{IBXXg}
\g^\mu \, V_{\mu\nu}^{\text{IB}} & \eq &   - \frac{ H_{\mu} }{p \cdot k} \, \g^\mu \, p_\nu   - \frac{ f_2 }{p \cdot k} \ds{k}\, p_\nu \nonumber \\
& & \ph{ - }  - \frac{ Z_{f_{+}} }{p \cdot k} \left( 2 \ds{p}_B - m_l \right)  p_\nu - \frac{ Z_{f_{-}} }{p \cdot k} \left( m_l \right)  p_\nu   + f_2 \, \g_\nu \nonumber \\
 & &   \ph{ - }  + \left(2 \ds{p}_B - m_l \right) \, Z_{f_+ \, \nu}      + \left(m_l \right) \, Z_{f_- \, \nu} \,,
 \eea
 with 
 \bea
 f_2 &\eq &  \left(f_+(t) - f_-(t)\right) \, .
\eea
The SD vector contributions can be written as \cite{1999PAN....62..975P}
\bea\label{sdveccur}
V_{\mu\nu}^{\text{SD}}  & \eq &  V_1 \left( k_\mu \, p_\nu - p \cdot k \, g_{\mu\nu}   \right) + V_2 \left( k_\mu \bve{p_l + p_\nu}_\nu  - k \cdot \bve{p_l + p_\nu} \, g_{\mu\nu} \right) \nonumber\\
 & & \ph{ - } + V_3 \left( k \cdot \bve{p_l + p_\nu} \, \bve{p_l + p_\nu}_\mu \, p_\nu - p \cdot k \, \bve{p_l + p_\nu}_\mu \, \bve{p_l + p_\nu}_\nu   \right) \nonumber \\
 & & \ph{ - } + V_4 \left( k \cdot \bve{p_l + p_\nu} \,  p_\mu \, p_\nu - p \cdot k \, p_\mu \, \bve{p_l + p_\nu}_\nu   \right) \, ,
\eea
where the Lorentz-invariant scalars $V_{i}$ are functions of the three independent scalar variables that can be constructed out of $p_B,p$, and $k$. The IB vector currents Eqns. (\ref{IBBBg}) and (\ref{IBXXg}) satisfy the Ward-identity Eqn. (\ref{IBWard}) and are constructed in such a way, that all infrared (IR) singular parts in the limit of $p \cdot k = 0$, or $p_B \cdot k = 0$, respectively, are contained within.

An alternative choice for the IB and SD separation used in this work is given by \cite{PhysRevD.72.094021} based on \cite{PhysRevD.2.542}. It can be obtained by a change of basis in Eqn. (\ref{sdveccur}), and correspondingly shifting terms of $\mathcal{O}(k)$ and higher into the SD contributions\footnote{Note that there is a subtlety in the used notation: IB and SD will always refer to $V_{\mu\nu}^{\text{IB}}$, and $V_{\mu\nu}^{\text{SD}} - A_{\mu\nu}^{\text{SD}}$, what should not be confused with the particular choice of the IB terms, Eqns. (\ref{IBBBg}) and (\ref{IBXXg}), which incorporates a number of \emph{structure-dependent} terms.}.

\subsection{Phenomenological model}\label{phenomodel}

So fare we did not introduce any model dependence and assuming one would have full knowledge of the effective Hamiltonian, that governs the semileptonic transition at $\mathcal{O}(\alpha \, \GF)$, one could now go ahead and perform the necessary calculations following the baselines for the real and virtual corrections outlined in the last two sections. However, the knowledge of the operators that connect the short-distance and long-distance domains, i.e. the SD contributions, is modest: The recent work of \cite{Becirevic:2009fy} addresses the real SD corrections to $B \to D \, l \, \nu \, \g$ decays by using lattice results of the $D^* \to D \, \g$ coupling to model the first dominant SD contributions, in the region of phase-space, where the $D^*$ is an on-shell intermediate resonance. The authors of \cite{PhysRevD.72.094021} discuss the matter for $B\to \pi \, l \, \nu \, \g$ decays, using soft-collinear effective theory (SCET) to isolate the expressions for the SD contributions in the soft-pion and hard-photon part of phase-space. The SD corrections to $B\to D^*_0 \, l \, \nu \, \g$ are unknown, but given the large decay widths of the $D^*_0$- and $D^*_1$-mesons, a considerable correction to the low-energy result of the pure IB prediction can be expected. This is a daunting starting point, but nonetheless allows one to examine the low-energy limit of the effective Hamiltonian, where SD contributions play only a minor role, and extrapolate these findings to the complete phase space. Note that this Ansatz -- which we describe as 'phenomenological model' in the following -- is superior to the use of all-purpose algorithms, that make use of the universality of electromagnetic corrections in the soft-limit: Our considerations are based on a (at last in principle) full process dependent next-to-leading order calculation valid near the soft-limit and extrapolated to complete phase space. 

\subsubsection{Virtual corrections}\label{virtualmodel}

In order to obtain finite predictions for the overall next-to-leading order rate, the second integral of Eqn. (\ref{npointseppv}) describing the short-distance interaction has to be regularized and renormalized to the Standard Model, all parameters describing the long-distance interaction are then renormalized automatically. This is exact as long as the complete Hamiltonian Eqn. (\ref{phenoweakhamiltonian}) with all higher order operators is used to calculate both the short-distance and the long-distance parts. In practice however, as discussed before, this is not possible since the higher order operators moderating the transition from long-distance to short-distance interactions are unknown. Hence supposing the matching scale $\mu_0$ is chosen in such a manner, that it effectively separates the long- and short-distance regimes, the long-distance corrections can be calculated using the leading operator of the effective Hamiltonian Eqn. (\ref{phenoweakhamiltonian}) and the known tree-level higher-order operators: in this phenomenological model the Standard Model and its effective degrees of freedom are described by scalar fields, what is justified at low scales. For the short-distance corrections $\mathcal{M}^1_{0,\,\text{sd}}$ the full Standard Model has to be invoked. This mere fact, however, directly leads to inconsistencies at the matching scale $\mu_0$, where both models should give the same answer. This renders the matching approximative and the impact of the choice of the matching scale can be studied by varying the matching scale $\mu_0$. The long-distance virtual corrections are further discussed along with the renormalization of the short-distance correction in Section \ref{virtualIB}. 

\subsubsection{Real corrections}\label{realmodel}

In order to obtain predictions for the real next-to-leading order corrections, we make use of the previously introduced separation into IB and SD contributions. We will neglect the largely unknown SD contributions. This is justified in the low-energy region of phase-space, which is dominant for the emission process. Only at high energies the emitted real photons resolves the bound state nature of the involved mesons, and the SD contributions become dominant over the IB ones. This reasoning, however, excludes intermediate excited states. In order to study the impact of such, the pure IB result presented in this work is compared to the findings of \cite{Becirevic:2009fy} and \cite{PhysRevD.72.094021} in Sec. \ref{realSDeffects}. Note that neglecting any SD contributions results in a model dependent choice of the IB contributions, what is discussed along with the construction of Eqns. (\ref{IBBBg}) and (\ref{IBXXg}) in Sec. \ref{realIB}.

%% file: nlo.tex
\section{Next-to-leading order calculation}\label{nlo}

In this section Eqns. (\ref{IBBBg}) and (\ref{IBXXg}) and all long-distance one-loop graphs are derived using the phenomenological model outlined in the previous section. In addition the approximative expression for the summed long- and short-distance next-to-leading order decay rate is obtained. 


\subsection{Real next-to-leading order corrections}\label{realIB}
\begin{figure}[h!]\begin{center}  
\vspace{0.5cm}
 \unitlength = 1mm
 \begin{fmfgraph*}(30,20)
 	\fmfleft{on1,on2,is,on3,on4}
	\fmfright{of0,of1,of2,fs,of3,of4,of5}
	\fmf{dashes,tension=2/3}{is,v0,v1,v2,fs}
	\fmf{plain,tension=1/3}{of4,v3,v1}
	\fmf{plain,tension=1/3}{of1,v4,v1}
	\fmf{fermion,tension=0}{of4,v1,of1}
         \fmf{photon, tension=0}{v0,of5}
         \fmfblob{0.1w}{v1}     
         	\fmflabel{$p_{B}$}{is}  
	\fmflabel{$k,\epsilon^*$}{of5}
	\fmflabel{$p$}{fs}  
	\fmflabel{$p_{l}$}{of4} 
	\fmflabel{$p_{\nu_{l}}$}{of1}  
	 \fmffreeze
	  \fmfdot{v0}     
	\fmflabel{\emph{a)}}{on4} 
\end{fmfgraph*} 
\hspace{1.cm}
 \begin{fmfgraph*}(30,20)
 	\fmfleft{on1,on2,is,on3,on4}
	\fmfright{of0,of1,of2,fs,of3,of4,of5}
	\fmf{dashes,tension=2/3}{is,v0,v1,v2,fs}
	\fmf{plain,tension=1/3}{of4,v3,v1}
	\fmf{plain,tension=1/3}{of1,v4,v1}
	\fmf{fermion,tension=0}{of4,v1,of1}
         \fmf{photon, tension=0}{v2,of3}
         \fmfblob{0.1w}{v1}     
	  \fmfdot{v2}     
	 \fmffreeze
	\fmflabel{\emph{b)}}{on4} 
\end{fmfgraph*}\\ \vspace{1.cm}
 \begin{fmfgraph*}(30,20)
 	\fmfleft{on1,on2,is,on3,on4}
	\fmfright{of0,of1,of2,fs,of3,of4,of5}
	\fmf{dashes,tension=2/3}{is,v0,v1,v2,fs}
	\fmf{plain,tension=1/3}{of4,v3,v1}
	\fmf{plain,tension=1/3}{of1,v4,v1}
	\fmf{fermion,tension=0}{of4,v1,of1}
         \fmf{photon, tension=0}{v1,of3}
         \fmfblob{0.1w}{v1}     
	 \fmffreeze
	\fmflabel{\emph {c)}}{on4} 
\end{fmfgraph*}
\hspace{1.cm}
 \begin{fmfgraph*}(30,20)
 	\fmfleft{on1,on2,is,on3,on4}
	\fmfright{of0,of1,of2,fs,of3,of4,of5}
	\fmf{dashes,tension=2/3}{is,v0,v1,v2,fs}
	\fmf{plain,tension=1/3}{of4,v3,v1}
	\fmf{plain,tension=1/3}{of1,v4,v1}
	\fmf{fermion,tension=0}{of4,v1,of1}
         \fmf{photon, tension=0}{v3,of3}
         \fmfblob{0.1w}{v1}     
	 \fmffreeze
	\fmflabel{\emph{d)}}{on4} 
\end{fmfgraph*}
\hspace{1.cm}
\end{center}
\caption{The real emission contributions are shown: \emph{a)} and \emph{b)} emission from point-like mesons, \emph{c)} from structure-dependent contributions and \emph{d)} from the lepton leg. In all shown diagrams the shaded circle represents the Fermi coupling.}   \label{legem}
 \end{figure}
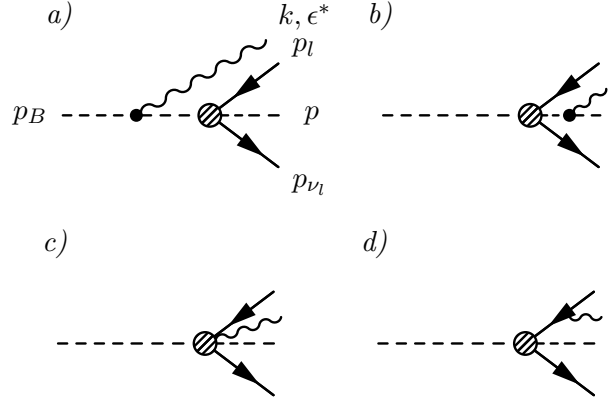  
 
The real emission from all external legs and structure-dependent contributions are depicted in Fig. \ref{legem}. With the photon coupling to the lepton leg, as given by Eqn. (\ref{eleccoupl}), the corresponding emission matrix element \emph{d)} is
 \bea\label{lepemissionmatrixel}
  \mathcal{M}^{\half}_1 & \eq &  i\,  e \, \frac{\GF}{\sqrt{2}} V_{\text{xb}} \bar u(p_\nu) \, \gamma^\mu \, \pl \,  \Bigg( - \frac{\ds{\epsilon}^* \ds{k}}{2 p_l \cdot k + \lambda^2}  + \frac{p_l \cdot \epsilon^*}{ p_l \cdot k + \half \lambda^2}\Bigg) v(p_l) \,  H_{\mu} \, ,
  \eea
  where $\epsilon^* = \epsilon^*(k)$ denotes the polarization vector of the real final state photon. It is further discussed along with the infrared regulator $\lambda$ in Sec. \ref{regirdiv}. 
  
  The emissions from the hadronic leg now have to be constructed in such a way that Eqn. (\ref{IBWard}) can be fulfilled. The intermediate propagator, \emph{a)}, with a four-momentum squared  $(p_B - k)^2$ generates a pole at $m_B$, which corresponds to an intermediate $B$-meson state. We isolate the contribution of this pole in the hadronic emission matrix element:
  \bea\label{hademissionmatrixelis}
     \mathcal{M}^{\half}_{1}   & \eq &i\,  e \frac{\GF}{\sqrt{2}} V_{\text{xb}}   \bar u(p_\nu) \, \gamma^\mu \, \pl \, v(p_l)  \, \left( - \frac{ p_B \cdot \epsilon^*}{p_B \cdot k + \half \lambda^2} \, H_{\mu} (p_B - k , p, t') - \tilde V_{\mu\nu} \, \epsilon^{* \, \nu}  \right) \, ,
 \eea
  with $t' \eq \bve{p_B - p - k}^2$. Similarly the intermediate propagator, \emph{b)}, with a four-momentum squared $(p+k)^2$ generate a pole at $m$, corresponding to an intermediate $X$-meson state. Isolating the pole in the hadronic emission matrix element results in
 \bea\label{hademissionmatrixelfs}
 \mathcal{M}^{\half}_1  & \eq & i\, e \frac{\GF}{\sqrt{2}} V_{\text{xb}}  \bar u(p_\nu) \, \gamma^\mu \,  \pl \, v(p_l) \, \left(- \frac{ p \cdot \epsilon^*(k)}{p \cdot k + \half \lambda^2} \, H_{\mu} (p_B , p + k, t')  - \tilde V_{\mu\nu} \, \epsilon^{* \, \nu}  \right)  \, .
\eea
The form factors in the hadronic currents $H_{\mu} (p_B , p + k, t')$ and $ H_{\mu} (p_B - k , p, t')$ are \emph{a priori} unrelated to the ones in the tree-level hadronic current $H_{\mu}$. They are
 \bea \label{hademcur1}
 H_{\mu} (p_B - k , p, t') & \eq & \bve{p_B + p}_\mu \, \tilde f_{+}(t') + \bve{p_B - p}_\mu \, \tilde f_{-}(t') - k_\mu \, \tilde f_{3}(t')  \, , \\ \label{hademcur2}
 H_{\mu} (p_B , p + k, t') & \eq & \bve{p_B + p}_\mu \, \tilde f_{+}(t') + \bve{p_B - p}_\mu \, \tilde f_{-}(t') + k_\mu \, \tilde  f_2(t') \, .
\eea 
with $\tilde f_\pm$ being the off-shell form factors. It is easy to see, that the off-shell form factors can only differ from the on-shell ones by contributions that are regular at $k=0$, which can be reabsorbed in the SD terms within $ \tilde V_{\mu\nu} $. This is true under the assumption that the limit $ \lim_{k \to 0 } \tilde f_\pm$ is well defined and $ \tilde V_{\mu\nu} $ does not possess any poles in $k=0$ . 
 Then 
\bea \label{wardidentitylimit}
 k_\nu \, V^{\mu\nu} & \eq & H'^\mu + k_\nu \, \tilde V^{\mu\nu} \,\, \eq \,\, H^\mu \, ,
\eea 
with $H'_\mu = H_{\mu} (p_B - k , p, t')$ or $H_{\mu} (p_B , p + k, t')$. Since $ k_\nu \, \tilde V^{\mu\nu} $ is regular, it cannot contain any terms proportional to $H^\mu$ and it must be
\bea\label{fpmhatexp}
\tilde f_{\pm}(t') & \eq & f_{\pm}(t) + k_\alpha \, Z_{f_{\pm}}^\alpha \, ,
\eea 
with $Z_{f_{\pm}}^\nu$ some arbitrary functional which we are free to choose. We construct $Z_{f_{\pm}}^\nu$ in such a way, that $\tilde f_{\pm}(t') = f_{\pm}(t')$ and the matrix element describes the emission diagrams \emph{a)} or \emph{b)}. The freedom of choosing $Z_{f_{\pm}}^\nu$ is nothing more than deciding what contributions we consider as a self-contained emission diagram. Another choice would just move contributions from \emph{a)} and \emph{b)} into \emph{c)} or vice versa. It remains to prove that  $\tilde V_{\mu\nu}$ can be constructed in such a way, that it possesses no poles. The Ward identity implies
\bea \label{wardidentitymagic}
 \lim_{k \to 0} \, k_\nu\, V^{\mu\nu} & \eq &  \lim_{k \to 0} H'^\mu + \lim_{k \to 0} k_\nu \, \tilde V^{\mu\nu} \,\, \eq \,\, H^\mu   \, .
\eea
 If $\tilde V^{\mu\nu}$ is regular at $\lim_{k \to 0} k_\nu \, \tilde V^{\mu\nu} = 0$ then our assumption is trivially true. We therefore assume, that $\tilde V^{\mu\nu}$ has a pole at $k=0$ and write $\lim_{k \to 0} k_\nu \, \tilde V^{\mu\nu} = \tilde V^{\mu}$:
\bea
 \lim_{k \to 0} \, k_\nu\, V^{\mu\nu} & \eq & \bve{p_B + p}^\mu \,  \lim_{k \to 0} \, \tilde f_{+}(t') + \bve{p_B - p}^\mu \,  \lim_{k \to 0} \, \tilde f_{-}(t')  + \tilde V^{\mu} \, , \\
 & \eq &  \bve{p_B + p}^\mu \,  f_{+}(t) + \bve{p_B - p}^\mu \,  f_{-}(t)  \, ,
\eea
where we made use of our premise, that the limit $ \lim_{k \to 0 } \tilde f$ is well defined and finite. When $ \lim_{k \to 0 } \tilde f_{\pm}(t') = f_{\pm}(t)$ it follows  $\tilde V^{\mu} = 0$ and our initial assumption is true. We therefore assume $ \lim_{k \to 0 } \tilde f_{\pm}(t') \neq f_{\pm}(t)$, then 
\bea
 \tilde V^\mu & \eq &  \bve{p_B + p}^\mu \, \left(  f_{+}(t) - \tilde f_{+}(t) \right) + \bve{p_B - p}^\mu \, \left(  f_{-}(t) -   \tilde f_{-}(t) \right) \, .
\eea
We can split $ \tilde V^{\mu\nu}$ into its singular and regular part: $ \tilde V^{\mu\nu} \eq  \tilde V_{\text{singular}}^{\mu\nu} +  \tilde V_{\text{regular}}^{\mu\nu}$, such that $\lim_{k \to 0} k_\mu \, \tilde  V_{\text{regular}}^{\mu\nu} = 0$ and $ \lim_{k \to 0}k_\mu \,  \tilde V_{\text{singular}}^{\mu\nu} = \tilde V_{\text{singular}}^{\mu}$. But then we can simply redefine our hadronic-current as
\bea
H''^\mu  & \eq &  H'^\mu + \, k_\nu \tilde V_{\text{singular}}^{\mu\nu}  \, , \quad \text{with form factors such that} \quad \lim_{k \to 0} \tilde f_{\pm}(t') = f_{\pm}(t) \, ,
\eea 
and our initial assumption is trivially true.  This proves that one can construct Eqns. (\ref{hademissionmatrixelis}) and (\ref{hademissionmatrixelfs}) in such a way, that $\tilde V_{\mu\nu}$ is regular at $k = 0$ and $\tilde f_{\pm}(t') = f_{\pm}(t')$ resulting in the emission graphs \emph{a)} and \emph{b)}. The premise that $\lim_{k \to 0} \tilde f_\pm$ is regular can always be fulfilled by shifting any singular contribution into $\tilde V_{\mu\nu}$. The latter must then contain a singular part itself and cancel the pole of  $\tilde f_\pm$ exactly, otherwise Eqn. (\ref{wardidentitylimit}) cannot be fulfilled, and gauge invariance would be spoilt. 

The authors of \cite{Gasser:2005uq} used another Ansatz to proof the splitting of Eqn. (\ref{fpmhatexp}) with $\tilde f_{\pm}(t') = f_{\pm}(t')$ for their studies of the radiative corrections to $K_{e3}$-decays by checking its validity at one-loop in Chiral-perturbation theory ($\chi$PT). Given this observation, the authors of \cite{Gasser:2005uq} conclude that such a separation is a general feature of QCD.
    
 It remains to explicitly construct the residual IB contributions through the Ward identity. Using Eqn. (\ref{fpmhatexp}), the Ward identity  and
 \bea
Z^\alpha_{f_\pm}(t,k)  =  \frac{\bve{p_B - p}^\alpha}{ k \cdot \bve{p_B - p} } \left( f_{\pm}(t') - f_{\pm}(t)  \right)     \, ,
\eea
the missing contributions can be shown to be
\bea
k_\nu \, \tilde V^{\mu\nu} & \eq & -  \bve{p_B + p - k}^\mu \, Z_{f_{+}}  - \bve{p_B - p - k}^\mu \, Z_{f_{-}} - k^\mu \, f_{3}(t) + k_\nu \, V^{\mu\nu}_{\text{SD}} \, , 
\eea
and
\bea
k_\nu \, \tilde V^{\mu\nu} & \eq & -  \bve{p_B + p + k}^\mu \, Z_{f_{+}}  - \bve{p_B - p - k}^\mu \, Z_{f_{-}} + k^\mu \, f_{2}(t) + k_\nu \, V^{\mu\nu}_{\text{SD}} \, .
\eea
Using the Dirac equation, four-momentum conservation and the inverse Ward-transformation $k^\nu \to \epsilon^{*\, \nu}$ result in the complete emission matrix elements:
\bea \label{summedmode1}
\sum \mathcal{M}^\half_1 & \eq &    i\,  e \, \frac{\GF}{\sqrt{2}} V_{\text{xb}} \bar u(p_\nu) \, P_{\text{R}}  \,  v(p_l)  \, \Bigg( - \frac{ H_{\mu}}{2 p_l \cdot k} \, \g^\mu \,  \left( \g^\nu \ds{k}  + 2 p_l^\rho \right)  - \frac{ H_{\mu} }{p_B \cdot k} \, \g^\mu \, \bve{p_B}_\nu   + \frac{ f_3 }{p_B \cdot k} \ds{k}\, \bve{p_B}_\nu \nonumber \\
& &  \ph{   i\,  e \, \frac{\GF}{\sqrt{2}} V_{\text{xb}} \bar u(p_\nu) \, P_{\text{R}}  \,  v(p_l)  \, \Bigg( - } - \frac{ Z_{f_{+}} }{p_B \cdot k} \left( 2 \ds{p} + m_l \right)  \bve{p_B}_\nu - \frac{ Z_{f_{-}} }{p_B \cdot k} \left( m_l \right)  \bve{p_B}_\nu   - f_3 \, \g_\nu \nonumber \\
 & & \ph{   i\,  e \, \frac{\GF}{\sqrt{2}} V_{\text{xb}} \bar u(p_\nu) \, P_{\text{R}}  \,  v(p_l)  \, \Bigg( - }    + \left(2 \ds{p} + m_l \right) \, Z_{f_+ \, \nu}      + \left(m_l \right) \, Z_{f_- \, \nu}  + \g^\mu \, \left( V_{\mu\nu}^{\text{SD}} -   A_{\mu\nu}^{\text{SD}} \right) \Bigg) \, \epsilon^{*\, \nu}(k)  \, , \nonumber \\
\eea
and
\bea \label{summedmode2}
\sum \mathcal{M}^\half_1 & \eq &    i\,  e \, \frac{\GF}{\sqrt{2}} V_{\text{xb}} \bar u(p_\nu) \, P_{\text{R}}  \,  v(p_l)  \, \Bigg( - \frac{ H_{\mu}}{2 p_l \cdot k} \, \g^\mu \,  \left( \g^\nu \ds{k}  + 2 p_l^\rho \right)  - \frac{ H_{\mu} }{p \cdot k} \, \g^\mu \, p_\nu   - \frac{ f_2 }{p \cdot k} \ds{k}\, p_\nu \nonumber \\
& &  \ph{   i\,  e \, \frac{\GF}{\sqrt{2}} V_{\text{xb}} \bar u(p_\nu) \, P_{\text{R}}  \,  v(p_l)  \, \Bigg( - } - \frac{ Z_{f_{+}} }{p \cdot k} \left( 2 \ds{p}_B - m_l \right)  p_\nu - \frac{ Z_{f_{-}} }{p \cdot k} \left( m_l \right)  p_\nu   + f_2 \, \g_\nu \nonumber \\
 & & \ph{   i\,  e \, \frac{\GF}{\sqrt{2}} V_{\text{xb}} \bar u(p_\nu) \, P_{\text{R}}  \,  v(p_l)  \, \Bigg( - }    + \left(2 \ds{p}_B - m_l \right) \, Z_{f_+ \, \nu}      + \left(m_l \right) \, Z_{f_- \, \nu}  +  \g^\mu \, \left( V_{\mu\nu}^{\text{SD}} -   A_{\mu\nu}^{\text{SD}} \right)  \Bigg) \, \epsilon^{*\, \nu}(k)  \, , \nonumber \\
\eea
respectively. 
Neglecting the $V_{\mu\nu}^{\text{SD}} -   A_{\mu\nu}^{\text{SD}} $ contributions results in an approximative description of the emission process, that is exact near the soft-photon limit, where SD contributions do not contribute: The matrix elements Eqn. (\ref{summedmode1}) and (\ref{summedmode2}) then treat the mesons as point-like particles with \emph{structure-dependent} corrections.

\subsection{IR divergencies}\label{regirdiv}

Following the standard prescription of QED, we introduce a small photon mass $\lambda$, that cancels exactly with the IR divergencies produced by the virtual diagrams. Regularizing these divergencies allow for the separate evaluation of the real and virtual matrix elements. The artificial dependence on $\lambda$ cancels when both are summed. Such a massive photon possesses a non-vanishing longitudinal polarization, what has to be taken into account. The three polarization vectors of a real photon emitted in the \bve{0,0,0,1} direction are 
\bea
 \epsilon_1^*(k) = \frac{1}{\sqrt{2}} (0,1,i,0) \, , \qquad \epsilon_2^*(k) = \frac{1}{\sqrt{2}} (0,1,-i,0) \,, \quad \, \text{and} \quad \, \epsilon_3^*(k) = \frac{1}{\lambda} (\sqrt{k^2 + \lambda^2},0,0,k) \, . \nonumber \\
\eea

\subsection{Virtual corrections}\label{virtualIB}

In this section, the short- and long-distance next-to-leading order virtual corrections are derived, following the splitting of Eqn. (\ref{npointseppv}) and the set of arguments of Sec. \ref{phenomodel}: the short-distance corrections are calculated within the Standard Model and the long-distance corrections modeled with the leading operator of Eqn. (\ref{phenoweakhamiltonian}) and the known higher order tree-level operators responsible for the $t$ dependence. The long-distance virtual diagrams are calculated in the on-shell renormalization scheme and at first regularized using dimensional regularization. We use the standard notation of Passarino and Veltman \cite{Passarino:1978jh} to express the arising loop-integrals as $A$, $B$, and $C$ functions, briefly summarized in App. \ref{appc}. The matching of the long-distance loop-integrals to the short-distance result is done via the Pauli-Villars prescription \cite{Pauli:1949zm} and the transformation of the long-distance loop-integrals from dimensional regularization to Pauli-Villars is further discussed in Sec. \ref{pvregularization}.

\subsubsection{Short-distance corrections and renormalization of $\GF$}\label{Sec:shortdistcorrGF}

The electroweak virtual corrections to Eqn. (\ref{partonlevelweakhamiltonian}) at parton level were calculated and renormalized to the Standard Model by \cite{Sirlin:1977sv,Sirlin:1981ie}. The author made use of the short-distance expansion of the Standard Model valid at high energies, such that no leading corrections from the strong interaction occur. Although the latter presents a neat feature, it is beyond next-to-leading-order and therefore beyond the order of correction we are interested to examine. The relevant tree-level matrix element in \cite{Sirlin:1977sv,Sirlin:1981ie} is given by the leading operator of  Eqn. (\ref{partonlevelweakhamiltonian}),
\bea\label{partontree}
 \mathcal{\hat M}^0_{0} & \eq &  - i \, \frac{\hat G_{\text{F}}}{\sqrt{2}} \, V_{\text{xb}} \, \left[  \bar u (p_\nu) \, \g^\mu \pl \, v(p_l) \right] \, h_\mu  \,  \quad \text{with} \quad \, h_\mu =  \bar x(q') \, \g_\mu \, \pl \, b(q) \,,
\eea
where the hats indicate not renormalized quantities. The full electroweak corrections to Eqn. (\ref{partontree}) involve the exchange of virtual photons, $W$- and $Z$-bosons, and Higgs scalars (since we are working in Feynman gauge). The electromagnetic diagrams relevant for the short- and long-distance photon loop-integral splitting of Eqn. (\ref{npointseppv}) are depicted in Fig. \ref{legvirtshort}: Diagrams \emph{a)} - \emph{c)} are inter-particle photon exchange diagrams between the lepton and the incoming or outgoing quark-line; diagrams \emph{d)} - \emph{f)} depict wave-function renormalizations; and diagrams \emph{g)} - \emph{i)} are corrections due to the coupling to the $W$-boson. The leading logarithm of the overall corrections is given by
\bea\label{sirlinresult}
 \mathcal{\hat M}^1_{0,\, \text{sd}} & \eq & \frac{\alpha \, \hat G_{\text{F}} }{4 \pi} \left[ 3 \ln \frac{m_W}{\mu_0} + 6 | \bar Q| \ln \frac{m_W}{\mu_0} - 3 | \bar Q| \ln \frac{m_W^2}{m_Z^2} + \dots   \right] \mathcal{\tilde M}^0_0 \, ,
\eea
with $\mathcal{\hat M}^0_0 =  \hat G_{\text{F}} \mathcal{ \tilde M}^0_0$, i.e. the bare leading order matrix element stripped off the bare Fermi coupling constant $\hat G_{\text{F}}$; it is $\bar Q = \pm \frac{1}{6}$ the averaged electromagnetic charge of the decaying quark line, and the ellipses denote further non-leading terms. Following the original prescription of \cite{Sirlin:1977sv,Sirlin:1981ie}, the arising ultraviolet divergence in Eqn. (\ref{sirlinresult}) has been regularized by the replacement  
\bea
 \frac{1}{k^2} \to \frac{1}{k^2 - \mu_0^2} \frac{m_W^2}{m_W^2 -k^2} \, ,
\eea
of the photon propagator, leading to a Pauli-Villars regularization, that diverges as the $W$-boson mass $m_W$ is taken to infinity. In addition a infrared regulator $\mu_0$ was introduced as a photon mass. The ultraviolet regularization in terms of the $W$-boson mass in Eqn. (\ref{sirlinresult}) is \emph{a priori} not problematic, as long as a renormalization procedure of the bare Fermi coupling  $\hat G_{\text{F}}$ involves a matching to a measured Fermi coupling obtained via the same regularization procedure, such that the explicit (but artificial) dependence in the leading logarithm of the $W$-boson mass cancels. Such a renormalization is calculated by the same author for the muon decay, 
\bea
 \GF & \eq & \hat G_{\text{F}} \left[ 1 + \frac{3\alpha}{8} \ln \frac{m_W^2}{m_Z^2} + \dots \right] \, ,
\eea
where $\GF$ is the measured Fermi decay constant. This leads to the desired short-distance virtual matrix element
\bea\label{m10sd}
 \mathcal{M}^1_{0,\,\text{sd}}& \eq & \mathcal{M}^0_0 \Big[ \frac{3}{4} \frac{\alpha}{\pi}\bve{1 + 2 | \bar Q|} \ln \frac{m_Z}{\mu_0} \Big] + \dots \, , 
\eea
where the ellipses denote further non-leading terms, and the leading matrix element $\mathcal{M}^0_0$ contains the renormalized Fermi coupling $\GF$. The necessity of the involved virtual photons to resolve the involved mesons of the decay usually leads to the choice of a hadronic scale for $\mu_0$: A choice often used for semileptonic $B$-meson decays is given by the $b$-quark mass in the $\overline{\text{MS}}$ scheme: $\mu_0 = m_b =  \overline m_b (\overline m_b)$. The logarithm in Eqn. (\ref{m10sd}) represents then the short-distance corrections due to virtual particle exchange ranging from energies between $m_b$ and $m_Z$. In order to obtain the desired separation of long- and short-distance scales in our model, we choose a lower matching scale and fix $\mu_0 = m_X$ for all decay modes. The logarithm in Eqn. (\ref{m10sd}) represents then the short-distance corrections due to virtual particle exchange ranging from $m_X$ to $m_Z$. 

\subsubsection{Next-to-next-to-leading order corrections at $\mathcal{O}( \alpha \, \alpha_s  \, \GF)$ }\label{Sec:strongcorrections}

QCD loop corrections change the form of the effective Hamiltonian of the matrix element Eqn. (\ref{partontree}) to 
\bea
 \mathcal{H}_{\text{eff.}} & \eq & \frac{\GF}{\sqrt{2}} \, V_{\text{xb}} \, C\left[\frac{m_W^2}{\mu^2}, \alpha_s(\mu^2)\right] \, \left[ \psi_\nu \,  \g^\mu \, \pl \, \psi_l \right] \, \hat h_\mu  \, ,
\eea
with $C\left[\frac{m_W^2}{\mu^2}, \alpha_s(\mu^2)\right] = 1 + \mathcal{O}\left(\alpha_s(\mu^2)\right)$. At one-loop level, assuming three colors and six flavors, the running of the strong coupling constant is given by $\alpha_s(\mu^2) = 4 \pi / \left(\beta_0 \, \ln \frac{\mu^2}{\Lambda_{QCD}^2}\right)$ with $\beta_0 = 7$. Then the correction due to strong interaction effects is approximatively \cite{Sirlin:1977sv,Sirlin:1981ie}
\bea\label{Eq:strongcorrestimator}
 \mathcal{M}_{0,\, \text{sd}}^{2} & \eq & \frac{3 \alpha}{4\pi^2} | \bar Q| \, I_{\alpha_s} \, \mathcal{M}_0^0 \, ,
\eea
with 
\bea
  I_{\alpha_s} & \eq &  - \int_{\mu_0^2}^{m_W^2} \, \ud k^2 \left( \frac{1}{k^2} \, \alpha_s(k^2)  \right) \, .
\eea
With $m_W = 80.34$ GeV, $\Lambda_{\text{QCD}} = 0.2$ GeV and $\mu_0 = m_b = 4.2$ GeV this leads to a correction of $-0.03\%$ to the transition probability predicted by Eqn. (\ref{m10sd}). Therefore for $\mu_0 = m_b$ strong correction effects, formally at next-to-next-to-leading order and $\mathcal{O}( \alpha \, \alpha_s  \, \GF)$ turn out to be completely negligible. The estimated corrections at  $\mathcal{O}(\GF \, \alpha \, \alpha_s)$   analogous to Eqn. (\ref{Eq:strongcorrestimator}) for $\mu_0$ for $B \to D \, l \, \nu \, (\g)$ and $B \to D^*_0 \, l \, \nu \, (\g)$ are summarized in Table \ref{table:mu0scale}. Note that one is free to choose a different matching scale $\mu_0$ for such corrections. Therefore non-perturbative corrections at next-to-next-to-leading order for $B \to \pi \, l \, \nu \, (\g)$ with $\mu_0 = m_\pi$ are not troublesome . 

  \begin{table}[h!]\begin{center}
\begin{tabular}{r ccc} 
   & $\mu_0$ & $\alpha_s(\mu_0)$ &   $\Delta \mathcal{O}( \alpha \, \alpha_s \, \GF)$ \\ \hline 
 $B^+ \to \bar D^0 \, l \, \nu \, (\g)$ & $m_{D^0}$ & 0.40 & -0.04\% \\  
 $B^0 \to D^+ \, l \, \nu \, (\g)$ & $m_{D^+}$ & 0.40 & -0.04\% \\ 
 $B^+ \to \bar D_0^{* \, 0} \, l \, \nu \, (\g)$ & $m_{D^{* \, 0}_0}$ & 0.36 &  -0.04\%  \\
 $B^0 \to D_0^{* \, +} \, l \, \nu \, (\g)$ & $m_{D^{* \, +}_0}$ & 0.36 &  -0.04\%  \\
\end{tabular}
 \caption{ The choice for the infrared regulator $\mu_0$ for $B^+ \to X^0 \, l \, \nu \, (\g)$ and  $B^0 \to X^- \, l \, \nu \, (\g)$ decays with $X = D$ or $D^*_0$ are listed. The value for $\alpha_s(\mu_0)$ was extracted with $\alpha_s(\mu^2) = 4 \pi / \left(\beta_0 \, \ln \frac{\mu^2}{\Lambda_{QCD}^2}\right)$ where $m_W = 80.34$ GeV and $\Lambda_{\text{QCD}} = 0.2$ GeV. $\Delta \mathcal{O}( \alpha \, \alpha_s\, \GF)$  is the correction in percent to the pure short-distance electroweak correction: all corrections due to $\mathcal{O}( \alpha \, \alpha_s\, \GF)$ effects are negligible small. }\label{table:mu0scale}
 \end{center}\end{table}

\subsubsection{Renormalization of masses and electromagnetic coupling}

\begin{figure}[h!]\begin{center}  
\vspace{1ex}
 \unitlength = 1mm
 \begin{fmfgraph*}(30,20)
 	\fmfleft{on1,on2,is,on3,on4}
	\fmfright{of0,of1,of2,fs,of3,of4,of5}
	\fmf{dashes,tension=2/3}{is,v0,v1,v2,fs}
	\fmf{plain,tension=1/3}{of4,v3,v1}
	\fmf{plain,tension=1/3}{of1,v4,v1}
	\fmf{fermion,tension=0}{of4,v1,of1}
  	 \fmffreeze
         \fmf{photon, tension=1/2,right}{v3,v0}
         \fmfblob{0.1w}{v1}     
         	\fmflabel{$p_{B}$}{is}  
	\fmflabel{$p$}{fs}  
	\fmflabel{$p_{l}$}{of4} 
	\fmflabel{$p_{\nu_{l}}$}{of1}  
	 \fmffreeze
	  \fmfdot{v0}     
	\fmflabel{\emph{a)}}{on4} 
\end{fmfgraph*} 
\hspace{1.cm}
 \begin{fmfgraph*}(30,20)
 	\fmfleft{on1,on2,is,on3,on4}
	\fmfright{of0,of1,of2,fs,of3,of4,of5}
	\fmf{dashes,tension=2/3}{is,v0,v1,v2,fs}
	\fmf{plain,tension=1/3}{of4,v3,v1}
	\fmf{plain,tension=1/3}{of1,v4,v1}
	\fmf{fermion,tension=0}{of4,v1,of1}
         \fmf{photon, tension=0,left}{v3,v2}
           	 \fmffreeze
         \fmfblob{0.1w}{v1}     
	  \fmfdot{v2}     
	\fmflabel{\emph{b)}}{on4} 
\end{fmfgraph*}  \\
\vspace{1.cm}
 \begin{fmfgraph*}(30,20)
 	\fmfleft{on1,on2,is,on3,on4}
	\fmfright{of0,of1,of2,fs,of3,of4,of5}
	\fmf{dashes,tension=2/3}{is,v0,v1,v2,fs}
	\fmf{plain,tension=1/3}{of4,v3,v1}
	\fmf{plain,tension=1/3}{of1,v4,v1}
	\fmf{fermion,tension=0}{v1,of1}
           	 \fmffreeze
         \fmfblob{0.1w}{v1}     
	\fmfv{decor.shape=circle,decor.filled=empty, decor.size=0.1w}{v3}    
	\fmflabel{\emph{c)}}{on4} 
\end{fmfgraph*} 
\hspace{1.cm}
 \begin{fmfgraph*}(30,20)
 	\fmfleft{on1,on2,is,on3,on4}
	\fmfright{of0,of1,of2,fs,of3,of4,of5}
	\fmf{dashes,tension=2/3}{is,v0,v1,v2,fs}
	\fmf{plain,tension=1/3}{of4,v3,v1}
	\fmf{plain,tension=1/3}{of1,v4,v1}
	\fmf{fermion,tension=0}{of4,v1,of1}
           	 \fmffreeze
         \fmfblob{0.1w}{v1}     
	\fmfv{decor.shape=circle,decor.filled=empty, decor.size=0.1w}{v2}    
	\fmflabel{\emph{d)}}{on4} 
\end{fmfgraph*}  
\hspace{1.cm}
 \begin{fmfgraph*}(30,20)
 	\fmfleft{on1,on2,is,on3,on4}
	\fmfright{of0,of1,of2,fs,of3,of4,of5}
	\fmf{dashes,tension=2/3}{is,v0,v1,v2,fs}
	\fmf{plain,tension=1/3}{of4,v3,v1}
	\fmf{plain,tension=1/3}{of1,v4,v1}
	\fmf{fermion,tension=0}{of4,v1,of1}
           	 \fmffreeze
         \fmfblob{0.1w}{v1}     
	\fmfv{decor.shape=circle,decor.filled=empty, decor.size=0.1w}{v0}    
	\fmflabel{\emph{e)}}{on4} 
\end{fmfgraph*} \\
\vspace{1.cm}
 \begin{fmfgraph*}(30,20)
 	\fmfleft{on1,on2,is,on3,on4}
	\fmfright{of0,of1,of2,fs,of3,of4,of5}
	\fmf{dashes,tension=2/3}{is,v0,v1,v2,fs}
	\fmf{plain,tension=1/3}{of4,v3,v1}
	\fmf{plain,tension=1/3}{of1,v4,v1}
	\fmf{fermion,tension=0}{of4,v1,of1}
           	 \fmffreeze
         \fmf{photon, tension=0,left}{v1,v3}
         \fmfblob{0.1w}{v1}     
	\fmflabel{\emph{f)}}{on4} 
\end{fmfgraph*} 
\hspace{1.cm}
 \begin{fmfgraph*}(30,20)
 	\fmfleft{on1,on2,is,on3,on4}
	\fmfright{of0,of1,of2,fs,of3,of4,of5}
	\fmf{dashes,tension=2/3}{is,v0,v1,v2,fs}
	\fmf{plain,tension=1/3}{of4,v3,v1}
	\fmf{plain,tension=1/3}{of1,v4,v1}
	\fmf{fermion,tension=0}{of4,v1,of1}
           	 \fmffreeze
         \fmf{photon, tension=0,right}{v1,v0}
         \fmfblob{0.1w}{v1}     
         	  \fmfdot{v0}     
	\fmflabel{\emph{g)}}{on4} 
\end{fmfgraph*} 
\hspace{1.cm}
 \begin{fmfgraph*}(30,20)
 	\fmfleft{on1,on2,is,on3,on4}
	\fmfright{of0,of1,of2,fs,of3,of4,of5}
	\fmf{dashes,tension=2/3}{is,v0,v1,v2,fs}
	\fmf{plain,tension=1/3}{of4,v3,v1}
	\fmf{plain,tension=1/3}{of1,v4,v1}
	\fmf{fermion,tension=0}{of4,v1,of1}
           	 \fmffreeze
         \fmf{photon, tension=0,right}{v2,v1}
         \fmfblob{0.1w}{v1}     
         	  \fmfdot{v2}     
	\fmflabel{\emph{h)}}{on4} 
\end{fmfgraph*} 
\end{center}
\caption{The long-distance one-loop diagrams for $B^+ \to X^0 \, l \, \nu$ and $B^0 \to X^- \, l \, \nu$ are shown: \emph{a)} and  \emph{b)} inter-quark photon exchange; \emph{c)} - \emph{e)} wave-function corrections; and \emph{f) - h)} structure-dependent corrections. }   \label{legvirt}
\end{figure}
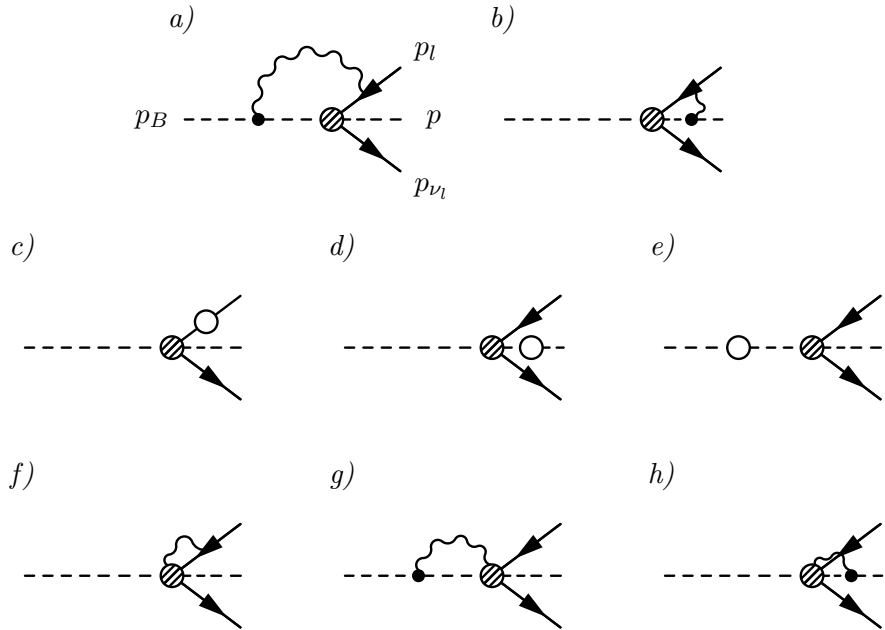

The corresponding long-distance correction to the short-distance result of Eqn. (\ref{m10sd}) is calculated in the following sections. We will work in the on-shell renormalization scheme, where the poles of the propagators are given by the phenomenological masses. In addition, we assume, that the electron charge is renormalized to its phenomenological measured value in a similar manner as the Fermi coupling constant. We will not work out the details of these renormalization steps, but only consider the diagrams that contribute to the virtual corrections of the $B \to X \, l \, \nu$ decay amplitude, as shown in Fig. (\ref{legvirt}). 


\subsubsection{Structure-dependent corrections to virtual diagrams}

For the long-distance loop calculations the meson will be treated as entirely point-like, neglecting all structure-dependent related corrections beyond the gauge fixing terms needed to fulfill the Ward-identity. Their coupling can be isolated in Eqns. (\ref{summedmode1}) and (\ref{summedmode2}) by setting $Z_{f_{\pm}} = 0$ and their contributions need to be included to obtain a consistent result with respect to gauge invariance. As long as $\mu_0 \backsim  m_X$ this assumption is justified, and the long-distnace loop integrals only receive additional corrections from higher order resonances through $B^{*/**/\dots}X\g$ and $X^{*/**/\dots}X\g$ coupling, which we neglect.

\newpage

\begin{figure}[h!]
\begin{center}  
\vspace{0.5cm}
 \unitlength = 1mm
 \begin{fmfgraph*}(30,20)
	\fmfleft{ll1,ll2,ql1,ql2}
	\fmfright{lr1,lr2,qr1,qr2}
	\fmf{plain,tension=5}{ql1,v0,v1,v2,v3,v4,v5,v6,qr2}
	\fmf{fermion,tension=5}{ql1,v3,qr2}
         \fmf{plain,tension=1}{lr1,w0,w1,w2,v3,w3,w4,w5,lr2}
         \fmf{fermion,tension=1}{lr1,v3,lr2}
         \fmf{photon,tension=0,right,label=$$}{v2,w2}
         \fmflabel{$q$}{ql1}  
         	\fmflabel{$q'$}{qr2}  
         	\fmflabel{$p_l$}{lr1}  
         	\fmflabel{$p_\nu$}{lr2}  
     	\fmflabel{$a)$}{ql2}  
         \fmfblob{0.1w}{v3}         
         	\fmfv{decor.shape=circle,decor.filled=full,decor.size=2thick}{v2}
         	\fmfv{decor.shape=circle,decor.filled=full,decor.size=2thick}{w2}
       \fmffreeze
\end{fmfgraph*} 
\hspace{1.25cm}
 \begin{fmfgraph*}(30,20)
	\fmfleft{ll1,ll2,ql1,ql2}
	\fmfright{lr1,lr2,qr1,qr2}
	\fmf{plain,tension=5}{ql1,v0,v1,v2,v3,v4,v5,v6,qr2}
	\fmf{fermion,tension=5}{ql1,v3,qr2}
         \fmf{plain,tension=1}{lr1,w0,w1,w2,v3,w3,w4,w5,lr2}
         \fmf{fermion,tension=1}{lr1,v3,lr2}
         \fmf{photon,tension=0,left,label=$$}{v4,w2}
	\fmflabel{$b)$}{ql2}  
         \fmfblob{0.1w}{v3}         
         	\fmfv{decor.shape=circle,decor.filled=full,decor.size=2thick}{w2}
         	\fmfv{decor.shape=circle,decor.filled=full,decor.size=2thick}{v4}
       \fmffreeze
\end{fmfgraph*}
\hspace{1.25cm}
 \begin{fmfgraph*}(30,20)
	\fmfleft{ll1,ll2,ql1,ql2}
	\fmfright{lr1,lr2,qr1,qr2}
	\fmf{plain,tension=5}{ql1,v0,v1,v2,v3,v4,v5,v6,qr2}
	\fmf{fermion,tension=5}{ql1,v3,qr2}
         \fmf{plain,tension=1}{lr1,w0,w1,w2,v3,w3,w4,w5,lr2}
         \fmf{fermion,tension=1}{lr1,v3,lr2}
         \fmf{photon,tension=0,left,label=$$}{v2,v4}
	\fmflabel{$c)$}{ql2}  
         \fmfblob{0.1w}{v3}         
         	\fmfv{decor.shape=circle,decor.filled=full,decor.size=2thick}{v2}
         	\fmfv{decor.shape=circle,decor.filled=full,decor.size=2thick}{v4}
       \fmffreeze
\end{fmfgraph*}
 \\
\vspace{1.cm}
 \begin{fmfgraph*}(30,20)
	\fmfleft{ll1,ll2,ql1,ql2}
	\fmfright{lr1,lr2,qr1,qr2}
	\fmf{plain,tension=5}{ql1,v0,v1,v2,v3,v4,v5,v6,qr2}
	\fmf{fermion,tension=5}{ql1,v3,qr2}
         \fmf{plain,tension=1}{lr1,w0,w1,w2,v3,w3,w4,w5,lr2}
         \fmf{fermion,tension=1}{lr1,v3,lr2}
  	\fmflabel{$d)$}{ql2}  
         \fmfblob{0.1w}{v3}      
         	\fmfv{decor.shape=circle,decor.filled=empty, decor.size=0.13w}{w1}       
       \fmffreeze
\end{fmfgraph*} 
\hspace{1.25cm}
 \begin{fmfgraph*}(30,20)
	\fmfleft{ll1,ll2,ql1,ql2}
	\fmfright{lr1,lr2,qr1,qr2}
	\fmf{plain,tension=5}{ql1,v0,v1,v2,v3,v4,v5,v6,qr2}
	\fmf{fermion,tension=5}{ql1,v3,qr2}
         \fmf{plain,tension=1}{lr1,w0,w1,w2,v3,w3,w4,w5,lr2}
         \fmf{fermion,tension=1}{lr1,v3,lr2}
    	\fmflabel{$e)$}{ql2}  
         \fmfblob{0.1w}{v3}         
         	\fmfv{decor.shape=circle,decor.filled=empty, decor.size=0.13w}{v5}       
       \fmffreeze
\end{fmfgraph*}
\hspace{1.25cm}
 \begin{fmfgraph*}(30,20)
	\fmfleft{ll1,ll2,ql1,ql2}
	\fmfright{lr1,lr2,qr1,qr2}
	\fmf{plain,tension=5}{ql1,v0,v1,v2,v3,v4,v5,v6,qr2}
	\fmf{fermion,tension=5}{ql1,v3,qr2}
         \fmf{plain,tension=1}{lr1,w0,w1,w2,v3,w3,w4,w5,lr2}
         \fmf{fermion,tension=1}{lr1,v3,lr2}
  	\fmflabel{$f)$}{ql2}  
         \fmfblob{0.1w}{v3}         
         	\fmfv{decor.shape=circle,decor.filled=empty, decor.size=0.13w}{v1}       
       \fmffreeze
\end{fmfgraph*}\\

\vspace{1.cm}
 \begin{fmfgraph*}(30,20)
	\fmfleft{ll1,ll2,ql1,ql2}
	\fmfright{lr1,lr2,qr1,qr2}
	\fmf{plain,tension=5}{ql1,v0,v1,v2,v3,v4,v5,v6,qr2}
	\fmf{fermion,tension=5}{ql1,v3,qr2}
         \fmf{plain,tension=1}{lr1,w0,w1,w2,v3,w3,w4,w5,lr2}
         \fmf{fermion,tension=1}{lr1,v3,lr2}
  	\fmflabel{$g)$}{ql2}  
         \fmfblob{0.1w}{v3}      
         \fmf{photon,tension=0,right,label=$$}{v3,w0}
                  	\fmfv{decor.shape=circle,decor.filled=full,decor.size=2thick}{w0}
       \fmffreeze
\end{fmfgraph*} 
\hspace{1.25cm}
 \begin{fmfgraph*}(30,20)
	\fmfleft{ll1,ll2,ql1,ql2}
	\fmfright{lr1,lr2,qr1,qr2}
	\fmf{plain,tension=5}{ql1,v0,v1,v2,v3,v4,v5,v6,qr2}
	\fmf{fermion,tension=5}{ql1,v3,qr2}
         \fmf{plain,tension=1}{lr1,w0,w1,w2,v3,w3,w4,w5,lr2}
         \fmf{fermion,tension=1}{lr1,v3,lr2}
    	\fmflabel{$h)$}{ql2}  
         \fmfblob{0.1w}{v3}         
         \fmf{photon,tension=0,left,label=$$}{v3,v6}
                  	\fmfv{decor.shape=circle,decor.filled=full,decor.size=2thick}{v6}
      \fmffreeze
\end{fmfgraph*}
\hspace{1.25cm}
 \begin{fmfgraph*}(30,20)
	\fmfleft{ll1,ll2,ql1,ql2}
	\fmfright{lr1,lr2,qr1,qr2}
	\fmf{plain,tension=5}{ql1,v0,v1,v2,v3,v4,v5,v6,qr2}
	\fmf{fermion,tension=5}{ql1,v3,qr2}
         \fmf{plain,tension=1}{lr1,w0,w1,w2,v3,w3,w4,w5,lr2}
         \fmf{fermion,tension=1}{lr1,v3,lr2}
  	\fmflabel{$i)$}{ql2}  
         \fmfblob{0.1w}{v3}         
         \fmf{photon,tension=0,right,label=$$}{v3,v0}
                  	\fmfv{decor.shape=circle,decor.filled=full,decor.size=2thick}{v0}       \fmffreeze
\end{fmfgraph*}

\end{center}
\caption{The electromagnetic short-distance one-loop photon diagrams for $b^+ \to x^0 \, l \, \nu$ and $b^0 \to x^- \, l \, \nu$ are shown: \emph{a)} -  \emph{c)} inter-particle photon exchange; \emph{d)} - \emph{f)} wave-function corrections; and \emph{g) - i)} coupling from external legs to the $W$-boson.}   \label{legvirtshort}
 \end{figure}
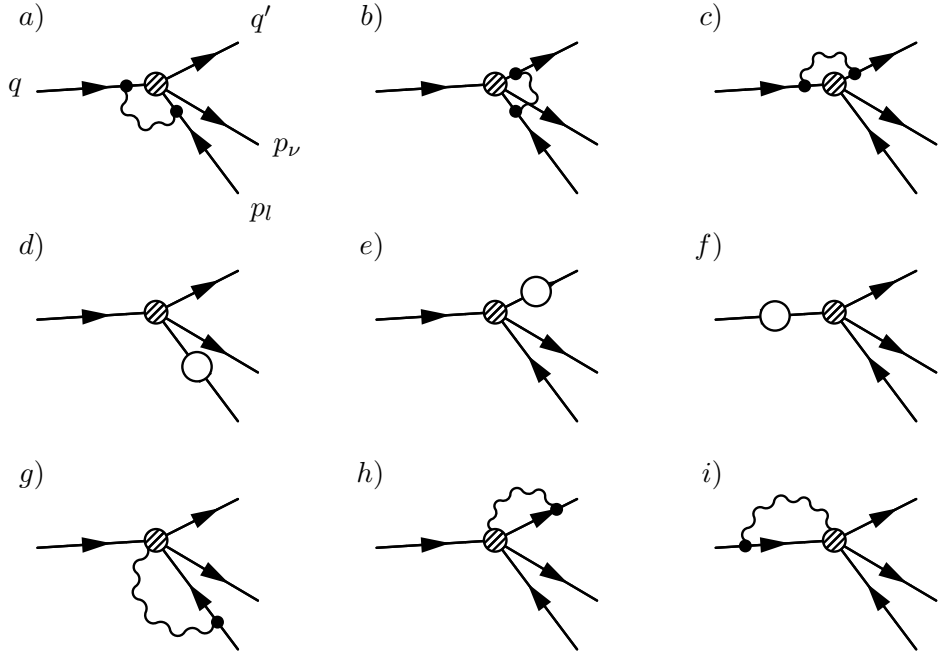  
 
\subsubsection{Long-distance wave function renormalization}

The fermionic long-distance wave function correction, Fig. \ref{legvirt} \emph{c)}, is given by 
 \bea\label{diractwopointderwvrenorm}
 \half \frac{\ud\Sigma (\ds{p}_l)}{\ud \ds{p}_l}   & \eq &  \half \frac{\alpha}{4\pi} \Big[  1 + 2 B_1(p_l^2;m_l^2,\lambda^2)  + 4 m_l^2 \dot B_1 (p_l^2;m_l^2,\lambda^2) + 8 m_l^2 \dot B_0(p_l^2;m_l^2,\lambda^2) \Big] \, , \nonumber \\
 \eea
 where $\Sigma (\ds{p}_l)$ is the fermionic self-energy correction. The long-distance wave function corrections of the scalar legs, Fig. \ref{legvirt} \emph{d)} and \emph{e)}, are given by
 \bea\label{scalartwopointderselfenergy}
 \half \frac{\ud \Sigma}{\ud p^2}  & \eq &\half \frac{\alpha}{4\pi} \Big[ 2 B_0(p^2;m^2,\lambda^2) + \bve{4 m^2 - \lambda^2} \dot B_0 (p^2;m^2,\lambda^2) \Big] \, ,
\eea
and
 \bea \label{scalartwopointderselfenergy2}
 \half \frac{\ud \Sigma}{\ud p_B^2}   & \eq & \half \frac{\alpha}{4\pi} \Big[ 2 B_0(p_B^2;m_B^2,\lambda^2) + \bve{4 m_B^2 - \lambda^2} \dot B_0 (p_B^2;m_B^2,\lambda^2) \Big] \, ,
\eea
respectively, with $\Sigma$ the scalar self-energy correction. The total correction due to long-distance wave function renormalization is then given by 
  \bea \label{m10wf1}
   \mathcal{M}^1_0 & \eq & \half \Bigg[  \frac{\ud\Sigma (\ds{p}_l)}{\ud \ds{p}_l}   + \frac{\ud \Sigma}{\ud p^2}   \Bigg] \, \mathcal{M}^0_0 \, , \label{m10wf1} 
\eea
or 
\bea
   \mathcal{M}^1_0 & \eq & \half \Bigg[  \frac{\ud\Sigma (\ds{p}_l)}{\ud \ds{p}_l}   + \frac{\ud \Sigma}{\ud p_B^2}   \Bigg] \, \mathcal{M}^0_0 \, .\label{m10wf2}
  \eea  
 The corresponding short-distance diagrams are given by Fig. \ref{legvirtshort} \emph{d)} - \emph{f)}.

\subsubsection{Long-distance inter-particle exchange}

The long-distance inter-particle exchanges, Figs. \ref{legvirt}  \emph{a)} and \emph{b)}, result in the following matrix element 
\bea\label{m10ip}
\mathcal{M}^1_0   & \eq & (+ \, e^2) \, \frac{1}{\sqrt{2}} \, G_{\text{F}} \, V_{\text{xb}} \, \big[\, \bar{u} (p_\nu) \, \rightproj \, \g_\mu  \, \Gamma^\mu \, v(p_l) \big] \, ,
\eea
with the correction tensor $ \Gamma^\mu$ as either
  \bea\label{pseudoscalarcorrectiontensor}
\Gamma^{\mu} & \eq &  \frac{i}{(4\pi)^2} \Big[ H^\mu \Big( - 4 \bve{p_B \cdot p_l} \Big) \Big] C_0(p_l^2, s, p_B^2; \lambda^2, m_l^2, m_B^2) \nonumber \\
 & & -  \frac{i}{(4\pi)^2} \Big[ \Big( f_{3}(t) \, p_l^\mu \Big) \bve{4 p_B \cdot p_l} + \big( H^\mu \big) \Big( 4 p_B \cdot p_l  + 2 m_l^2 - 2 \ds{p}_B \ds{p}_l) \Big)  \Big] C_1(p_l^2, s, p_B^2; \lambda^2, m_l^2, m_B^2) \nonumber \\
  & & -  \frac{i}{(4\pi)^2} \Big[ \Big( f_{3}(t) \, p_B^\mu \Big) \bve{4 p_B \cdot p_l} + \big( H^\mu \big) \Big( 2 p_B \cdot p_l + 2 m_B^2 \Big) \Big] C_2(p_l^2, s, p_B^2; \lambda^2, m_l^2, m_B^2)   \nonumber \\
  & &  + \frac{i}{(4\pi)^2} \Big[ \Big( - f_{3}(t) \, p_l^\mu \Big)\Big( 4 p_B \cdot p_l  + 2 m_l^2 - 2 \ds{p}_B \ds{p}_l \Big) \Big] C_{11}(p_l^2, s, p_B^2; \lambda^2, m_l^2, m_B^2) \nonumber \\
  & &  + \frac{i}{(4\pi)^2} \Big[ \Big( - f_{3}(t) \, p_B^\mu \Big) \Big( 2 p_B \cdot p_l + 2 m_B^2 \Big)  \Big] C_{22}(p_l^2, s, p_B^2; \lambda^2, m_l^2, m_B^2) \nonumber \\
  & & + \frac{i}{(4\pi)^2} \Big[ \Big( - f_{3}(t) \, p_l^\mu \Big) \Big(  2 p_B \cdot p_l + 2 m_B^2 \Big)  \nonumber \\
   & & \phantom{+ \frac{i}{(4\pi)^2} \Big[ \,} + \Big( - f_{3}(t) \, p_B^\mu \Big) \Big( 4 p_B \cdot p_l  + 2 m_l^2 - 2 \ds{p}_B \ds{p}_l \Big)  \Big] C_{12}(p_l^2, s, p_B^2; \lambda^2, m_l^2, m_B^2)  \nonumber \\
   & & + \frac{i}{(4\pi)^2} \Big[ \Big( -f_3(t) \Big) \Big( 2\,p_l^\mu + 2 \gamma^\mu \ds{p}_B  \Big) \Big] C_{00}(p_l^2, s, p_B^2; \lambda^2, m_l^2, m_B^2) \nonumber \\
   & & + \frac{i}{(4\pi)^2} \Big[ - H^\mu +  f_{3}(t)  \, p_l^\mu \Big] B_{0}(s;m_l^2, m_B^2)  + \frac{i}{(4\pi)^2} \Big[ f_{3}(t) \bve{p_l - p_B}^\mu \Big] B_{1}(s;m_l^2, m_B^2) \,, \nonumber \\
\eea
with $s=\bve{p_l - p_B}^2$, or
 \bea\label{pseudoscalarcorrectiontensor2}
\Gamma^{\mu} & \eq & - \frac{i}{(4\pi)^2} \Big[ H^\mu \Big( - 4 \bve{p \cdot p_l} \Big) \Big] C_0(p_l^2, s, p^2; \lambda^2, m_l^2, m^2) \nonumber \\
& &  +  \frac{i}{(4\pi)^2} \Big[ \Big( f_{2}(t) \, p_l^\mu \Big) \bve{ - 4 p \cdot p_l} + \big( H^\mu \big) \Big( 4 p \cdot p_l  - 2 m_l^2 - 2 \ds{p} \ds{p}_l) \Big)  \Big] C_1(p_l^2, s, p^2; \lambda^2, m_l^2, m^2) \nonumber \\
& &  - \frac{i}{(4\pi)^2} \Big[ \Big( f_{2}(t) \, p^\mu \Big) \bve{ - 4 p \cdot p_l} + \big( H^\mu \big) \Big(2 m^2 - 2 p \cdot p_l \Big)  \Big] C_2(p_l^2, s, p^2; \lambda^2, m_l^2, m^2) \nonumber \\
& & - \frac{i}{(4\pi)^2} \Big[ \Big(  f_{2}(t) \, p_l^\mu \Big) \Big( 4 p \cdot p_l - 2 m_l^2 - 2 \ds{p} \ds{p}_l \Big) \Big] C_{11}(p_l^2, s, p^2; \lambda^2, m_l^2, m^2) \nonumber \\
& & - \frac{i}{(4\pi)^2} \Big[ \Big(  f_{2}(t) \, p^\mu \Big) \Big( 2 m^2 - 2 p \cdot p_l \Big) \Big] C_{22}(p_l^2, s, p^2; \lambda^2, m_l^2, m^2) \nonumber \\
&& + \frac{i}{(4\pi)^2} \Big[ \Big(  f_{2}(t) \, p_l^\mu \Big) \Big( 2 m^2 - 2 p \cdot p_l \Big) \nonumber \\
& & \phantom{ - \frac{i}{(4\pi)^2} \Big[ + } +  \Big(  f_{2}(t) \, p^\mu \Big) \Big( 4 p \cdot p_l - 2 m_l^2 - 2 \ds{p} \ds{p}_l \Big) \Big] C_{12}(p_l^2, s, p^2; \lambda^2, m_l^2, m^2) \nonumber \\
& & - \frac{i}{(4\pi)^2} \Big[ \Big(  f_{2}(t) \Big) \Big(- 2 \, p_l^\mu + 2 \g^\mu \ds{p}  \Big) \Big] C_{00}(p_l^2, s, p^2; \lambda^2, m_l^2, m^2) \nonumber \\
& & - \frac{i}{(4\pi)^2} \Big[  H^\mu + f_2(t) \, p_l^\mu \Big]  B_0(s;m_l^2, m^2)  + \frac{i}{(4\pi)^2} \Big[   f_2(t) \bve{p_l + p}^\mu \Big]  B_1(s;m_l^2, m^2) \,  , \nonumber \\
\eea
with $s = \bve{p_l + p}^2$, respectively. The corresponding short-distance diagrams are given by Fig. \ref{legvirtshort} \emph{a)} - \emph{c)}.

\subsubsection{Structure-dependent contributions}
 
  The long-distance one-loop diagram between the remaining structure-dependent contributions in the IB coupling and the external legs, Figs. \ref{legvirt} \emph{f} - \emph{h}, are given by  
\bea\label{m10vertex}
\mathcal{M}^1_0  & \eq & (e^2) \frac{G_{\text{F}}}{\sqrt{2}} \, V_{\text{xb}} \big[ \bar u(p_\nu) P_R  \, \Lambda \,  v(p_l) \big] \,,
\eea
with the correction matrix $ \Lambda $, which is either
  \bea\label{vertexloopisfinal}
  \Lambda & \eq  & \frac{i}{(4\pi)^2}  \Big[  2 \ds{p}_l \, f_{3}(t) \Big] \, B_0(m_l^2; \lambda^2, m_l^2)+  \frac{i}{(4\pi)^2} \big[ 2 \ds{p}_B \, f_{3}(t) \big] B_0(m_B^2; \lambda^2, m_B^2) \nonumber \\
    & &  + \frac{i}{(4\pi)^2} \frac{1}{m_l^2} \Big[  \ds{p}_l \, f_{3}(t) \Big] \, A_0(m_l^2) + \frac{i}{(4\pi)^2} \frac{1}{2 m_B^2} \big[ - \ds{p}_B \, f_3(t) \big] A_0(m_B^2) \nonumber \\
     &&  + \frac{i}{(4\pi)^2}  \Big[ \ds{p}_l \, f_{3}(t) \Big]  \,,
 \eea
or 
  \bea\label{vertexloopfsfinal}
 \Lambda& \eq  & \frac{i}{(4\pi)^2}  \Big[ - 2 \ds{p}_l \, f_{2}(t) \Big] \, B_0(m_l^2; \lambda^2, m_l^2)+  \frac{i}{(4\pi)^2} \big[ 2 \ds{p} \, f_{2}(t) \big] B_0(m^2; \lambda^2, m^2) \nonumber \\
    & &  + \frac{i}{(4\pi)^2} \frac{1}{m_l^2} \Big[ -  \ds{p}_l \, f_{2}(t) \Big] \, A_0(m_l^2) + \frac{i}{(4\pi)^2} \frac{1}{2 m^2} \big[ - \ds{p} \, f_2(t) \big] A_0(m^2) \nonumber \\
     &&  + \frac{i}{(4\pi)^2}  \Big[ \ds{p}_l \, f_{2}(t) \Big]  \,.
 \eea
The corresponding short-distance diagrams are given by Fig. \ref{legvirtshort} \emph{g)} - \emph{i)}.
 
 \subsection{Pauli-Villars regularization}\label{pvregularization}
 
 It remains to transform the long-distance loop-expressions from dimensional regularization into the Pauli-Villars regularization scheme. This can be done by a simple set of replacement rules and counter terms. The Pauli-Villars prescription \cite{Pauli:1949zm} introduces a massive photon field with opposite norm into the theory via
 \bea
 \mathcal{L}_{\text{PV}} & \eq &  \frac{1}{4} \tilde F^{2} - \mu_0^2 \, \tilde A^2 \, .
\eea
 Taking the limit of $\mu_0 \to \infty$ decouples the field from the rest of the theory and gives raise to a UV divergency growing as $\sim \ln \mu_0^2$, which is exactly the desired opposite behavior of the short-distance result. The massive field suppresses the exchange of virtual particles with energies larger than $\mu_0$. 
 
\subsubsection{Wave function counter terms in Pauli-Villars}

 Eqn. (\ref{diractwopointderwvrenorm}) has to be modified via
  \bea\label{diracpvreptsim}
   B_0(m_l^2,m_l^2,\lambda^2) & \to &  B_0(m_l^2;m_l^2,\lambda^2) -  B_0(m_l^2;m_l^2,\mu_0^2) \, , \nonumber \\
   \dot{B}_i(m_l^2,m_l^2,\lambda^2) & \to &  \dot{B}_i(m_l^2,m_l^2,\lambda^2) -  \dot{B}_i(m_l^2,m_l^2,\mu_0^2) \, ,
\eea 
with $i=0,1$. Adding the counter-term
\bea
 \frac{\ud \Sigma(\ds{p}_l)_{\text{PV terms}}}{\ud \ds{p}_l} & = &  - \frac{\alpha}{4 \pi }\,  ,
\eea
results in the desired transformation. Similarly, Eqn. (\ref{scalartwopointderselfenergy}) replaces
\bea\label{scalarpvreptsim}
   B_0(m^2,m^2,\lambda^2) & \to &  B_0(m^2;m^2,\lambda^2) -  B_0(m^2;m^2,\mu_0^2) \, , \nonumber \\
   \dot{B}_0(m^2,m^2,\lambda^2) & \to &  \dot{B}_0(m^2,m^2,\lambda^2) -  \dot{B}_0(m^2,m^2,\mu_0^2) \, ,
\eea 
with a counter-term 
\bea\label{scalarpvreptsimcounterterm}
 \frac{\ud \Sigma(p^2)_{\text{PV terms}}}{\ud p^2} & \eq & \frac{\alpha}{4 \pi} \Big[  \mu_0^2 \, \dot{B}_0(m^2,m^2,\mu_0^2)\Big] \, . 
\eea
The rules for Eqn. (\ref{scalartwopointderselfenergy2}) can be obtained by replacing $m^2 \to m_B^2$ in Eqns. (\ref{scalarpvreptsim}) and (\ref{scalarpvreptsimcounterterm}).

\subsubsection{Inter-particle exchange diagrams in Pauli-Villars}

Eqn. (\ref{pseudoscalarcorrectiontensor}) has to be modified via
\bea
 C_i(p_l^2, s, p_B^2; \lambda^2, m_l^2, m_B^2) & \to & C_i(p_l^2, s, p_B^2; \lambda^2, m_l^2, m_B^2) -  C_i(p_l^2, s, p_B^2; \mu_0^2, m_l^2, m_B^2) \, , \nonumber \\
 C_{ij}(p_l^2, s, p_B^2; \lambda^2, m_l^2, m_B^2) & \to & C_{ij}(p_l^2, s, p_B^2; \lambda^2, m_l^2, m_B^2) -  C_{ij}(p_l^2, s, p_B^2; \mu_0^2, m_l^2, m_B^2) \, , \nonumber \\
  B_{i}(s;m_l^2, m_B^2) & \to & 0 \, ,
\eea
with $i,j=0,1,2$. Adding the counter terms
\bea
\Gamma^\mu_{\text{PV terms}} & \eq &  - \frac{i}{\bve{4\pi}^2} \Big[ H^\mu \,  \mu_0^2 \Big] C_0(p_l^2, s, p_B^2; \mu_0^2, m_l^2, m_B^2)  \nonumber \\
&& +  \frac{i}{\bve{4\pi}^2} \Big[ f_3(t) \, \bve{p_l}^\mu \,  \mu_0^2 \Big] C_1(p_l^2, s, p_B^2; \mu_0^2, m_l^2, m_B^2)  \nonumber \\
&& +  \frac{i}{\bve{4\pi}^2} \Big[ f_3(t) \, \bve{p_B}^\mu \,  \mu_0^2 \Big] C_2(p_l^2, s, p_B^2; \mu_0^2, m_l^2, m_B^2)  \, .
\eea
results in the desired transformation. For Eqn. (\ref{pseudoscalarcorrectiontensor2}) one modifies 
\bea
 C_i(p_l^2, s, p^2; \lambda^2, m_l^2, m^2) & \to & C_i(p_l^2, s, p^2; \lambda^2, m_l^2, m^2) -  C_i(p_l^2, s, p^2; \mu_0^2, m_l^2, m^2) \, , \nonumber \\
 C_{ij}(p_l^2, s, p^2; \lambda^2, m_l^2, m^2) & \to & C_{ij}(p_l^2, s, p^2; \lambda^2, m_l^2, m^2) -  C_{ij}(p_l^2, s, p^2; \mu_0^2, m_l^2, m^2) \, , \nonumber \\
  B_{i}(s;m_l^2, m^2) & \to & 0 \, ,
\eea
with $i,j=0,1,2$ and adds the counter terms
\bea
\Gamma^\mu_{\text{PV terms}} & \eq &  \frac{i}{\bve{4\pi}^2} \Big[ H^\mu \,  \mu_0^2 \Big] C_0(p_l^2, s, p^2; \mu_0^2, m_l^2, m_B^2)  \nonumber \\
&& -  \frac{i}{\bve{4\pi}^2} \Big[ f_2(t) \, \bve{p_l}^\mu \,  \mu_0^2 \Big] C_1(p_l^2, s, p^2; \mu_0^2, m_l^2, m^2)  \nonumber \\
&& +  \frac{i}{\bve{4\pi}^2} \Big[ f_2(t) \, \bve{p}^\mu \,  \mu_0^2 \Big] C_2(p_l^2, s, p^2; \mu_0^2, m_l^2, m^2)  \, .
\eea

\subsubsection{Structure-dependent contributions in Pauli-Villars}

Eqn. (\ref{vertexloopisfinal}) is modified according to
\bea
  B_{i}(m_l^2;\lambda^2, m_l^2) & \to &  B_{i}(m_l^2;\lambda^2, m_l^2) -  B_{i}(m_l^2;\mu_0^2, m_l^2) \, ,\nonumber \\
  B_{i}(m_B^2;\lambda^2, m_B^2) & \to &  B_{i}(m_B^2;\lambda^2, m_B^2) -  B_{i}(m_B^2;\mu_0^2, m_B^2) \, , 
\eea
with $i = 0,1$. Adding the counter-term 
\bea
 \Lambda_{\text{PV terms}} & \eq & - \frac{i}{(4\pi)^2} \Big[ \ds{p}_l \, f_3(t) \Big] \, ,
\eea
results in the desired transformation. Eqn. (\ref{vertexloopfsfinal}) has to be modified according to
\bea
  B_{i}(m_l^2;\lambda^2, m_l^2) & \to &  B_{i}(m_l^2;\lambda^2, m_l^2) -  B_{i}(m_l^2;\mu_0^2, m_l^2) \, , \nonumber \\
  B_{i}(m^2;\lambda^2, m^2) & \to &  B_{i}(m^2;\lambda^2, m^2) -  B_{i}(m^2;\mu_0^2, m^2) \, , 
\eea
with $i = 0,1$, and a counter term of
\bea
 \Lambda_{\text{PV terms}} & \eq & -  \frac{i}{(4\pi)^2} \Big[ \ds{p}_l \, f_2(t) \Big] \, .
\eea

 \subsection{Next-to-leading order differential and total decay rate}\label{rateformulas}
 
Summing all virtual long-distance corrections for $B^+ \to X^0 \, l \, \nu$ or $B^0 \to X^- \, l \, \nu$, given by Eqns. (\ref{m10ip}), and either (\ref{m10wf1}) or (\ref{m10wf2}), and (\ref{m10vertex}), and the short-distance result Eqn. (\ref{m10sd}), yields the approximative virtual next-to-leading order differential decay rate. It is given by
\bea\label{virtnlorate}
\ud \Gamma^0_0 + \ud \Gamma^1_0 & \eq & \frac{1}{64 \, \pi^3 m_B} \Bigg(  \big| \mathcal{M}^0_0 \big|^2 + 2 \sum \mathcal{M}^0_0 \mathcal{M}^1_0 + 2 \big| \mathcal{M}^0_0 \big|^2 \Big[  \frac{3}{4} \frac{\alpha}{\pi}\bve{1 + 2 |\bar Q|} \ln \frac{m_Z}{\mu_0}\Big]  \Bigg) \, \ud E \, \ud E_l  \, . \nonumber \\
\eea
Similarly the real corrections with Eqn. (\ref{summedmode1}) or (\ref{summedmode2}) result in the approximative real next-to-leading order differential decay rate:
\bea\label{realnlorate}
  \ud \Gamma^{\half}_1 & \eq & \frac{1}{ (2 \pi)^{12} E \, E_l \, E_\nu \, E_k  } \delta^{(4)}\bve{m_B - p - p_l - p_\nu  -k}    \big| \sum \mathcal{M}^\half_1 \big|^2 \,\, \ud^3 p \,\, \ud^3 p_l \,\, \ud^3 p_\nu \,\, \ud^3 k \, , \nonumber \\
\eea
with $E_k = k^0$ and $E_\nu = p_\nu^0$. Integrating Eqns. (\ref{virtnlorate}) and (\ref{realnlorate}) yields the approximative next-to-leading order total decay rate. Comparing with the total tree-level decay rate, the integral over phasespace of Eqn. (\ref{treeleveldifferentialrate}), yields the correction factor $\delta_{\text{total}}$ due to next-to-leading order effects: it contains all short- and long-distance next-to-leading order corrections.
 It is 
\bea\label{deltasddeltald}
 \G^0_0 + \G^1_0 + \G^\half_1 & \eq &  \Big( 1 + \delta_{\text{total}}  \Big) \, \G^0_0 \,\, \eq \,\, \Big( 1 + \delta_{\text{sd}} + \delta_{\text{ld}} \Big) \, \G^0_0 \, , 
\eea
where $\delta_{\text{total}} = \delta_{\text{sd}} + \delta_{\text{ld}}$, with the short-distance contribution $\delta_{\text{sd}} =  \frac{2\alpha}{\pi} \ln \frac{m_Z}{\mu_0} $ from Eqn. (\ref{m10sd}), and $\delta_{\text{ld}}$ denotes the long-distance corrections.

%% file: numerics.tex
\section{Numerical evaluation}\label{numerics}

In order to evaluate Eqns. (\ref{virtnlorate}) and (\ref{realnlorate}), we developed the Monte Carlo generator \texttt{BLOR} (\emph{$B$-meson leading order radiative corrections}), which is derived from the code of \texttt{KLOR} (\emph{Kaon leading order radiative corrections} - written by Troy Andre \cite{Andre:2004fs}). It incorporates all form-factor models of App. \ref{appa} and the next-to-leading order matrix elements derived in the previous section. The loop-integrals are evaluated numerically using the software package \texttt{Looptools} \cite{PhysRevD.80.076010} and the regularization procedure can be changed between dimensional regularization, Pauli-Villars and an approximative Euclidian cut-off method. 
The phase-space integration occurs in two steps: the total tree-level and next-to-leading order rates are calculated to determine $\delta_{\text{ld}}$ and the IR cut-off dependent radiative fraction of:
\bea \label{radfrac}
  \frac{\G^0_0 + \G^1_0}{ \G^\half_1} \, .
\eea
This is done using the VEGAS algorithm described in \cite{numericalrecipes}. Subsequently the radiative fraction Eqn. (\ref{radfrac}) is used to mix events generated from the real and virtual next-to-leading order matrix elements using standard Monte Carlo techniques. In order to optimize the event generation procedure, moderately pre-tuned probability density functions of the integration variables in question are used.
In addition to the implemented next-to-leading order calculations, the \texttt{BLOR} Monte Carlo generator can be interfaced with the approximative next-to-leading order package \texttt{PHOTOS} \cite{Barberio:1990ms,Barberio:1993qi}. The code of \texttt{BLOR}  is publicly accessible and can be found at \cite{blorcode}. The particle masses and couplings implemented therein are summarized in Table \ref{pdgpars}. 

\vspace{2cm}

 \begin{table}[hb!]\begin{center}
\begin{tabular}{rc} 
 Particle & Mass   \\ \hline 
$ m_{\Upsilon(4S)} $            & $10.5794$ GeV$/c^2$ \\
$ \Gamma_{\Upsilon(4S)} $            & $20.5$ MeV$/c^2$ \\
 $  m_{B^{+}} $ & $5.27913$ GeV$/c^2$ \\
 $  m_{B^{0}} $ & $5.27950$ GeV$/c^2$ \\
 $  m_{D^{+}} $ & $1.86950$ GeV$/c^2$ \\
 $  m_{D^{0}} $ & $1.86484$ GeV$/c^2$ \\
 $  m_{\pi^{+}} $ & $0.13957$ GeV$/c^2$ \\
 $  m_{\pi^{0}} $ & $0.13498$ GeV$/c^2$ \\
 $  m_{D^{* \,+}_0} $ & $2.40300$ GeV$/c^2$ \\
 $  m_{D^{* \, 0}_0} $ & $2.35200$ GeV$/c^2$ \\ \hline
 \end{tabular} \hspace{1.5cm}
\begin{tabular}{rc} 
  Parameter &  Value   \\ \hline 
$ \lambda  $            &$10^{-7}$  GeV$/c^2$ \\
 $ \alpha $ & $0.00729735039$ \\
 $ \GF $ &   $1.16637 \cdot 10^{-5}$ $\bve{\hbar c}^3$GeV${}^{-2}$ \\ \hline
\end{tabular}
 \caption{Particle masses and couplings used in the simulation. All values are taken from \cite{Amsler:2008zzb}.} \label{pdgpars}
 \end{center}\end{table}

%% file: results.tex
\section{Predictions}\label{results}

In the following, we quote the value of $\delta_{\text{total}}$ calculated from the next-to-leading order model and try to assess its uncertainty due to the approximative character of the matching procedure and missing real and virtual SD contributions for $B \to D \, l \, \nu \, (\g)$, $B \to \pi \, l \, \nu \, (\g)$, and $B \to D^*_0 \, l \, \nu \, (\g)$. Then, we present the main result of this study: normalized kinematical distributions for the final state lepton, meson, and photon. For a comparison of our findings with the next-to-leading order algorithms of \texttt{PHOTOS} and \texttt{PHOTONS++} \cite{Schonherr:2008av}, we refer the reader to \cite{resultpaper}. 

\subsection{Uncertainties of $\delta_{\text{total}}$}\label{Sec:uncertainties}

Uncertainties to the next-to-leading order corrections come from five sources: the numerical evaluation ($\sigma_{\text{numerical}}$), the matching procedure between long- and short-distance result ($\sigma_{\text{matching}}$), the limitations of the effective theory describing the complete phase-space adequately due to neglecting the $V_{\mu\nu}^{\text{SD}} -   A_{\mu\nu}^{\text{SD}}$ contributions ($\sigma_{\text{SD}}$), next-to-next-to-leading order electromagnetic effects ($\sigma_{\text{nnlo}}$), and last but not least model dependent uncertainties. The total uncertainty by the first four sources is given by
\bea
 \sigma^2_{\text{total}} & \eq &  \sigma^2_{\text{numerical}} +\sigma^2_{\text{nnlo}} + \sigma^2_{\text{matching}} + \sigma^2_{\text{SD}}  \ \, ,
\eea
It is not possible to assess the model dependent uncertainties due to the negligence of the unknown higher-order operators of Eqn. (\ref{phenoweakhamiltonian}), therefore the stated uncertainties should be interpreted as uncertainties that can be assessed within the model itself. Without full knowledge of the full effective theory, no deeper and more satisfying assessment can be made. In the following the four sources of uncertainties that can be assessed are discussed. 

\subsubsection{Numerical uncertainty of the integration}

The total integration error of $\delta_{\text{total}}$ is given by
\bea
 \sigma_{\text{numerical}} & \eq & \Big( \sum^m_{i=1} \frac{1}{\sigma_i^2} \Big)^{-\half}\, ,
\eea
where $\sigma_i$ is the error of the $i$th \texttt{VEGAS} evaluation, and $m$ the number of total evaluation steps. 

\subsubsection{Corrections due to next-to-next-to-leading order electromagnetic corrections}

The next-to-next-to-leading order (nnlo) electromagnetic corrections are estimated as
\bea
 \sigma_{\text{nnlo}} & \eq & \alpha \, \Big(  \delta_{\text{sd}} +  \delta_{\text{ld}} \Big) \, .
\eea

\subsubsection{Uncertainty due to the matching procedure}\label{limitations}

The uncertainty due to the matching procedure is studied by varying the matching scale $\mu_0$ within reasonable bounds. By doing so, one varies with what contributions of the effective Hamiltonian Eqn. (\ref{phenoweakhamiltonian}) the 'intermediate region' with respect to the short- and long-distance regime are modeled. This unveils a shift in $\delta_{\text{total}} = \delta_{\text{total}}(\mu_0)$ and we choose to vary the matching scale within $\half \mu_0$ and $2 \mu_0$. The resulting estimator for the matching uncertainty is then calculated as 
\bea
  \sigma_{\text{matching}} & \eq & \half  \big|  \delta_{\text{total}} (\mu_0) -  \delta_{\text{total} }  (\half \mu_0 ) \big| + \half \big|   \delta_{\text{total}} (\mu_0)  - \delta_{\text{total}} (2 \mu_0 )  \big| \, .
\eea
Note that for a matching scale choosen at $2 \mu_0$, the phenomenological model for the long-distance contributions is beyond its expected validity; whereas fixing the matching scale at $\half \mu_0$ short- and long-distance corrections should represent a reasonable description. The dependence of the matching scale $\mu_0$ on the overall correction $\delta_{\text{total}}$ are summarized in Tables \ref{matchD}, \ref{matchpi}, and \ref{matchDs}.

\subsubsection{ Corrections due to real SD contributions for $B \to D \, l \, \nu \, \g$ decays}\label{realSDeffects}

\begin{figure}[h!]\begin{center}  
\vspace{1.5cm}
 \unitlength = 1mm
 \begin{fmfgraph*}(30,20)
 	\fmfleft{on1,on2,is,on3,on4}
	\fmfright{of0,of1,of2,fs,of3,of4,of5}
	\fmf{dashes,tension=2/3}{is,v0,v1,v2,fs}
	\fmf{plain,tension=1/3}{of4,v3,v1}
	\fmf{plain,tension=1/3}{of1,v4,v1}
	\fmf{fermion,tension=0}{of4,v1,of1}
         \fmf{photon, tension=0}{v0,of5}
         \fmf{dbl_wiggly,tension=0}{v0,v1}
         \fmfblob{0.1w}{v1}     
         	\fmflabel{$p_{B}$}{is}  
	\fmflabel{$k,\epsilon^*$}{of5}
	\fmflabel{$p$}{fs}  
	\fmflabel{$p_{l}$}{of4} 
	\fmflabel{$p_{\nu_{l}}$}{of1}  
	 \fmffreeze
	  \fmfdot{v0}     
	\fmflabel{\emph{a)}}{on4} 
\end{fmfgraph*} 
\hspace{1.cm}
 \begin{fmfgraph*}(30,20)
 	\fmfleft{on1,on2,is,on3,on4}
	\fmfright{of0,of1,of2,fs,of3,of4,of5}
	\fmf{dashes,tension=2/3}{is,v0,v1,v2,fs}
	\fmf{plain,tension=1/3}{of4,v3,v1}
	\fmf{plain,tension=1/3}{of1,v4,v1}
	\fmf{fermion,tension=0}{of4,v1,of1}
         \fmf{photon, tension=0}{v2,of3}
         \fmf{dbl_wiggly,tension=0}{v2,v1}
         \fmfblob{0.1w}{v1}     
	  \fmfdot{v2}     
	 \fmffreeze
	\fmflabel{\emph{b)}}{on4} 
\end{fmfgraph*}
\hspace{1.cm}
\end{center}
\caption{The SD one-particle contributions to the real emission are shown: \emph{a)} and \emph{b)} emission from a $B^*B\g$ and $D^*D\g$ vertex, respectively.}   \label{poles}
 \end{figure}
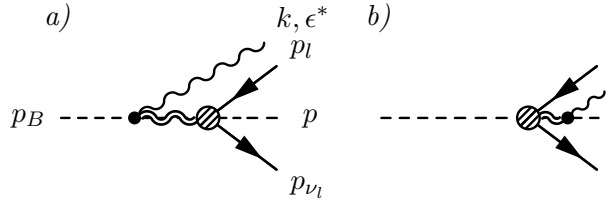

For $B \to D \, l \, \nu \, \g$ decays, the dominant corrections from hadronic emissions lie in the soft-photon part of phase space and are due to IB contributions. The additional SD contributions in this region of phase-space are well approximated by emissions from the two lowest excited states, $B^*$ and $D^*$, respectively. These intermediate emissions are depicted in Fig. (\ref{poles}) and their contributions to the SD vector and axial-vector coupling are
\bea\label{vmasdexp}
V_{\mu\nu}^{\text{SD}} -   A_{\mu\nu}^{\text{SD}}  & = &  \frac{i \, \bra  D | J_{\nu}^{\text{em}} | D^*\ket \, \bra D^* | V_{\mu} - A_{\mu}| B \ket }{  \bve{p+ k}^2 - m_{D^*}^2 } + \frac{i \, \bra  D |  V_{\mu}- A_{\mu}| B^*\ket \, \bra B^* | J_{\nu}^{\text{em}} | B \ket }{  \bve{p- k}^2 - m_{B^*}^2 } \, + \dots , \nonumber \\
\eea
where the ellipses denote contributions from higher excitations and further correction terms. The emission from the $D^*$-line is expected to be dominant over $B^*$-pole contribution, since the latter can only occur off-shell and the large mass of the $B^*$ excitation suppresses emissions considerably. The intermediate $D^*$ is kinematically allowed to be on-shell in certain regions of phase-space. As proposed by \cite{Becirevic:2009fy} in these on-shell regions one can use the lattice $D^*D\g$ coupling of \cite{Becirevic:1166910} to estimate the first term of Eqn. (\ref{vmasdexp}) as
\bea \label{btodgest}
& & \frac{ e  \, \varepsilon_{\nu\rho\alpha\beta} \, \bve{p+ k}^\alpha \, p^\beta \, g_{DD^*\g} }{2 p \cdot k + \Delta m_{D^*}^2}   \times  \Big( 2 \, i \, \varepsilon^{\mu\nu\alpha'\beta'} \, \bve{p_B}_{\alpha'} \bve{p+k}_{\beta'} \, g(t') - g^{\mu\nu} f(t')  \nonumber \\
        & & \ph{ \frac{ e  \, \varepsilon_{\nu\rho\alpha\beta} \, \bve{p+ k}^\alpha \, p^\beta \, g_{DD^*\g} }{2 p \cdot k + \Delta m_{D^*}}   \times  \Big( - }   - \bve{p_B + p + k}_\mu \bve{p_B - p - k}_\nu \, a_{+}(t') \nonumber \\
        & &  \ph{ \frac{ e  \, \varepsilon_{\nu\rho\alpha\beta} \, \bve{p+ k}^\alpha \, p^\beta \, g_{DD^*\g} }{2 p \cdot k + \Delta m_{D^*}}   \times  \Big(  - }   - \bve{p_B - p - k}_\mu \bve{p_B + p + k}_\nu \, a_{-}(t') \Big) \,,
\eea
where $g_{D^*D\g}$ denotes the corresponding effective $D^*D\g$ coupling, $\Delta m_{D^*}^2 = m_D^2 - m^2_{D^*} + i \, m_{D^*} \, \G_{D^*}$ with $\G_{D^*}$ the $D^*$ width, and $f$, $g$, and $a_{\pm}$ the on-shell $B \to D^*$ form factors, further discussed in App. \ref{appa}. Assuming the dominant contributions occur near the on-shell value of the propagator, the additional SD contributions to the hadronic $B \to D \, l \, \nu \, \g$ emissions can be estimated by including Eqn. (\ref{btodgest}) into the emission matrix element. The resulting shifts in $\delta_{\text{total}}$ are summarized in Table \ref{SDterms}, where $\delta_{\text{total + SD approx.}}$ denotes the total predicted correction gained by including Eqn. (\ref{btodgest}). The difference to $\delta_{\text{total}}$ is taken as uncertainty $ \sigma_{\text{SD}}$ due to real SD emissions:
\bea
 \sigma_{\text{SD}} & \eq &  \large|\delta_{\text{total}} -  \delta_{\text{total + SD approx.}}\large| \, .
\eea 
Contributions from higher scalar and tensor charmed resonances at $\mathcal{O}(\GF \, \alpha)$ to real $B \to D \, l \, \nu \, \g$ decays are forbidden by parity, angular momentum and spin conservation. Higher vector charmed resonances, however, can contribute at $\mathcal{O}(\GF \, \alpha)$.

   \begin{table}[h!]\begin{center}
\begin{tabular}{llll}
           					&  $\delta_{\text{total}}$ & $ \delta_{\text{total + SD approx.}}$ & $\sigma_{\text{SD}}$  \\ \hline
    $B^0 \to D^{+} \, l \, \nu \, (\g)$        &  $0.02223(6)$      &     $0.02225(7)$           &   $0.00002$      \\
    $B^+ \to \bar D^{0} \, l \, \nu  \, (\g)$       &  $0.01463(5)$      &     $0.01627(6)$           &   $0.00158$     \\
 \hline
\end{tabular}
 \caption{
 The effect of including the approximative SD emissions on $ \delta_{\text{ld}}$ for $B \to D \, e \, \nu \, (\g)$ is shown. The difference between the charged and uncharged charmed final states originates from the effective coupling $g_{D^*D\g}$: it is $g_{ D^{*\,+}D^+\g} = - 0.1(7)$ and $g_{D^{*\,0}D^0\g} = 2.7(1.2)$ from \cite{Becirevic:1166910}. 
  } \label{SDterms}
 \end{center}\end{table}

 \subsubsection{Corrections due to real SD contributions for $B \to \pi \, l \, \nu \, \g$}\label{Sec:realSDcontributions}

\begin{figure}[PhT!]\begin{center}  
\vspace{0.5cm}
 \unitlength = 1mm
 \includegraphics[width=0.6\textwidth]{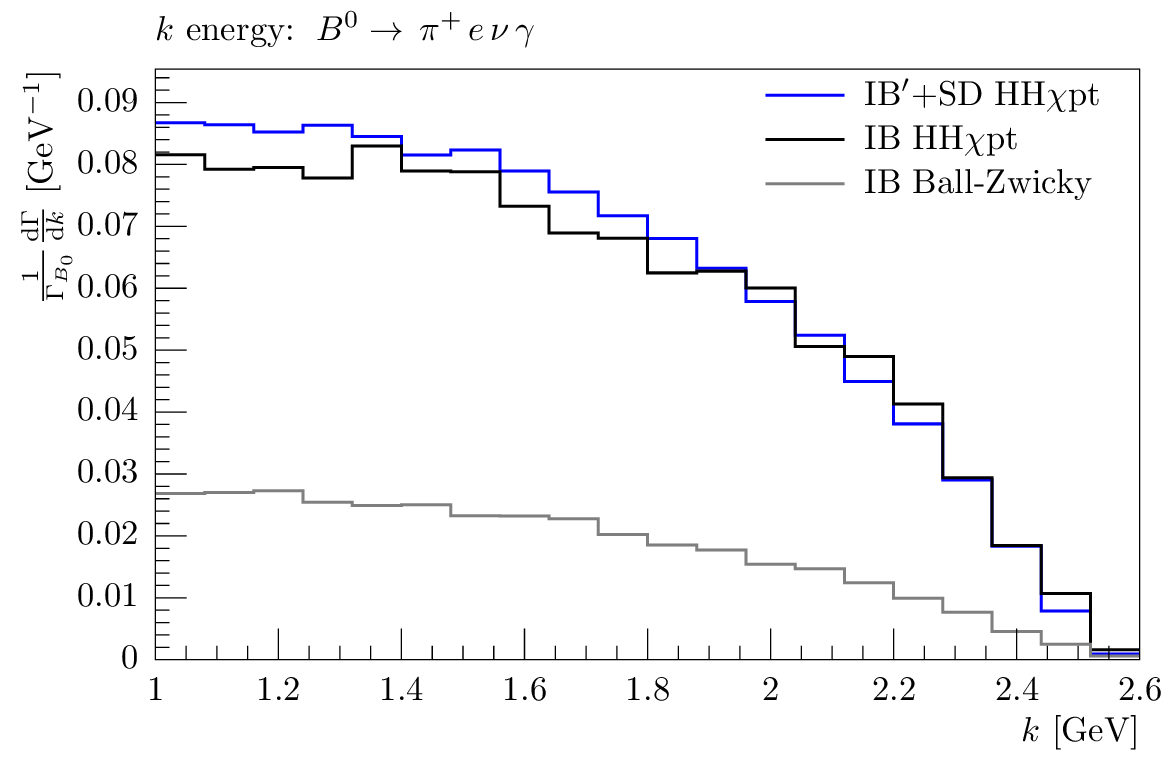} \\
 \caption{    
 The differential branching fractions of the $B^0 \to \pi^+ \, e \, \nu \, \g$ decay is shown: blue shows the result of \cite{PhysRevD.72.094021}, black is the prediction based on the IB contributions with HH$\chi$pt form factors, and grey is the prediction based on the IB contributions with Ball-Zwicky form factors. 
 }   \label{factversusIBGraph}
 \end{center}
 \end{figure}

For $B \to \pi \, l \, \nu \, \g$ the real IB result can be compared with the findings of \cite{PhysRevD.72.094021}: The authors calculated the real IB and SD corrections in the framework of SCET and in the phase-space region of soft pions and hard photons: $E_\pi < 0.5$ GeV, $k > 1.0$ GeV, and $\theta_{e\g} > 5^{o}$. In this region a significant influence of SD contributions can be expected, since the first light intermediate resonance receive on-shell contributions at $k \approx 0.8$ GeV. Since the splitting of \cite{PhysRevD.72.094021} in IB and SD contributions differs from our construction by not incorporating corrections of  $\mathcal{O}(k)$ and beyond into their IB term, the comparison also offers interesting insight into the predictive power of including such terms. In order to distinguish this we write IB$'$ in the following when referring to their choice. The calculation of \cite{PhysRevD.72.094021} uses the leading order heavy hadron chiral perturbation theory (HH$\chi$pt) form factor predictions:
\bea\label{hhchptff}
 f_{\pm}(E_\pi) & \propto & \pm  \frac{1}{E_\pi + \Delta} \, 
\eea
with $E_\pi = p^0$, and $\Delta = m_{B^*} - m_B = 50$ MeV. The radiative form factors in our formulation of the IB emissions are thus given by $f_\pm (E_\pi + E_k)$ for a charged final-state pion, whereas in the IB$'$ choice, Eqn. (\ref{hhchptff}) is used as presented. The partial and differential branching fractions from both calculations are compared in Table \ref{factversusIB1} and Fig. \ref{factversusIBGraph}: the pure IB based prediction based on the model dependent choice in this work, reproduces the photon energy spectrum well. The authors of  \cite{PhysRevD.72.094021} also quote the results of the partial branching fraction gained from the IB$'$ terms alone, and from including the leading terms from Low's theorem ($k^{-1}$ and $k^0$), denoted as IB$''$ in the following. The partial branching fractions for the IB$'$ and  IB$''$ calculations are stated in Table \ref{factversusIB2}, and differ significantly from the IB$'$+SD and the IB predictions.
   \begin{table}[h!]\begin{center}
\begin{tabular}{c|cccc}
   Cuts & & $\mathcal{BF}(B^0 \to \pi^+ \, e \, \nu \, \g)_{\text{cut}}$ &   \\
    & IB$'$ + SD HH$\chi$pt   \cite{PhysRevD.72.094021} & IB  HH$\chi$pt     & IB  Ball-Zwicky   \\ \hline
    $E_\g> 1.0$ GeV & & \\
    $E_\pi < 0.5$ GeV & $1.2 \, {}^{+2.2}_{-0.6} \, \times 10^{-6}$ & $1.16(1) \, \times 10^{-6}$   &   $0.35(1) \, \times 10^{-6}$ \\  
   $\theta_{e\g} > 5^{o}$ & &\\ 		\hline
\end{tabular}
 \caption{
The partial branching fractions $\mathcal{BF}(B^0 \to \pi^+ \, e \, \nu \, \g)_{\text{cut}}$ for the IB$'$+SD and IB calculation are stated. The errors in the parentheses are numerical.  } \label{factversusIB1}
 \end{center}\end{table}
 
Therefore including terms of $\mathcal{O}(k^2)$ and beyond into the IB contribution via Eqn. (\ref{summedmode2}) reproduces the main features of the photon energy spectra and provides a partial branching fraction close to the result of the full IB$'$ + SD calculation. 

    \begin{table}[h!]\begin{center}
\begin{tabular}{c|cccc}
   Cuts & & $\mathcal{BF}(B^0 \to \pi^+ \, e \, \nu \, \g)_{\text{cut}}$ &   \\
    & IB$'$  HH$\chi$pt   \cite{PhysRevD.72.094021}   & IB$''$  HH$\chi$pt \cite{PhysRevD.72.094021}   \\ \hline
    $E_\g> 1.0$ GeV & & \\
    $E_\pi < 0.5$ GeV & $2.8 \, \times 10^{-6}$   &  $2.4 \, \times 10^{-6}$ \\  
   $\theta_{e\g} > 5^{o}$ & &\\ 		\hline
\end{tabular}
 \caption{
The partial branching fractions $\mathcal{BF}(B^0 \to \pi^+ \, e \, \nu \, \g)_{\text{cut}}$ for the IB$'$ and IB$''$ based predictions.  
  } \label{factversusIB2}
 \end{center}\end{table}
 
By varying the HH$\chi$pt  IB result of the partial branching fraction to the extremal values within the error band of the result from \cite{PhysRevD.72.094021}, one  obtains the estimator for the uncertainty $\sigma_{\text{SD}}$ due to SD effects in $B^0 \to \pi^+ \, l \, \nu \, \g$ in this region of phase-space, as
 \bea\label{sd2btopi}
  \sigma_{\text{SD}} & \eq &  \large|\delta_{\text{total IB}} -  \delta_{\text{total IB + SD}}\large| \quad \eq \quad 0.0048 \, ,
 \eea 
 We interprete this as an estimator for the uncertainty for the entire phase space. Comparing the HH$\chi$pt prediction of $\delta_{\text{total}}$ and the partial branching fraction with the result obtained with the Ball-Zwicky form factors (see App. \ref{appa}), reveals a deviation due to their different parametrization and normalization. The predictions for  $\delta_{\text{total}}$ are listed in Table \ref{factversusIB3}, and the differential and partial branching fractions are depicted in Fig. \ref{factversusIBGraph} and listed in Table \ref{factversusIB1}. The deviation in $\delta_{\text{total}}$ between HH$\chi$pt  and Ball-Zwicky is small with respect to $\sigma_{\text{SD}}$. This suggests that Eqn. (\ref{sd2btopi}) can be used as a valid estimator for the missing SD contributions for the $\delta_{\text{total}}$ prediction based on the Ball-Zwicky form factors. 

 
       \begin{table}[h!]\begin{center}
\begin{tabular}{rll}
$B^0 \to \pi^+ \, e \, \nu \, (\g)$  &   HH$\chi$pt    &   Ball-Zwicky \\ \hline
   $1 + \delta_{\text{sd}}+ \delta_{\text{ld}}$    &  $0.0562(1)$  & $0.0555(1) $    \\ \hline
\end{tabular}
 \caption{
The prediction for $1 + \delta_{\text{sd}}+ \delta_{\text{ld}}$ for HH$\chi$pt and Ball-Zwicky form factors for $B^0 \to \pi^+ \, e \, \nu \, (\g)$ are shown. The uncertainties in the parentheses are numerical. 
  } \label{factversusIB3}
 \end{center}\end{table}
 
 For the radiative $B^+ \to \pi^0 \, l \, \nu \,  \g$ decay no calculations exists that estimates the real SD contributions, therefore no uncertainty due to such corrections can be assigned to our IB based prediction.

\subsubsection{Corrections due to real SD contributions from $B \to D_0^* \, l \, \nu \, \g$}

For $B \to D_0^* \, l \, \nu  \, \g$ decays significant corrections due to SD contributions can be expected: the short lifetime of the $D_0^*$-meson results in a large decay width, which causes an overlap with the broad charged $D^{* \, +}_{1}$- and the narrow uncharged $D^{* \, 0}_1$-states. We are not aware of any lattice results, which examine the coupling between these two charmed meson resonances, the $ D_0^*$-meson and the photon, or of any work that could help estimating this uncertainty.

%

\subsection{Predictions for $\delta_{\text{total}}$}

The predictions for $1+\delta_{\text{sd}} + \delta_{\text{ld}}$ and $1/ \sqrt{1+\delta_{\text{sd}} + \delta_{\text{ld}}}$ for $B \to D \, l \, \nu \, (\g)$ are listed in Table \ref{ldsdpredward1}. The assigned uncertainty interval estimates possible further contributions due to real SD terms, matching uncertainties, numerical uncertainties, and next-to-next-to-leading order corrections -- within the limitation addressed in Sec. \ref{Sec:uncertainties}. The mass of the $D-$meson allows a reasonable separation of short- and long-distance domains. Table \ref{ldsdpredward2} summarizes the predictions of $1+\delta_{\text{sd}} + \delta_{\text{ld}}$ and $1 / \sqrt{1+\delta_{\text{sd}} + \delta_{\text{ld}}}$ for $B \to \pi \, l \, \nu \, (\g)$. The assigned uncertainty interval estimates the matching uncertainties, numerical uncertainties, and next-to-next-to-leading order corrections. In addition for $B^0 \to \pi^+ \, l \, \nu \, (\g)$ the real SD corrections were estimated. Table \ref{ldsdpredward3} eventually summarizes the predictions of $1+\delta_{\text{sd}} + \delta_{\text{ld}}$ and $1 / \sqrt{1+\delta_{\text{sd}} + \delta_{\text{ld}}}$  for $B \to D^*_0 \, l \, \nu \, (\g)$. The assigned uncertainty interval estimates the matching uncertainties, numerical uncertainties, and next-to-next-to-leading order corrections.  In addition the mass of the $D^*_0-$meson allows a reasonable separation of the short- and long-distance separation. But current knowledge does not allow to make any assessment for real SD corrections. 

The inverse of $\sqrt{1+\delta_{\text{sd}} + \delta_{\text{ld}}}$ is the relevant correction to the CKM matrix elements $|V_{\text{cb}}|$ and  $|V_{\text{ub}}|$ gained from measured decay rates. In summary, the source of the dominant uncertainty in our presented approach originate the matching procedure. In addition unknown uncertainties due to the neglegence of unknown SD contributions and model dependences arise. 

 The predicted corrections break as expected the isospin symmetry of the semileptonic decay process, hence the dominant real and virtual corrections scale with the mass of the charged meson in the decay.

  \begin{table}[h!]\begin{center}
\begin{tabular}{r lll} 
   & $1 + \delta_{\text{sd}}$ + $\delta_{\text{ld}}$ & $1 / \sqrt{1 + \delta_{\text{sd}}+ \delta_{\text{ld}}}$    \\ \hline 
    $B^0 \to D^- \, e^+ \, \nu_e \, (\g)$ & $1.0222(1 \pm 2 \pm 17 \pm 1)$  & $0.9891(1 \pm 1 \pm 4 \pm 1)$  \\
    $B^0 \to D^- \, \mu^+ \, \nu_\mu \, (\g)$ & $1.0222(1 \pm 2 \pm17 \pm 1)$  & $0.9891(1 \pm 1 \pm 4 \pm 1)$  \\
    $B^+ \to \bar D^0 \, e^+ \, \nu_e \, (\g)$  & $1.0146(1 \pm 1 \pm 39 \pm 16 )$  & $0.9928(1 \pm 1 \pm 10 \pm 4)$  \\
    $B^+ \to \bar D^0 \, \mu^+ \, \nu_\mu \, (\g)$ & $1.0147(1 \pm 1 \pm 39 \pm 16)$  & $0.9927(1 \pm 1 \pm 10 \pm 4)$  \\ \hline
\end{tabular}
 \caption{Predictions for $\delta_{\text{total}}$ for $B \to D \, l \, \nu \, (\g)$ are listed. The uncertainties in the parentheses are numerical, next-to-next-to-leading order, matching,  and due to neglected real SD contributions, respectively. }\label{ldsdpredward1}
 \end{center}\end{table}
 \vspace{-7ex}
   \begin{table}[h!]\begin{center}
\begin{tabular}{r lll} 
   & $1 + \delta_{\text{sd}}$ + $\delta_{\text{ld}}$ & $1 / \sqrt{1 + \delta_{\text{sd}}+ \delta_{\text{ld}}}$    \\ \hline 
    $B^0 \to \pi^- \, e^+ \, \nu_e \, (\g)$     & $1.0555(1 \pm 4 \pm 148 \pm 48)$  & $0.9734(1 \pm 1 \pm 33 \pm 10)$  \\
    $B^0 \to \pi^- \, \mu^+ \, \nu_\mu \, (\g)$ & $1.0545(1 \pm 4 \pm 136 \pm 48)$  & $0.9738(1 \pm 1 \pm 31 \pm 11)$  \\
    $B^+ \to \pi^0 \, e^+ \, \nu_e \, (\g)$     & $1.0411(1 \pm  3 \pm 100)$  & $0.9801(1 \pm 1 \pm 23)$  \\
    $B^+ \to \pi^0 \, \mu^+ \, \nu_\mu \, (\g)$ & $1.0401(1 \pm  3 \pm 89)$  & $0.9805(1 \pm 1 \pm 21)$   \\ \hline
\end{tabular}
 \caption{Predictions for $\delta_{\text{total}}$ for $B \to \pi \, l \, \nu \, (\g)$ are listed. The uncertainties in the parentheses are numerical, next-to-next-to-leading order, matching, and for $B^0 \to \pi^+ \, l \, \nu \, (\g)$ due to real SD contributions, respectively. }\label{ldsdpredward2}
 \end{center}\end{table}
  \vspace{-5ex} 
   \begin{table}[h!]\begin{center}
\begin{tabular}{r lll} 
   & $1 + \delta_{\text{sd}}$ + $\delta_{\text{ld}}$ & $1 / \sqrt{1 + \delta_{\text{sd}}+ \delta_{\text{ld}}}$    \\ \hline 
    $B^0 \to D^{*\,-}_0 \, e^+ \, \nu_e \, (\g)$    & $1.0224(1 \pm 2 \pm 10)$  & $0.9890(1 \pm 1 \pm 2)$  \\
    $B^0 \to D^{*\,-}_0 \, \mu^+ \, \nu_\mu \, (\g)$ & $1.0226(1 \pm 2 \pm 10)$  & $0.9889(1 \pm 1 \pm 2)$  \\
    $B^+ \to \bar D^{*\,0}_0 \, e^+ \, \nu_e \, (\g)$   & $1.0142(1 \pm 1 \pm 35)$  & $0.9930(1 \pm 1 \pm 8)$  \\
    $B^+ \to \bar D^{*\,0}_0 \, \mu^+ \, \nu_\mu \, (\g)$ & $1.0144(1 \pm 1 \pm 35)$  & $0.9929(1 \pm 1 \pm 8)$  \\ \hline
\end{tabular}
 \caption{Predictions for $\delta_{\text{total}}$ for $B \to  D^{*}_0 \, l \, \nu \, (\g)$ are listed. The uncertainties in the parentheses are numerical, next-to-next-to-leading order, and matching, respectively. }\label{ldsdpredward3}
 \end{center}\end{table}

\newpage

\subsubsection{Corrections to $|V_{\text{cb}}|$}

The lepton mass for $l = e$ or $\mu$ only has a negligible effect on the value of the total decay rate. One can therefore assume that
\bea
\G^0_0(B^+ \to X^0 \, e \, \nu) = \G^0_0(B^+ \to X^0 \, \mu \, \nu) \qquad \text{and} \qquad \G^0_0(B^0 \to X^+ \, e \, \nu) = \G^0_0(B^0 \to X^+ \, \mu \, \nu) \,, \nonumber \\
\eea
and average the results of $1+\delta_{\text{sd}} + \delta_{\text{ld}}$ and $\sqrt{1+\delta_{\text{sd}} + \delta_{\text{ld}}}$ in Table \ref{ldsdpredward1} from both lepton channels to decrease the numerical uncertainty of the integration. The SD and next-to-next-to-leading order uncertainties then are correlated by $100\%$. This yields an averaged correction of $2.22(17)\%$ to the total  $B^0 \to D^+ \, l \, \nu$ decay rate in comparison to the tree-level decay rate. The increased total decay rate translates into a decrease of the extracted value of  $|V_{\text{cb}}|$  from measurements by $-1.09(4)\%$ in comparison to extractions based on the tree-level calculation. Similarly,  the $B^+ \to \bar D^0 \, l \, \nu$ total decay rate increases by $1.46(43)\%$, translating into a decrease of $-0.72(10)\%$ of $|V_{\text{cb}}|$.

Analyses often study $B$-mesons from the decay chain $\Upsilon(4S)\to B \, \bar B$ with $B$ either $B^+$ or $B^0$. In order to reduce the statistical uncertainty, and assuming isospin, the $B^+ \to \bar D^0 \, l \, \nu$ and $B^0 \to D^+ \, l \, \nu$ decay rates can be combined. At tree-level, this combined total decay rate is given by
\bea\label{totalaverageddecratehqet}
\half \G^0_0(B \to D \, l \, \nu) & \eq &  f_{+-} \,  \G^0_0(B^+ \to D^0 \, l \, \nu) +  f_{00} \,  \G^0_0(B^0 \to D^+ \, l \, \nu) \, ,
\eea 
with
\bea
f_{+-} \, = \, \mathcal{BF}(\Upsilon(4S) \to B^+ \, B^-) \qquad \text{and} \qquad f_{00} \, = \, \mathcal{BF}(\Upsilon(4S) \to B^0 \, \bar B^0) \, ,
\eea
with the branching fractions of the $\Upsilon(4S)$ resonance either decaying into a $B^+ \, B^-$ or $ B^0 \, \bar B^0$ pair. Including the isospin conserving parton-level correction in Eqn. (\ref{m10sd}) yields
 \bea
 \bve{1 + \delta_{\text{sd}}} \,  \G^0_0(B \to D \, l \, \nu)  & \propto &  |V_{\text{cb}}|^2 \, ,
 \eea
 with $\delta_{\text{sd}} =1.430\%$, resulting in a correction of $-0.707\%$ to the extracted value of $ |V_{\text{cb}}|$.
Including the isospin breaking contributions from the long-distance corrections  of Table \ref{ldsdpredward1} result in
\bea\label{properdeltahqet}
 \bve{1 + \delta_{\text{sd}} + \delta^+_{\text{ld}} } \,  f_{+-} \,  \G^0_0(B^+ \to \bar D^0 \, l \, \nu) +  \bve{1 + \delta_{\text{sd}} + \delta^0_{\text{ld}} } \,   f_{00} \,  \G^0_0(B^0 \to D^+ \, l \, \nu)  & \propto &  |V_{\text{cb}}|^2 \, , \nonumber \\
\eea
  where $ \delta^+_{\text{ld}}$ and $ \delta^0_{\text{ld}}$ either denote the long-distance next-to-leading order correction of  $B^+ \to D^0 \, l \, \nu$ or $B^0 \to D^+ \, l \, \nu$ decays, respectively. The total correction is
     \bea
   \bve{1 + \delta_{\text{total}}} \,  \G^0_0(B \to D \, l \, \nu)  & \propto &  |V_{\text{cb}}|^2 \, ,
 \eea
  with
  \bea
 \delta_{\text{total}} & \eq & \bve{ \delta_{\text{sd}} + \delta^+_{\text{ld}} } \, f_{+-}\,  \frac{\G^0_0(B^0 \to D^+ \, l \, \nu)}{\G^0_0(B^+ \to D^0 \, l \, \nu)} + \bve{ \delta_{\text{sd}} + \delta^0_{\text{ld}} } \, f_{00}\,  \frac{\G^0_0(B^+ \to D^0 \, l \, \nu)}{\G^0_0(B^0 \to D^+ \, l \, \nu)} \, .\nonumber \\
  \eea
 Assuming isospin symmetry at tree-level and averaging over both lepton modes, and with $\frac{f_{+-}}{f_{00}} = 1.065(26)$ from \cite{Amsler:2008zzb} yields an overall correction of $\delta_{\text{total}} = 1.86(29)\%$ due to short- and long-distance next-to-leading order effects\footnote{The uncertainty is propagated assuming a correlation of $100\%$ between the SD and next-to-next-to-leading order uncertainties of each channel and a correlation of $100\%$ between the isospin doublet. Further the uncertainties of $f_{+-}$ and $f_{00}$  are correlated by $-100\%$.}. This translates in a reduction of $-0.91(7)\%$ to the extracted value of $|V_{\text{cb}}|$. 
 
  \begin{table}[h!]\begin{center}
\begin{tabular}{r lll} 
   & $1 + \delta_{\text{sd}}$ + $\delta_{\text{ld}}$ & $1 / \sqrt{1 + \delta_{\text{sd}}+ \delta_{\text{ld}}}$    \\ \hline 
    $B^0 \to D^- \, l \, \nu$ & $1.0222(17)$  & $0.9891(4)$  \\
    $B^+ \to \bar D^0 \, l \, \nu$  &  $1.0146(43)$& $0.9928(10)$   \\
    $B \to D \, l  \, \nu$  & $1.0186(29)$ & $0.9909(7)$ \\ \hline
\end{tabular}\vspace{0.5cm}

\begin{tabular}{r ll} 
   & $1 + \delta_{\text{sd}}$ + $\delta_{\text{ld}}$ & $1 / \sqrt{1 + \delta_{\text{sd}}+ \delta_{\text{ld}}}$     \\ \hline 
    $B^0 \to \pi^- \, l \, \nu$      & $1.0550(150) $  &  $0.9736(34)  $ \\
\end{tabular}\vspace{0.5cm}

 \caption{Averaged integration results for  $1 + \delta_{\text{sd}}$ + $\delta_{\text{ld}}$ and $\sqrt{1 + \delta_{\text{sd}}+ \delta_{\text{ld}}}$ for $B \to D \, l \, \nu$ and $B^0 \to \pi^+ \, l \, \nu$: The uncertainties in the parentheses are the sum of  numerical, next-to-next-to-leading order, matching, and due to missing real SD contributions. The result for  $B^0 \to \pi^+ \, l \, \nu$ should be used with care.}\label{ldsdpredwardaveraged}
 \end{center}\end{table}
 
%
  

\subsubsection{Corrections to $|V_{\text{ub}}|$}

Averaging the leptonic results yields for the $B^0 \to \pi^+ \, l \, \nu$ total decay rate an increase by $5.50(1.50)\%$, translating into a decrease of $-2.64(34)\%$ of extracted values of $|V_{\text{ub}}|$ from measurements, as summarized in Table \ref{ldsdpredwardaveraged}. 

\subsection{Predictions for next-to-leading-order differential rates}

In the following, the normalized differential decay rates for the tree-level and next-to-leading order decays for the lepton, meson, and photon are shown:
\bea
 \frac{1}{\G} \, \frac{\ud \G}{\ud p_l} \,, \qquad \frac{1}{\G} \frac{\ud \G}{\ud p} \,, \qquad \text{and} \qquad \frac{1}{\G} \, \frac{\ud \G}{\ud k} \, ,
\eea
with $\G \eq \G^0_0$ or $\G \eq \G^0_0 + \G^1_0 + \G^{\half}_1$. All decay rates are presented in the experimentally important rest frame of the $\Upsilon(4S) \to B \, \bar B$ decay. 

\begin{figure}[Pht!]\begin{center}  
\vspace{0.5cm}
 \unitlength = 1mm
  \includegraphics[width=0.49\textwidth]{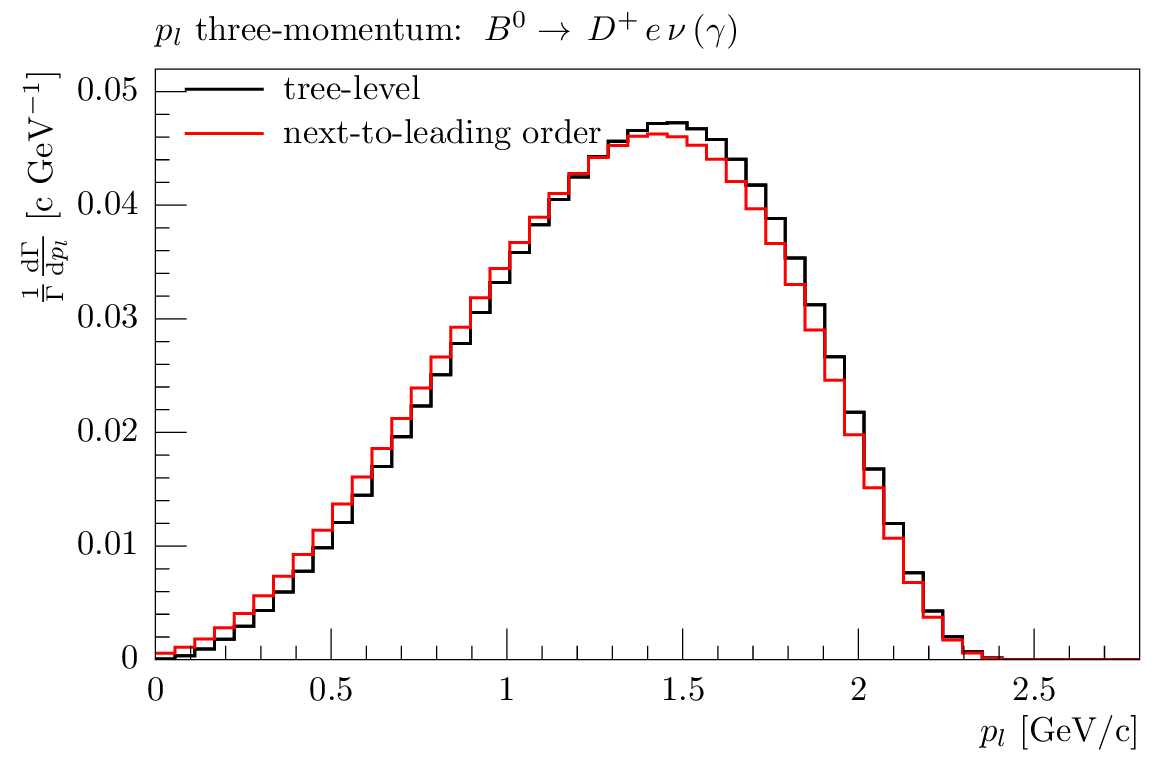}
  \includegraphics[width=0.49\textwidth]{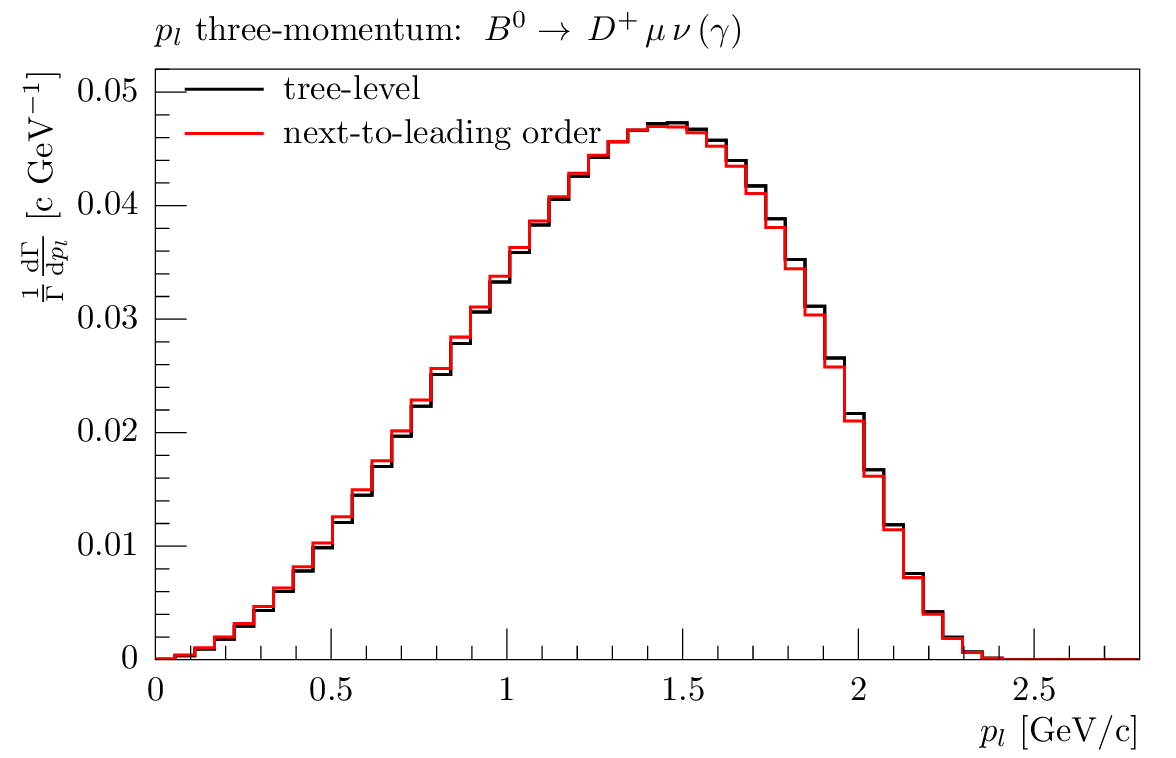} \\
  \includegraphics[width=0.49\textwidth]{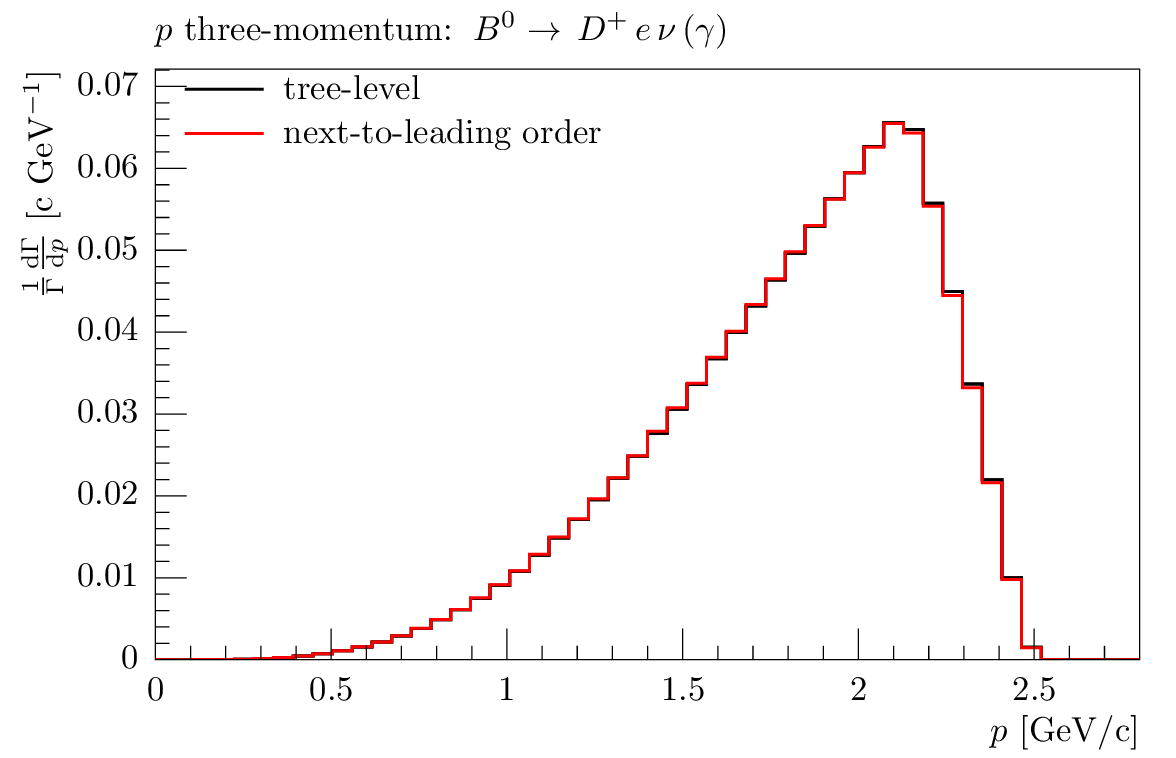} 
  \includegraphics[width=0.49\textwidth]{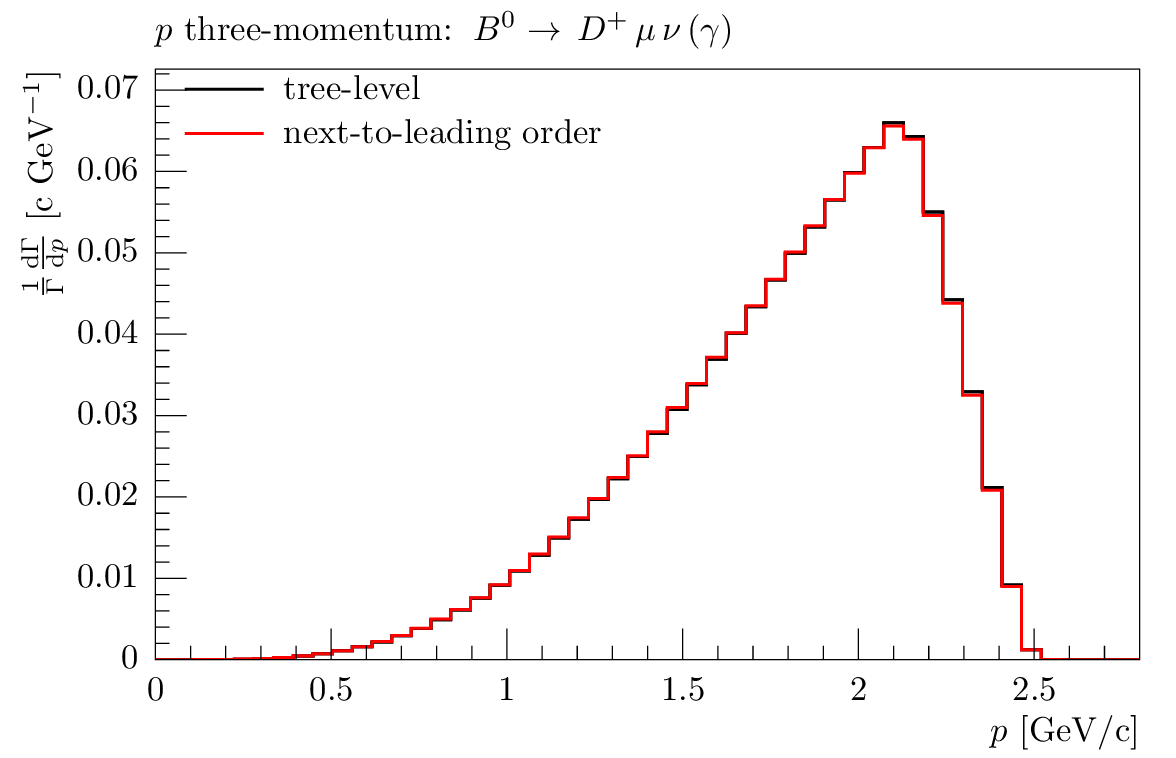} \\
  \includegraphics[width=0.49\textwidth]{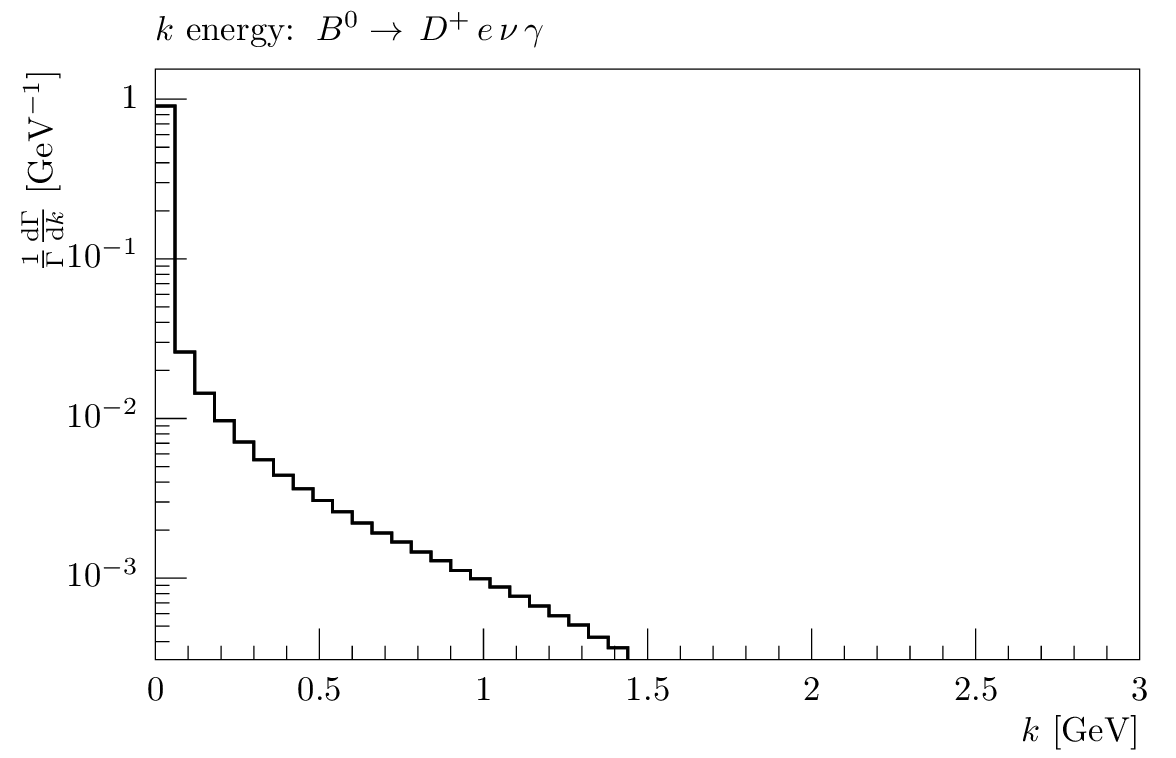} 
  \includegraphics[width=0.49\textwidth]{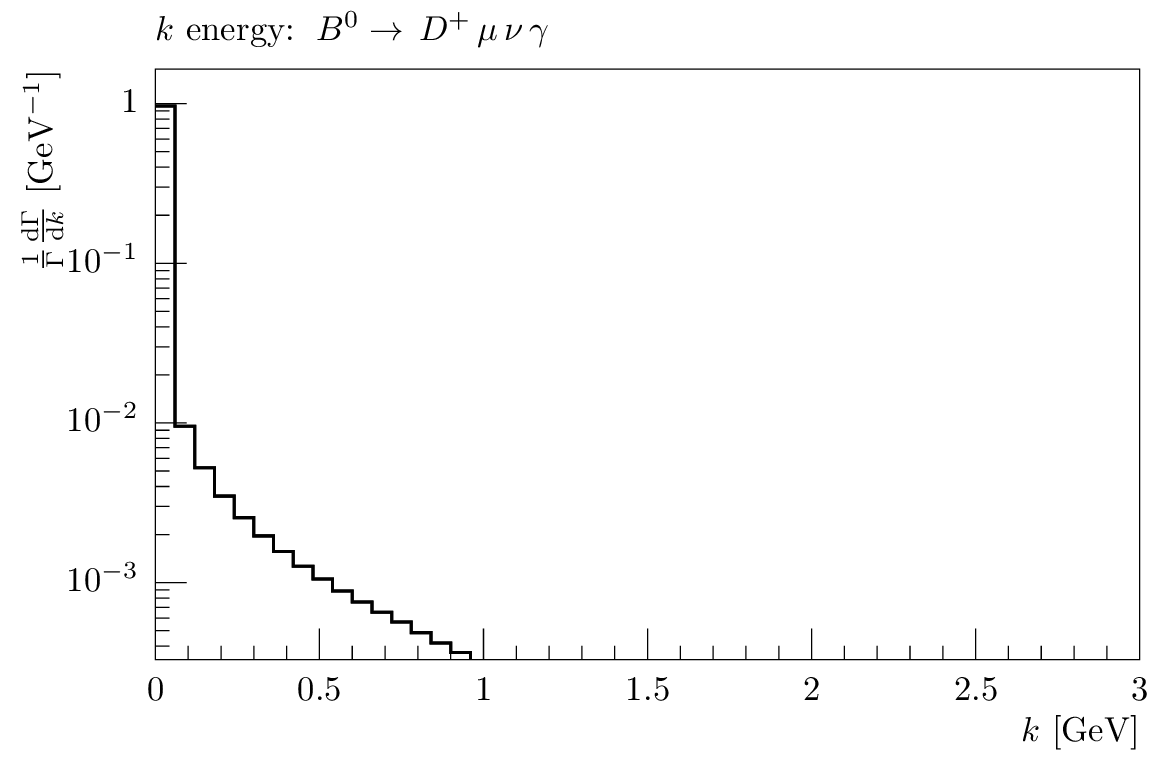} 
 \caption{ $B^0 \to \bar D^+ \, l \, \nu \, (\g)$: The predicted lepton and $D$ three-momentum magnitudes, $p_l$ and $p$, for tree-level and next-to-leading order and the logarithmic photon energy $k$ distribution are shown.   }   \label{hqetpartialratepred1}
 \end{center}
 \end{figure}

\begin{figure}[Pht!]\begin{center}  
\vspace{0.5cm}
 \unitlength = 1mm
 \includegraphics[width=0.49\textwidth]{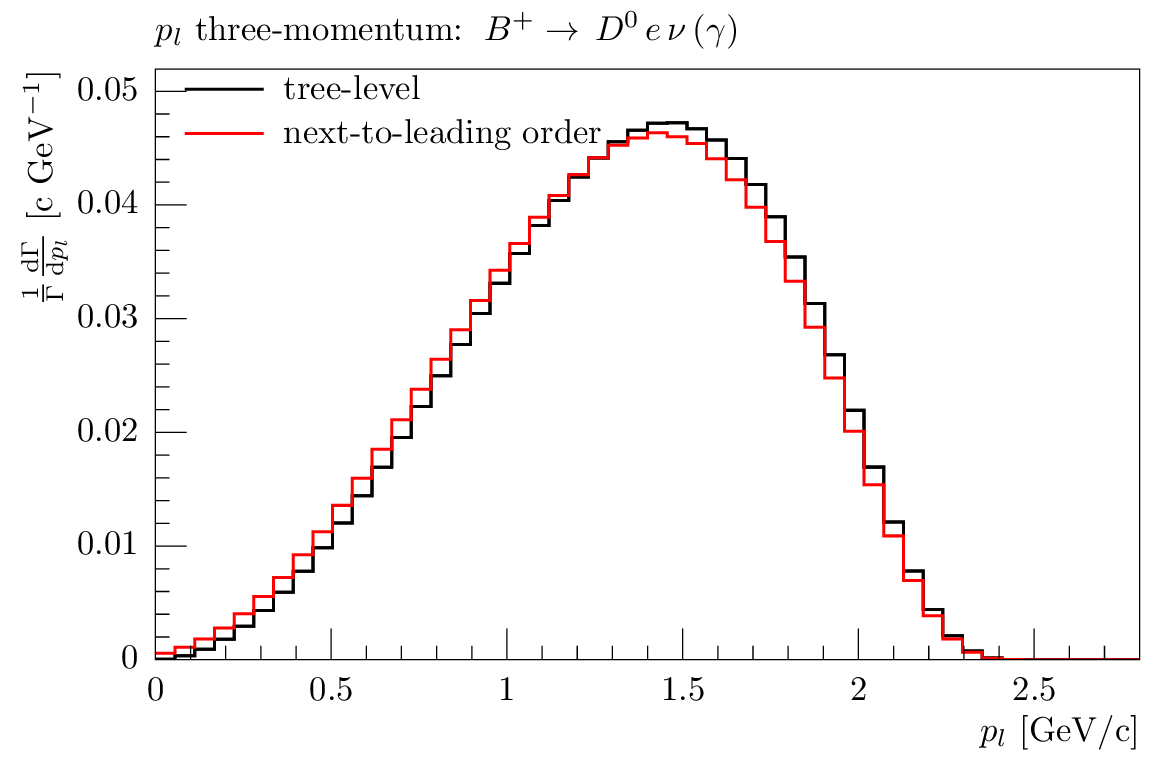} 
 \includegraphics[width=0.49\textwidth]{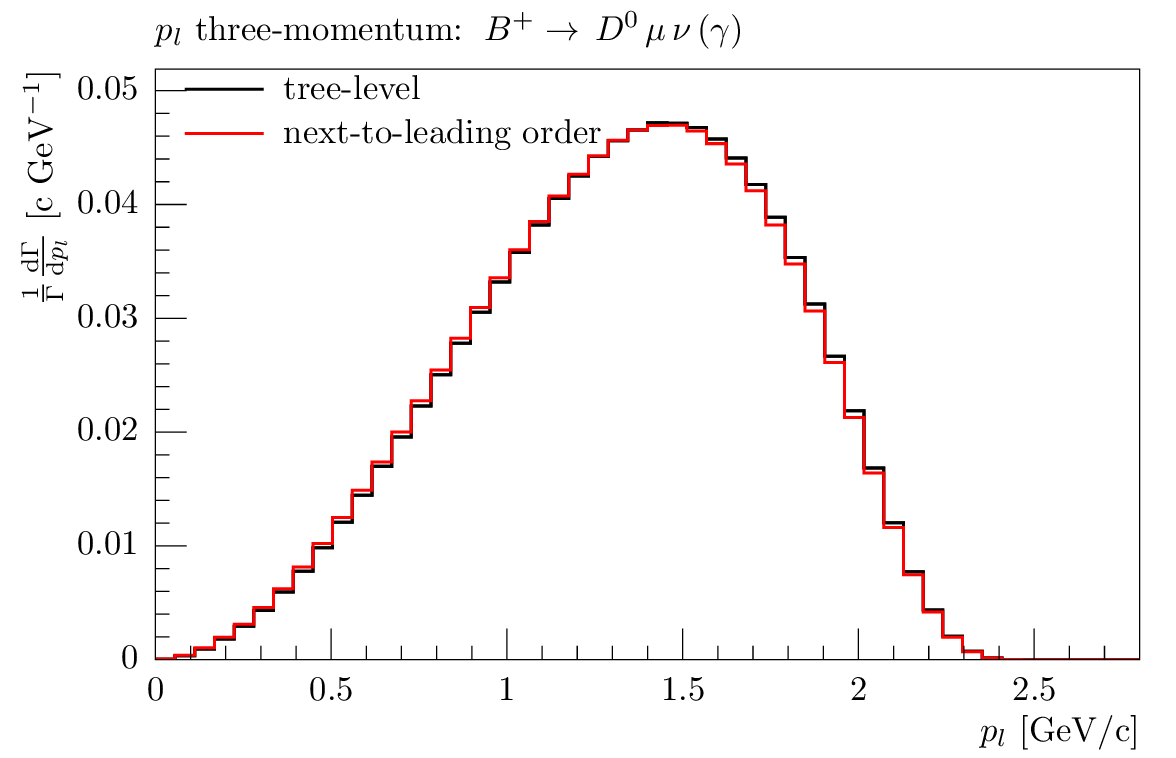} \\
 \includegraphics[width=0.49\textwidth]{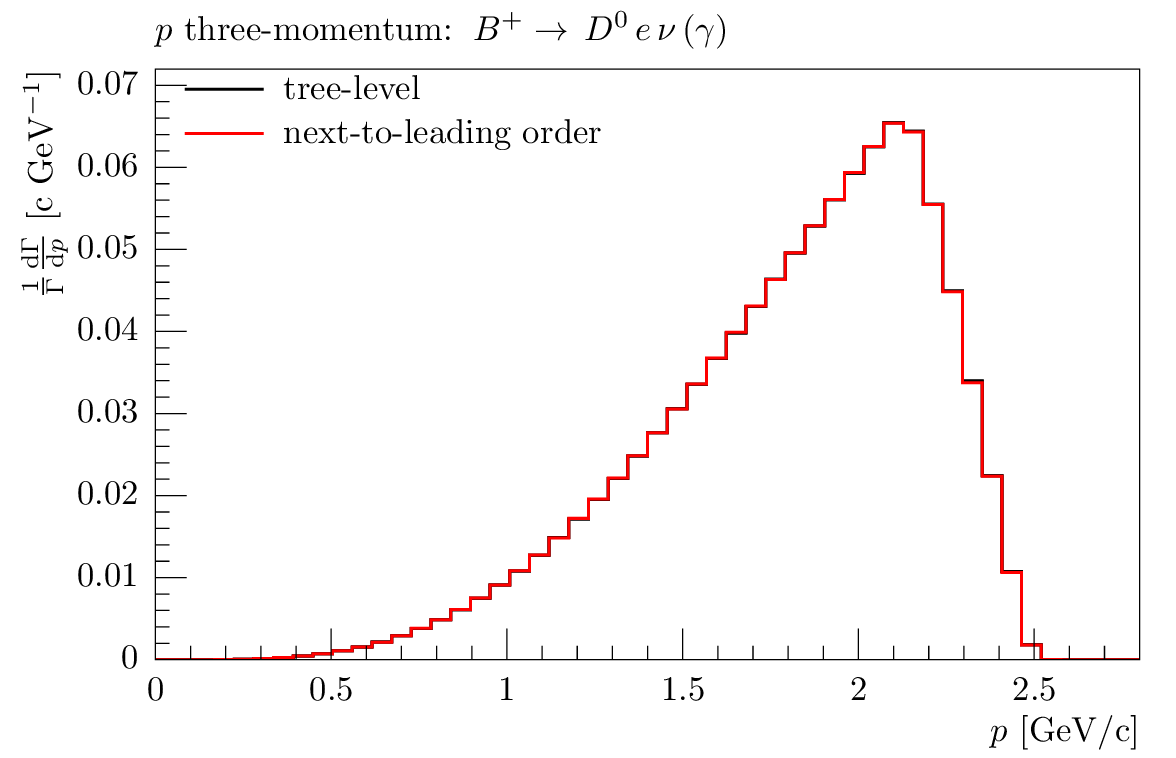} 
 \includegraphics[width=0.49\textwidth]{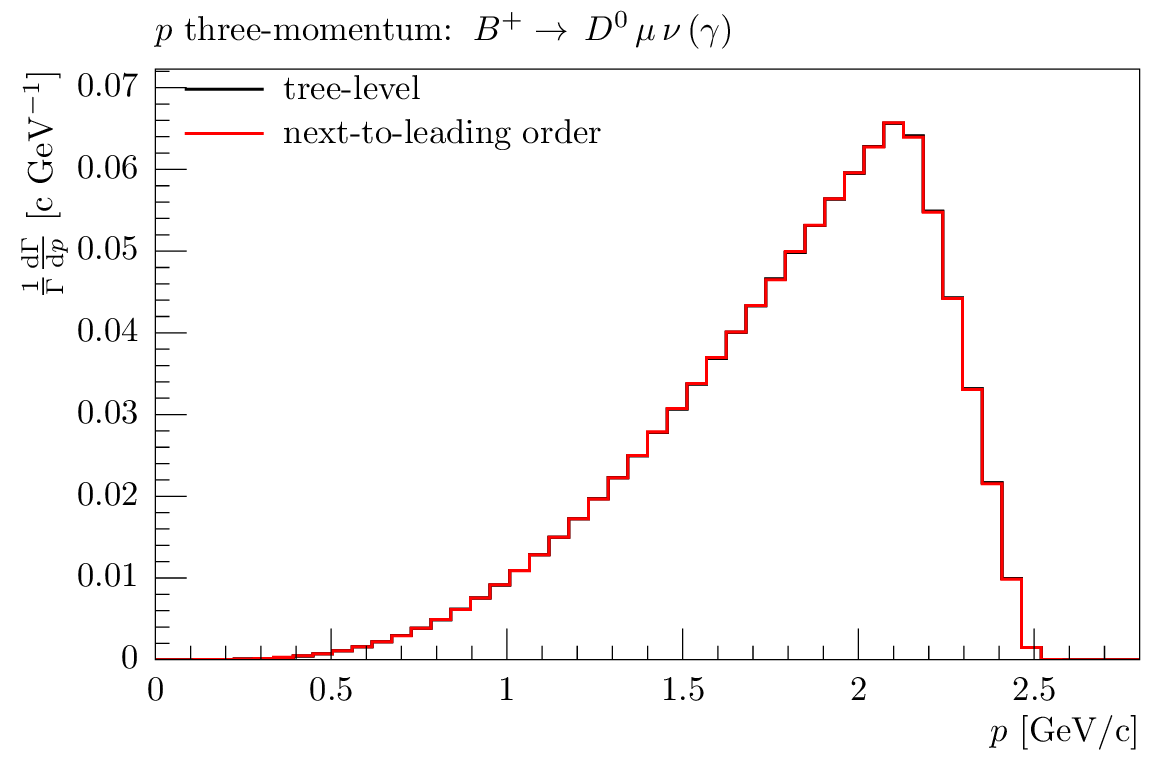} \\
 \includegraphics[width=0.49\textwidth]{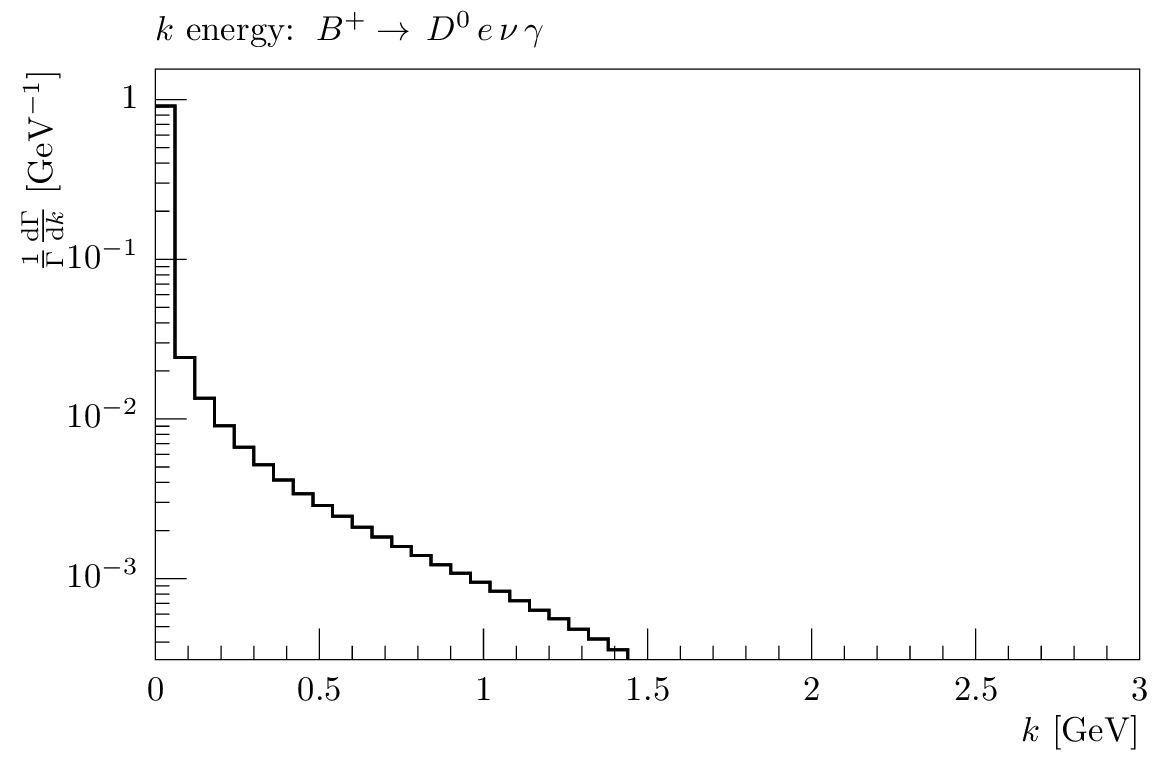} 
 \includegraphics[width=0.49\textwidth]{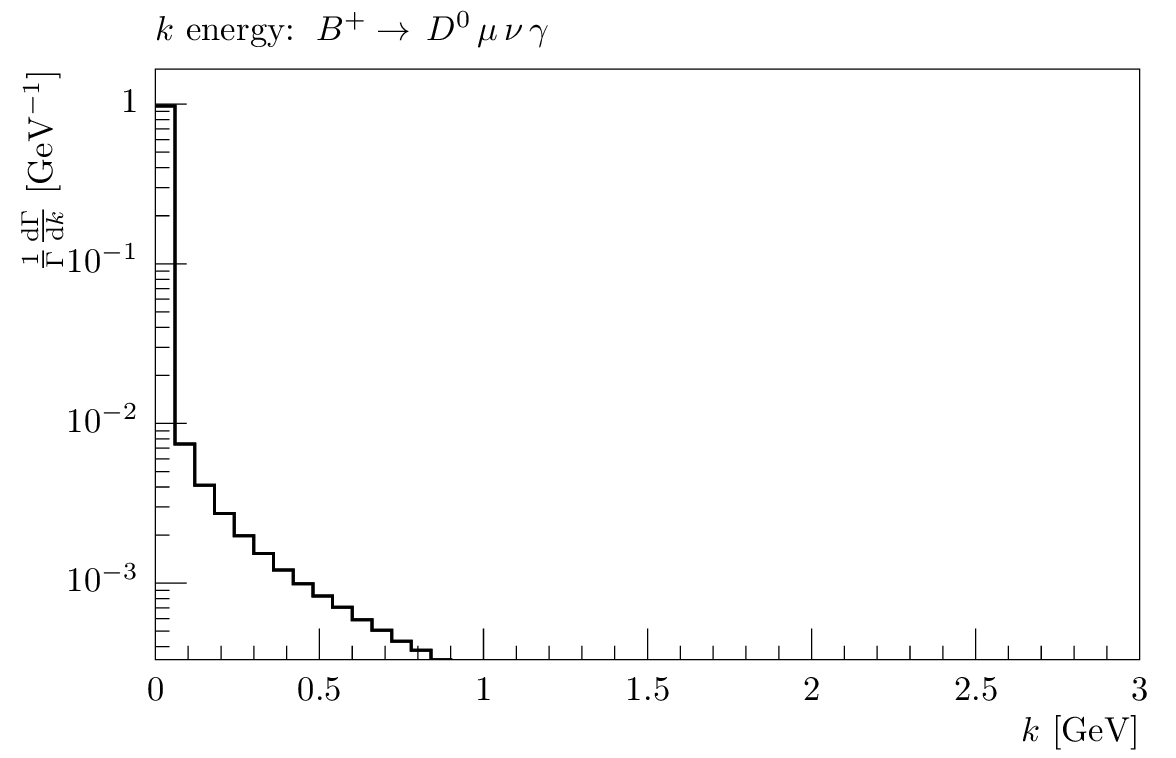} 
 \caption{ $B^+ \to \bar D^0 \, l \, \nu\, (\g)$: The predicted lepton and $D$ three-momentum magnitudes, $p_l$ and $p$, for tree-level and next-to-leading order and the logarithmic photon energy $k$ distribution are shown.  }   \label{hqetpartialratepred3}
 \end{center}
 \end{figure}

Fig. \ref{hqetpartialratepred1} depict the results for $B^0 \to \bar D^+ \, e \, \nu \, (\g)$ and  $B^0 \to \bar D^+ \, \mu \, \nu \, (\g)$ decays. Fig. \ref{hqetpartialratepred3} depict correspondingly the $B^+ \to \bar D^0 \, e \, \nu \, (\g)$ and  $B^+ \to \bar D^0 \, \mu \, \nu \, (\g)$ decays. The very massive $D$-meson only receives small corrections due to radiative effects, whereas the electron three-momentum distributions are notably shifted towards smaller three-momentum. The correction on the muon three-momentum distribution are smaller, due to its larger mass in comparison to the electron.

\begin{figure}[Pht!]\begin{center}  
\vspace{0.5cm}
 \unitlength = 1mm
 \includegraphics[width= 0.49\textwidth]{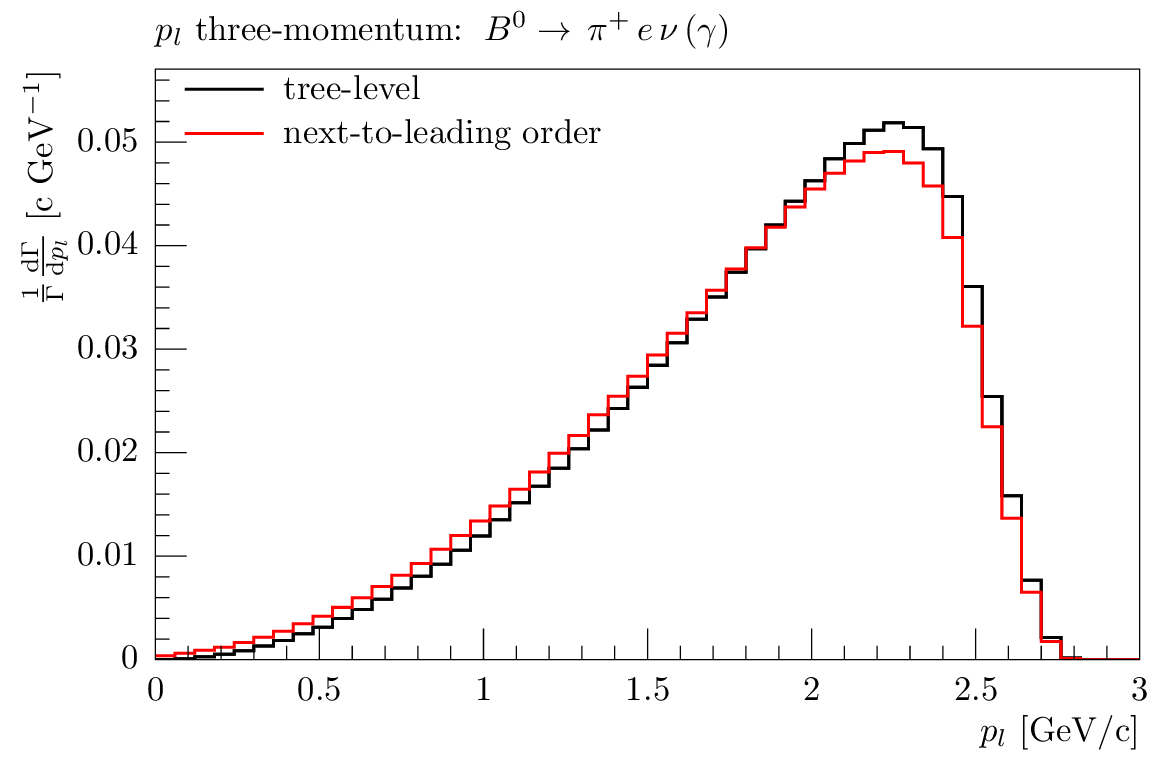} 
 \includegraphics[width= 0.49\textwidth]{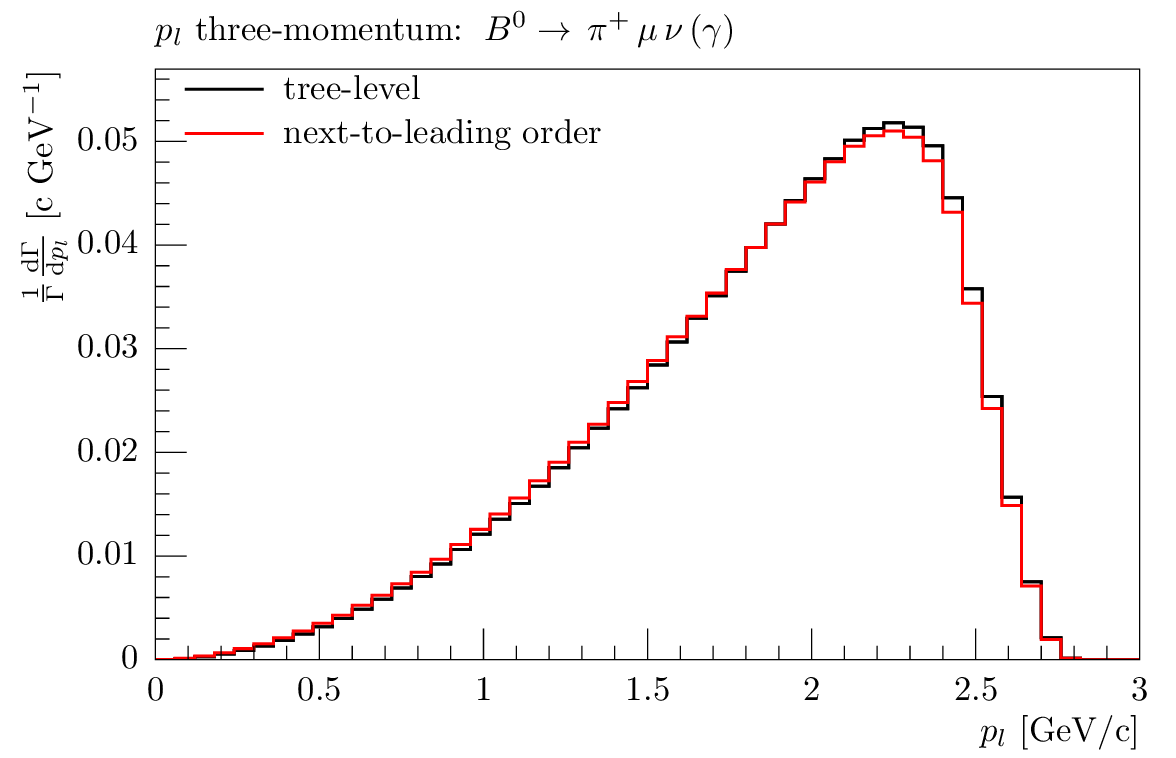} \\
 \includegraphics[width= 0.49\textwidth]{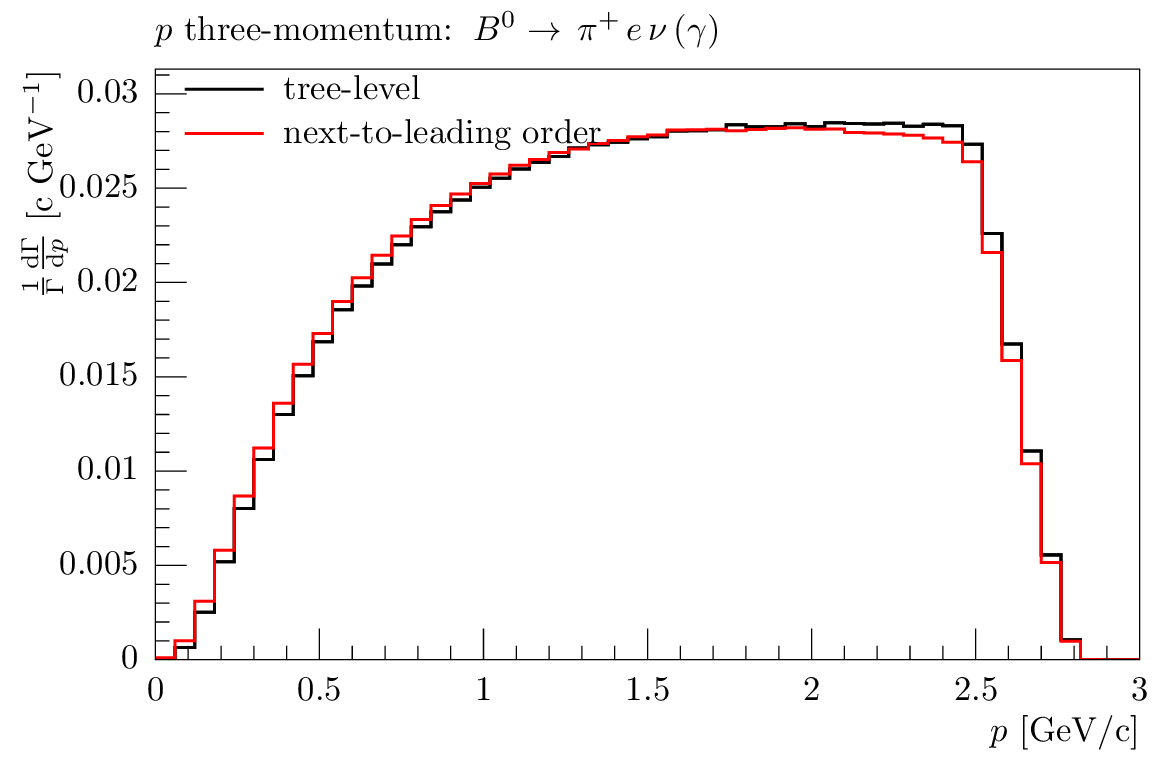} 
 \includegraphics[width= 0.49\textwidth]{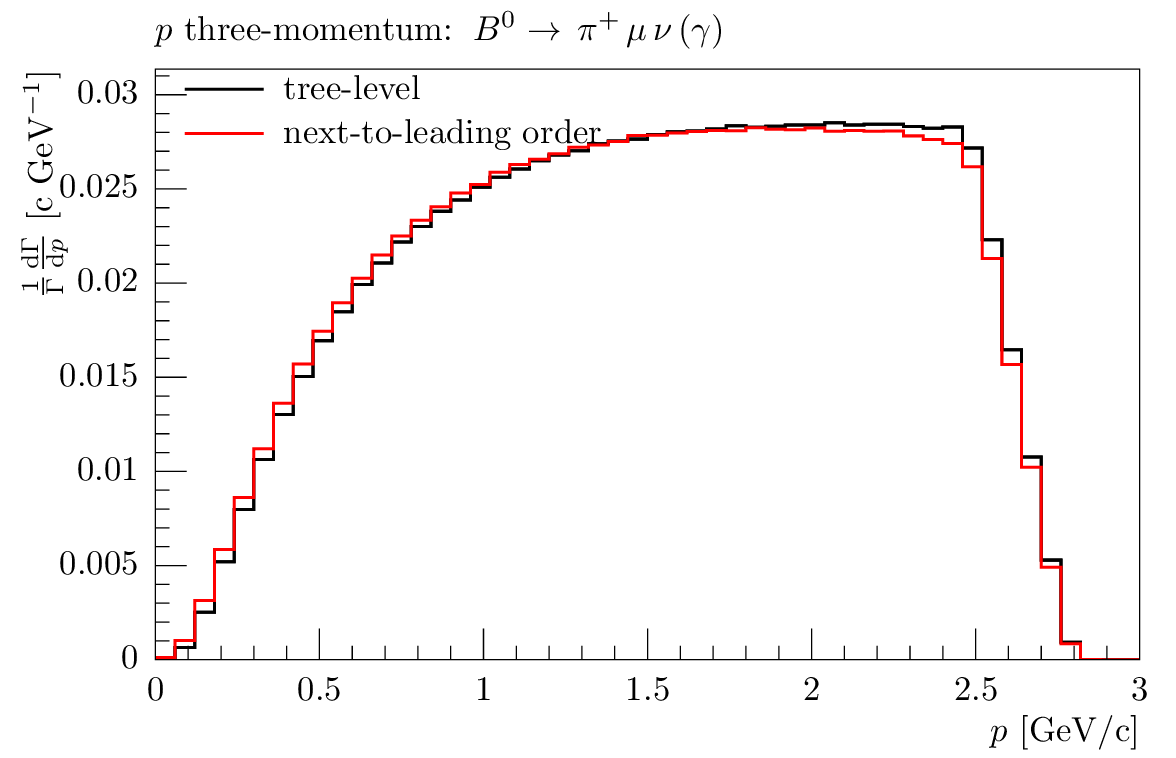} \\
 \includegraphics[width= 0.49\textwidth]{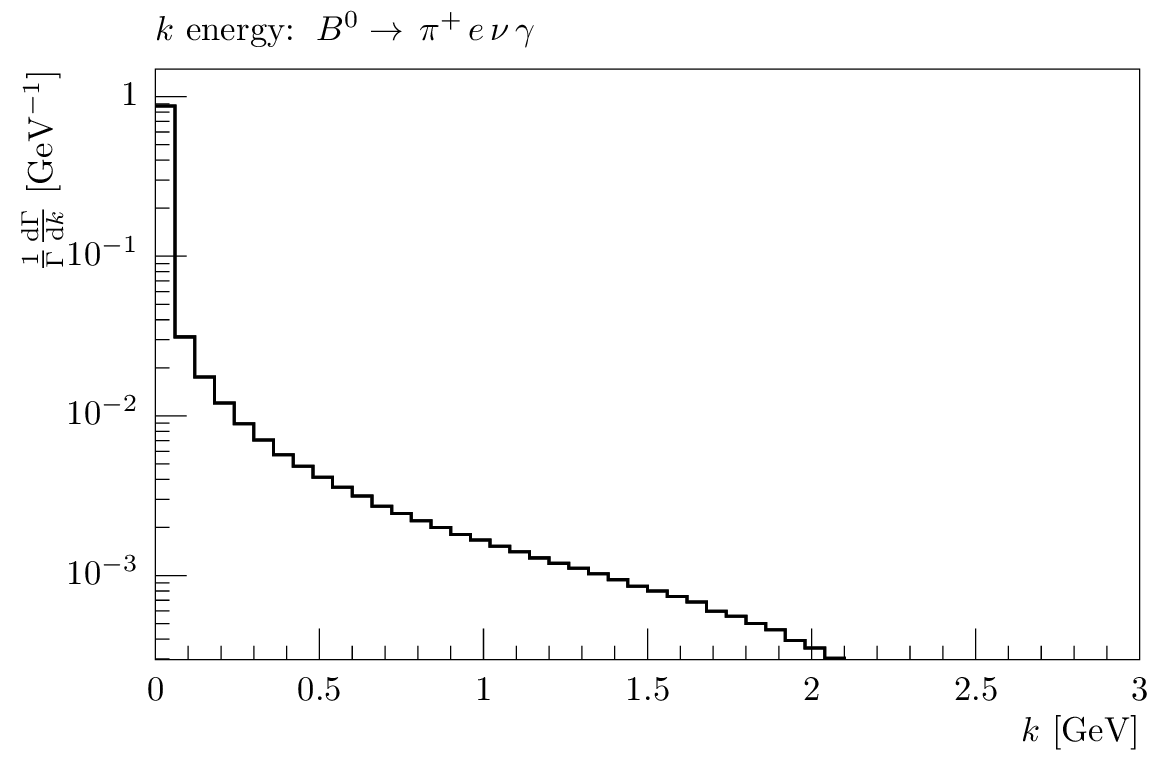} 
 \includegraphics[width= 0.49\textwidth]{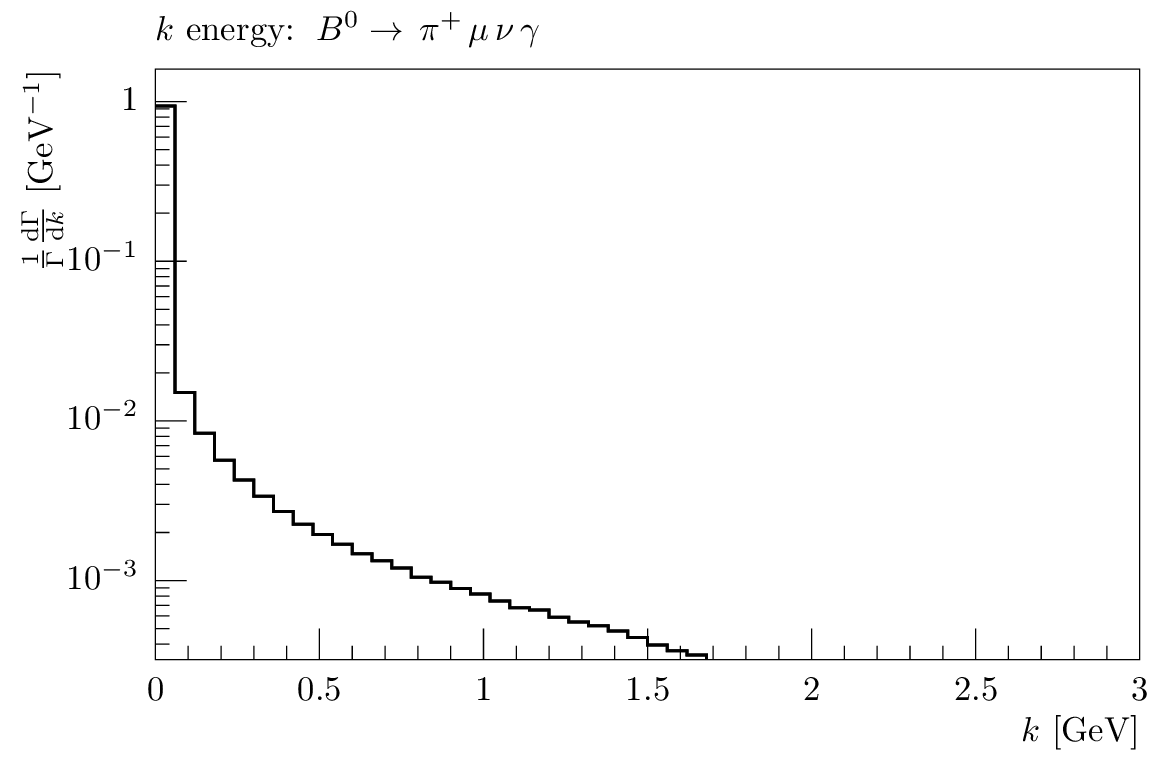} 
 \caption{ $B^0 \to \pi^+ \, l \, \nu \, (\g)$: The predicted lepton and $\pi$ three-momentum magnitudes, $p_l$ and $p$, for tree-level and next-to-leading order and the logarithmic photon energy $k$ distribution are shown.   }   \label{bzpartialratepred1}
 \end{center}
 \end{figure}

\begin{figure}[Pht!]\begin{center}  
\vspace{0.5cm}
 \unitlength = 1mm
 \includegraphics[width= 0.49\textwidth]{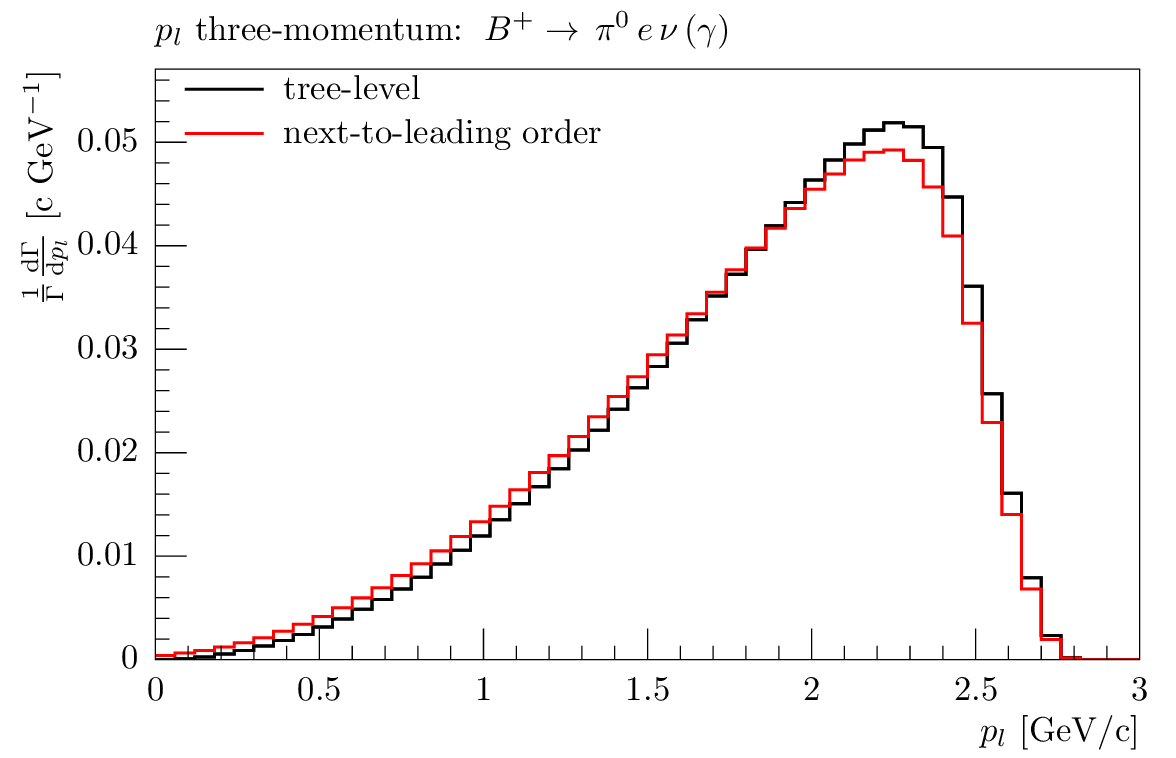} 
 \includegraphics[width= 0.49\textwidth]{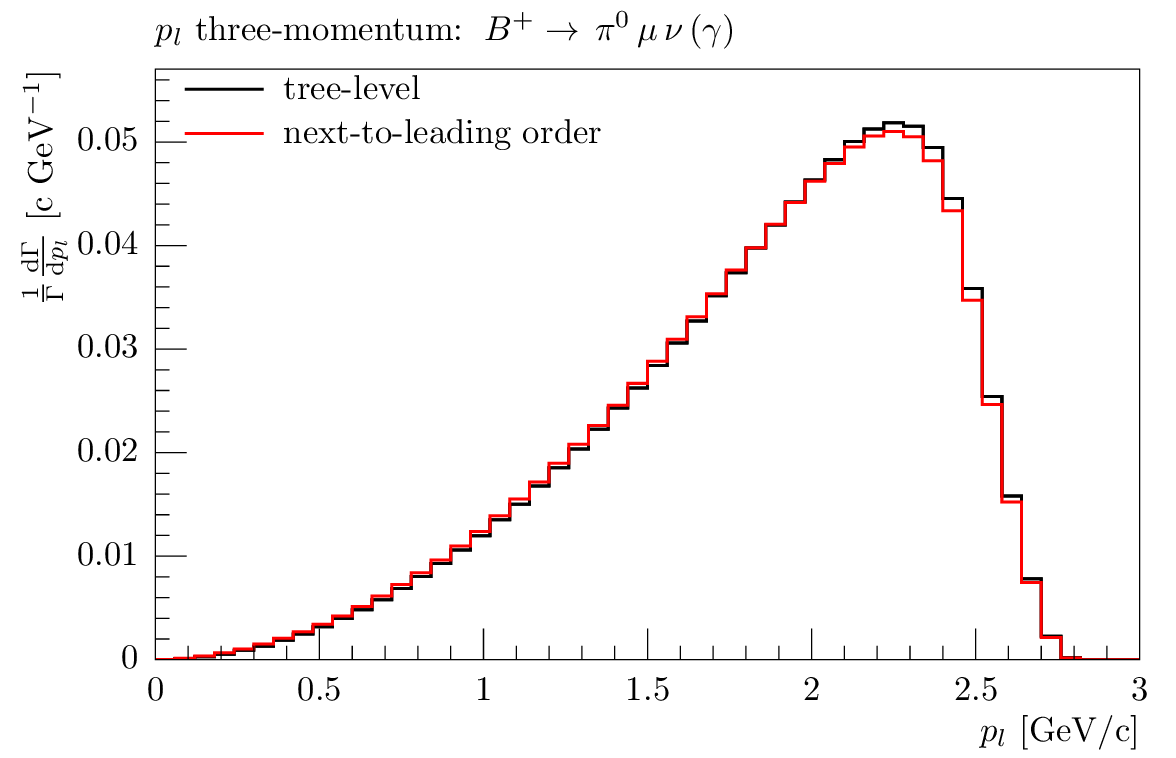} \\
 \includegraphics[width= 0.49\textwidth]{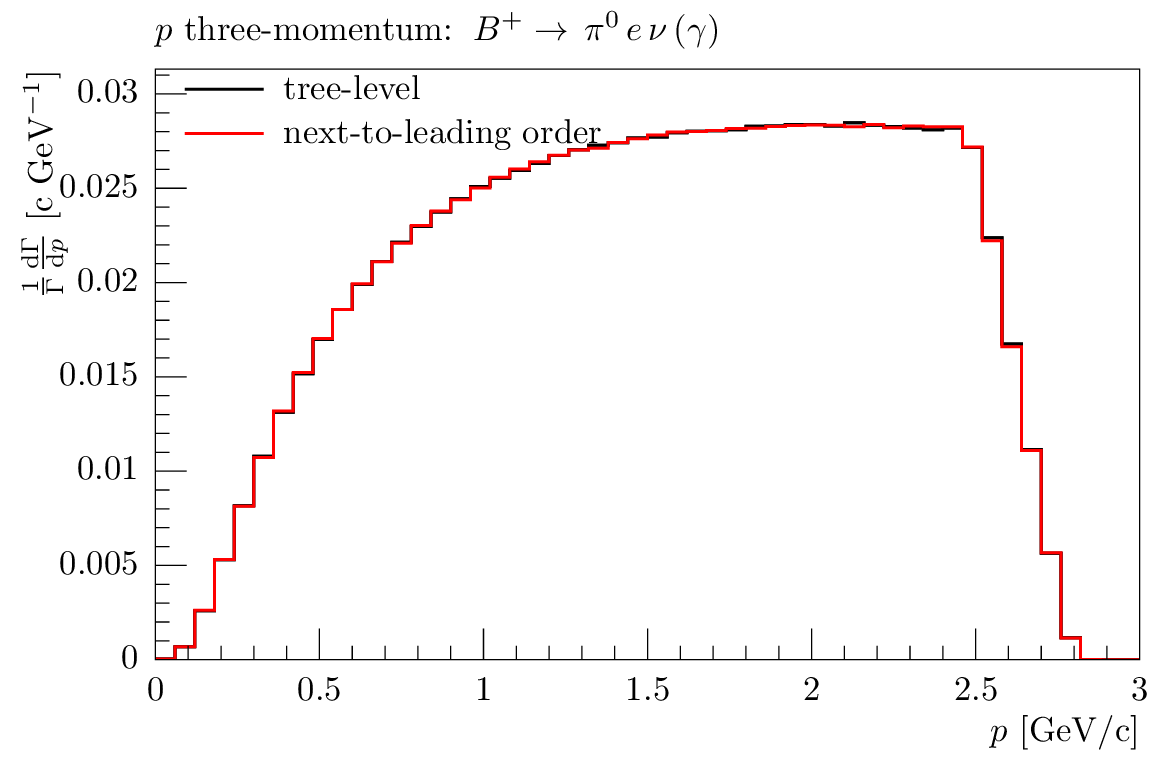} 
 \includegraphics[width= 0.49\textwidth]{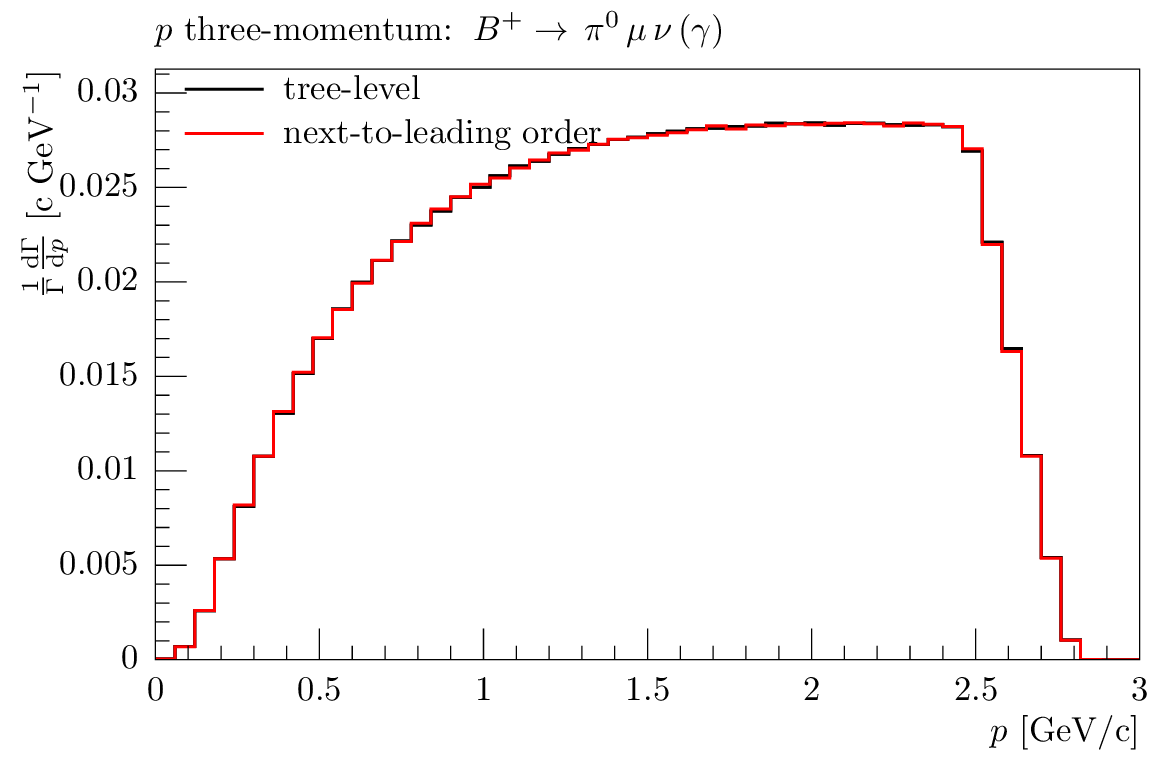} \\
 \includegraphics[width= 0.49\textwidth]{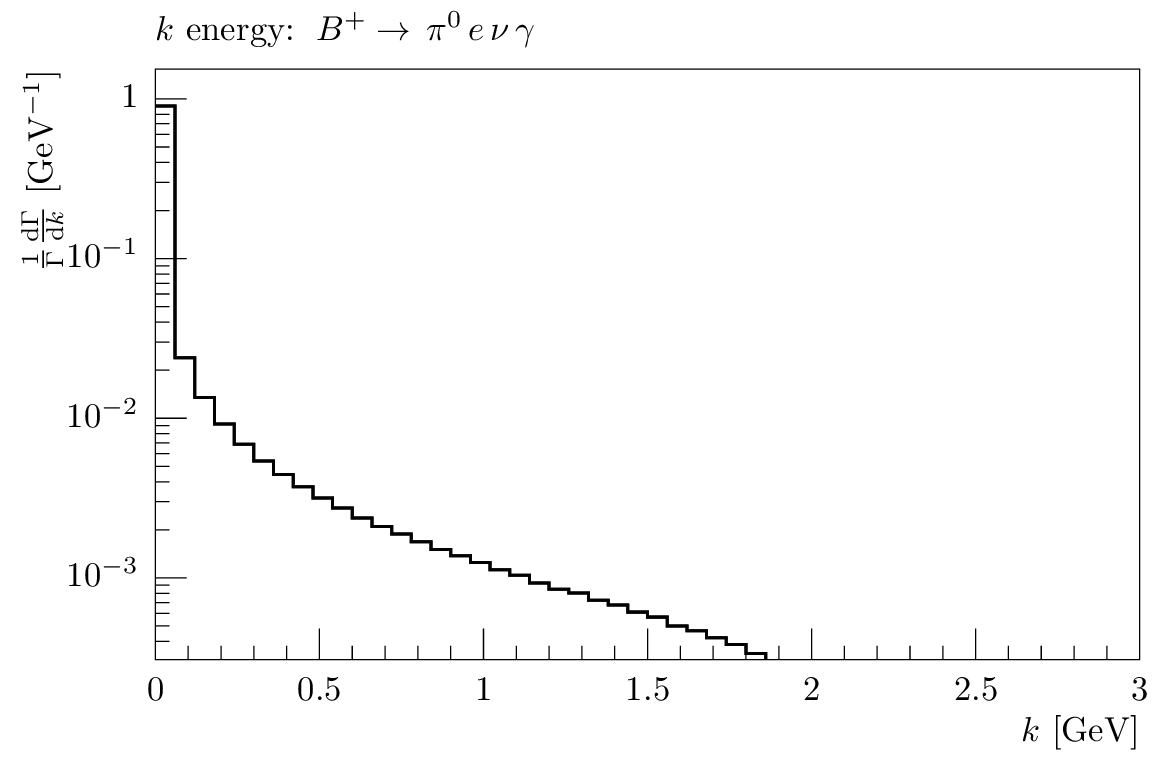} 
 \includegraphics[width= 0.49\textwidth]{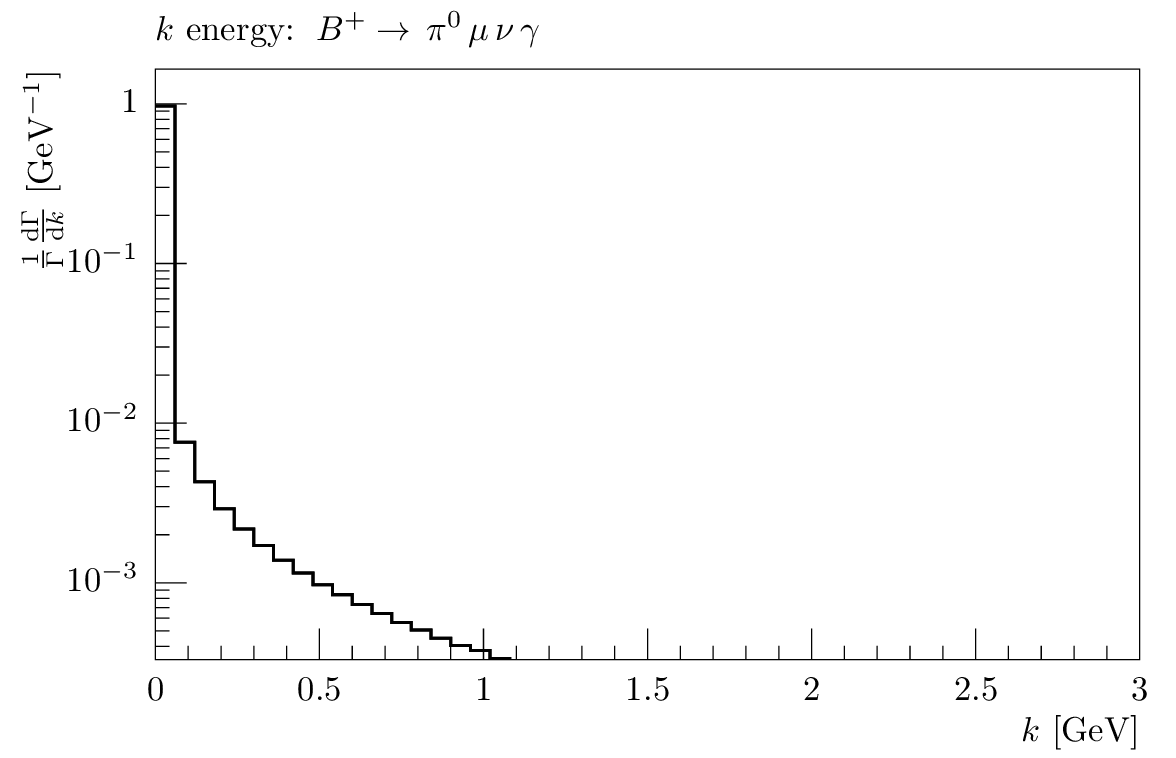} 
 \caption{ $B^+ \to \pi^0 \, l \, \nu\, (\g)$:  The predicted lepton and $\pi$ three-momentum magnitudes, $p_l$ and $p$, for tree-level and next-to-leading order and the logarithmic photon energy $k$ distribution are shown.   }   \label{bzpartialratepred3}
 \end{center}
 \end{figure}

Fig. \ref{bzpartialratepred1}  depict the predictions for the $B^0 \to \pi^+ \, e \, \nu \, (\g)$ and  $B^0 \to \pi^+ \, \mu \, \nu \, (\g)$ decay modes. The charged pion final state radiates considerably and receives corrections in its shape. This is caused by the dependence on the radiative four-momentum transfer squared $t' = \bve{p_B - p - k}^2$ of the IB terms. The electron three-momentum distributions are shifted towards lower momenta, and the muon three-momentum distributions receive small corrections. Fig. \ref{bzpartialratepred3} show the $B^+ \to \pi^0 \, e \, \nu \, (\g)$ and  $B^+ \to \pi^0 \, \mu \, \nu \, (\g)$ predictions. The uncharged pion receives negligible corrections from the radiative four-momentum transfer squared $t' = \bve{p_B - p -k}^2$ due to the large mass of the $B$-meson, that suppresses considerably real IB next-to-leading order corrections.

\begin{figure}[Pht!]\begin{center}  
\vspace{0.5cm}
 \unitlength = 1mm
 \includegraphics[width= 0.49\textwidth]{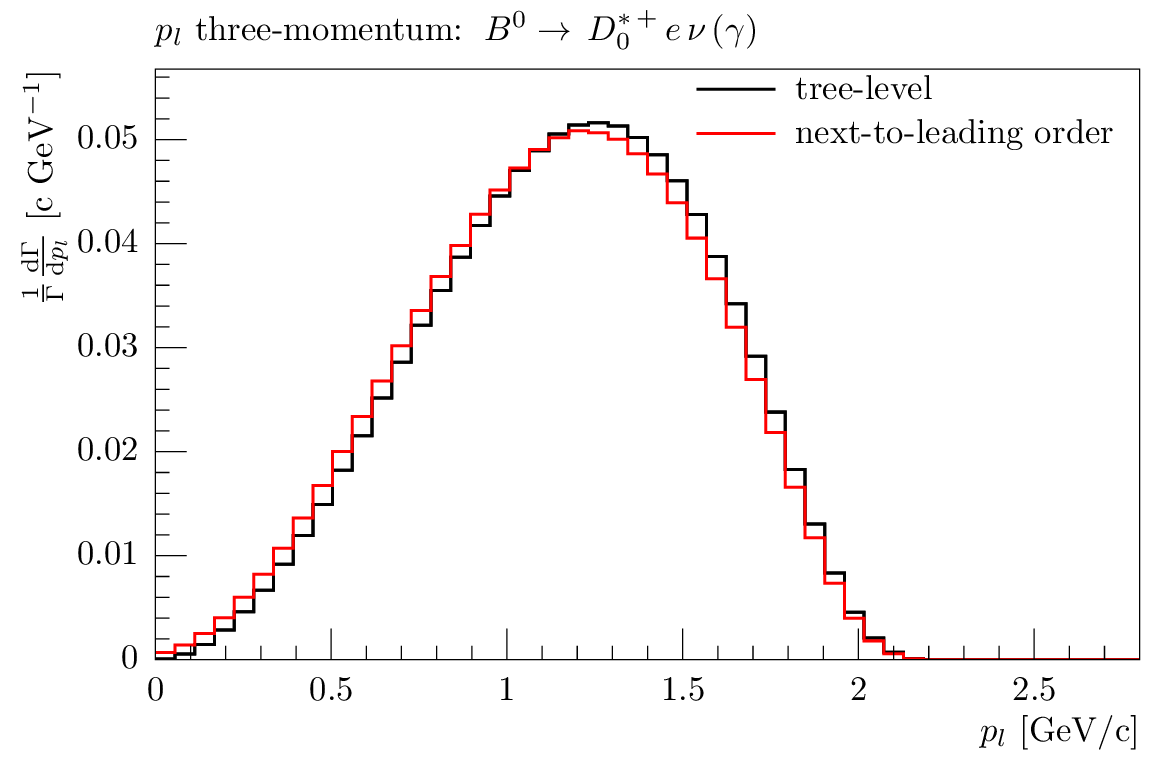} 
 \includegraphics[width= 0.49\textwidth]{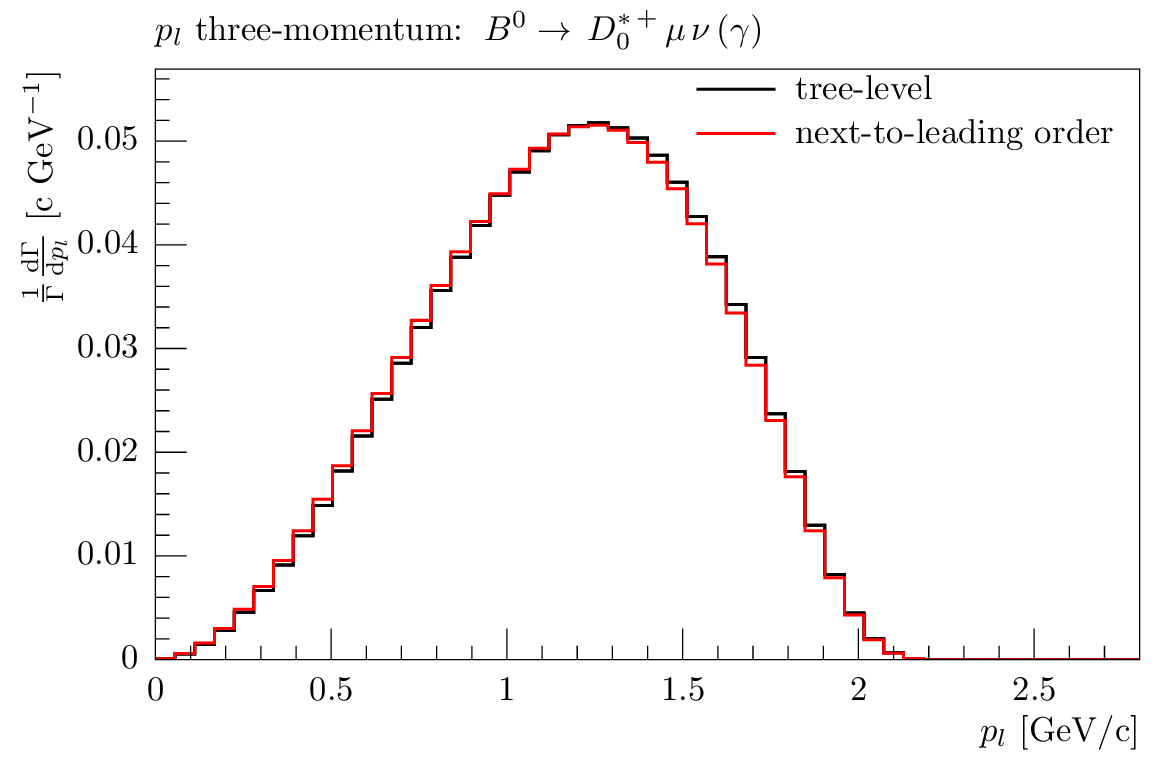} \\
 \includegraphics[width= 0.49\textwidth]{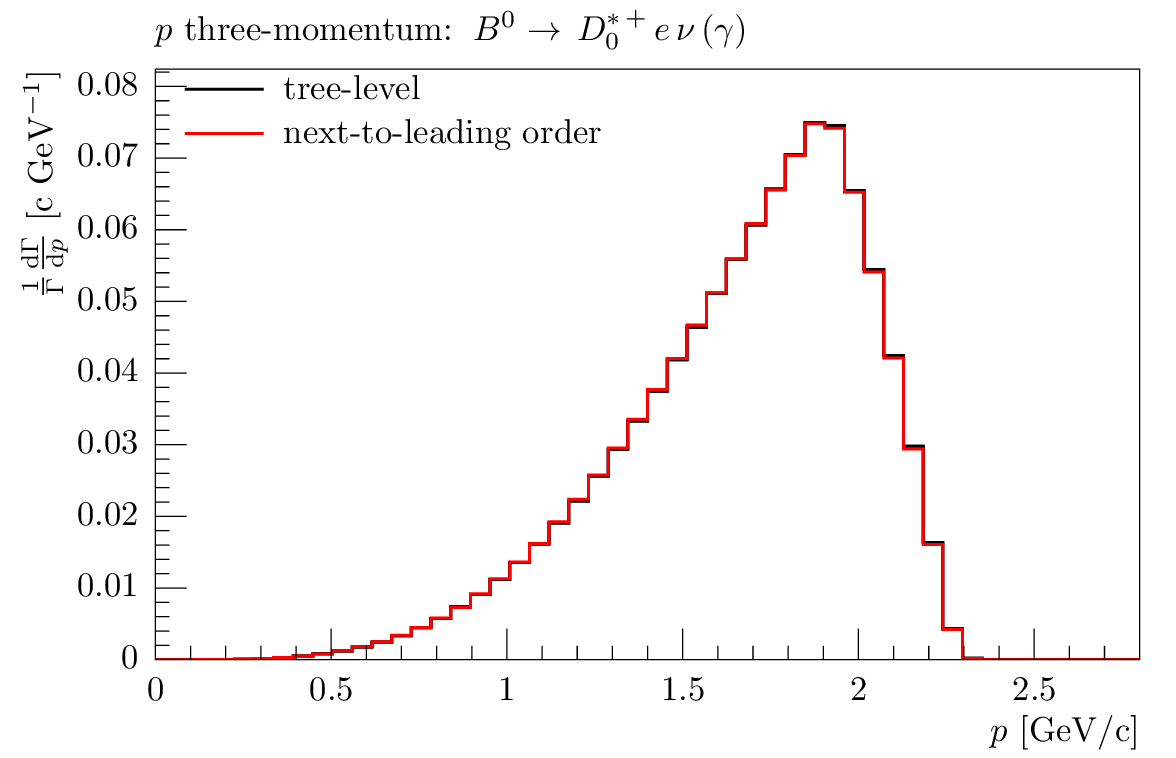} 
 \includegraphics[width= 0.49\textwidth]{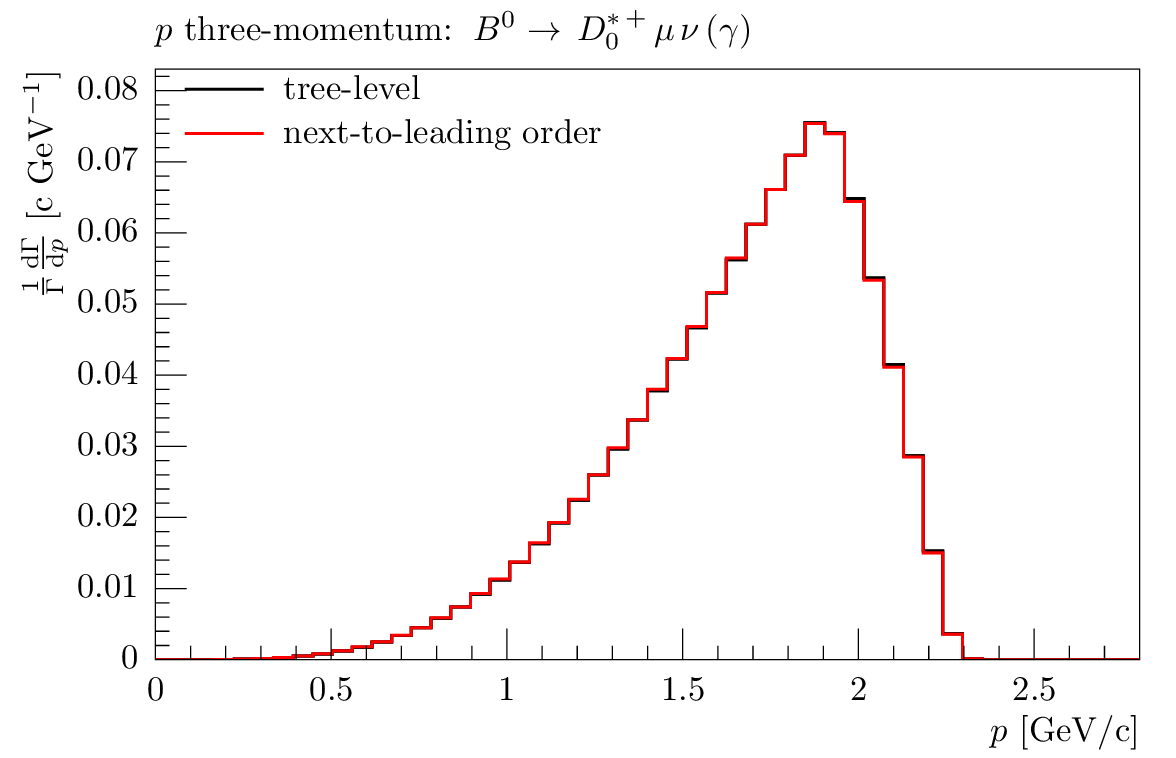} \\
 \includegraphics[width= 0.49\textwidth]{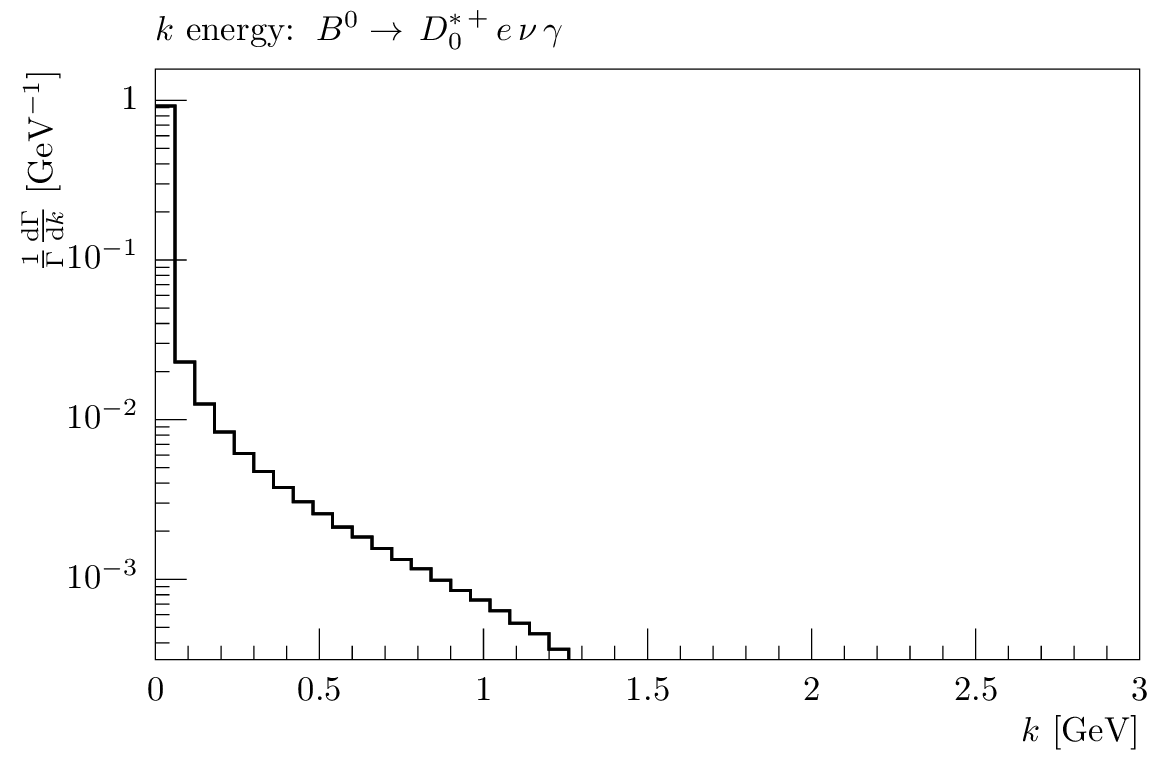} 
 \includegraphics[width= 0.49\textwidth]{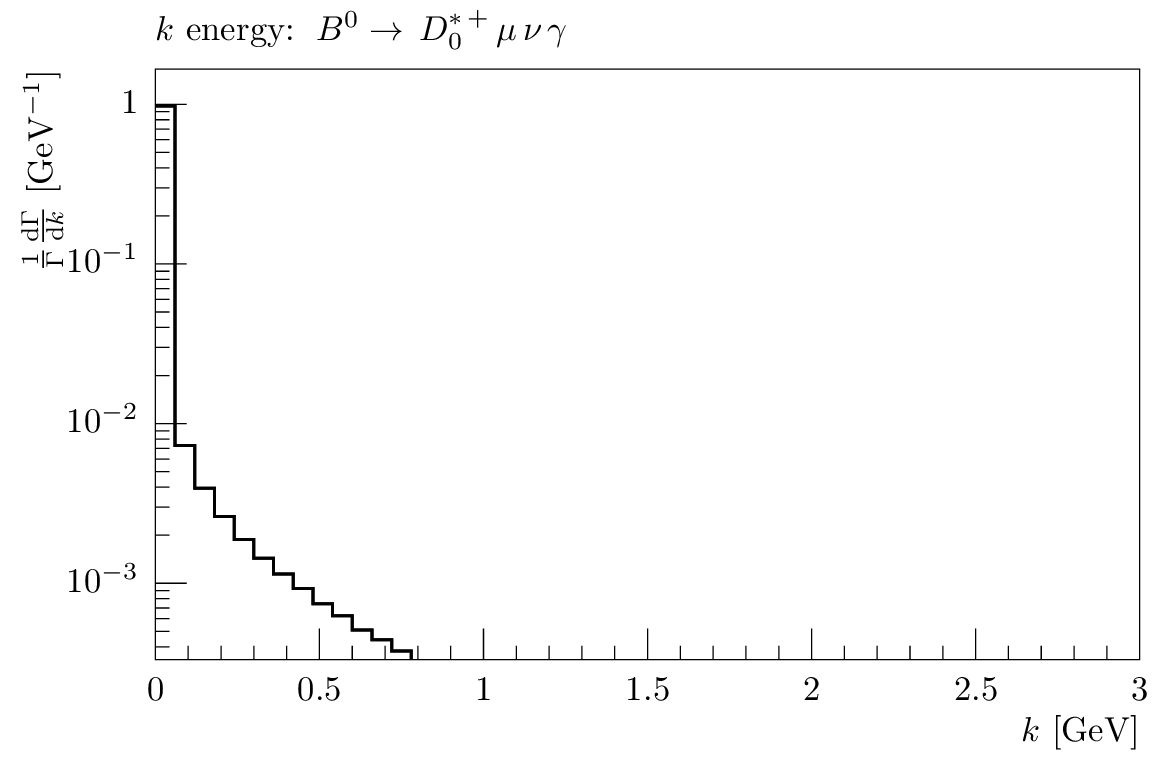} 
 \caption{ $B^0 \to \bar D^{*\, +}_0 \, l \, \nu\, (\g)$: The predicted lepton and $D^*_0$ three-momentum magnitudes, $p_l$ and $p$, for tree-level and next-to-leading order and the logarithmic photon energy $k$ distribution are shown.  }   \label{llswpartialratepred1}
 \end{center}
 \end{figure}

\begin{figure}[Pht!]\begin{center}  
\vspace{0.5cm}
 \unitlength = 1mm
 \includegraphics[width= 0.49\textwidth]{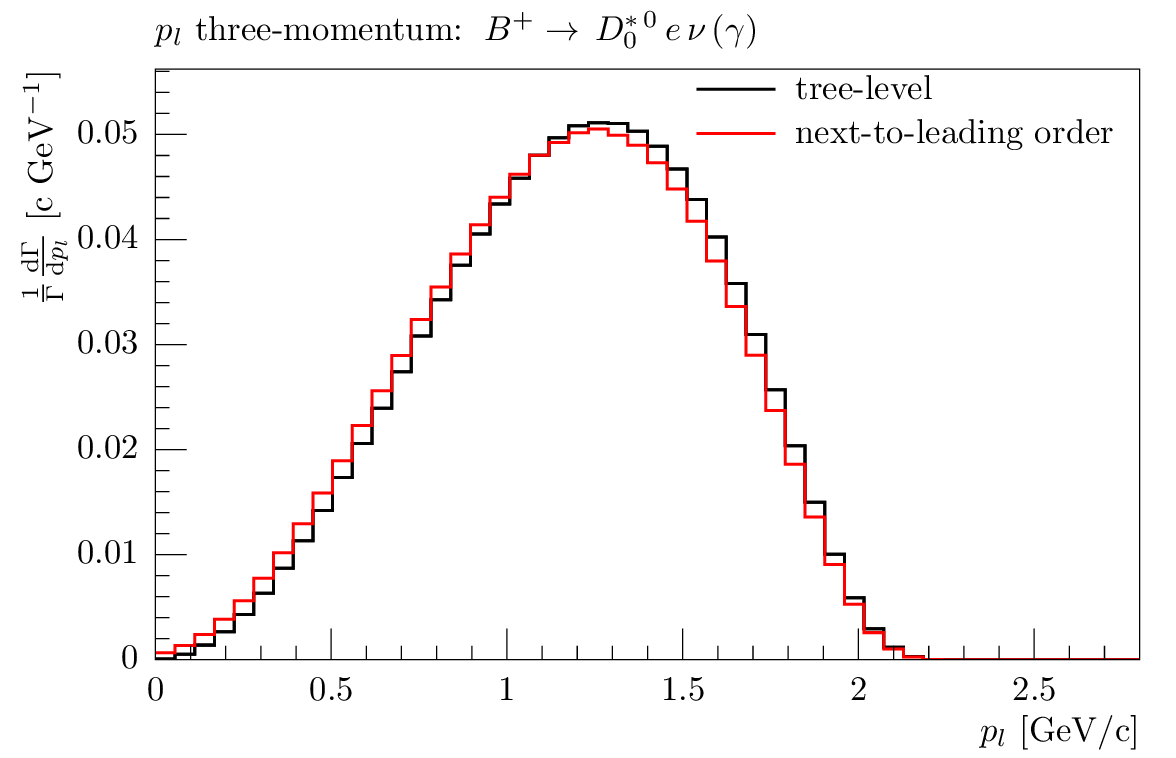} 
 \includegraphics[width= 0.49\textwidth]{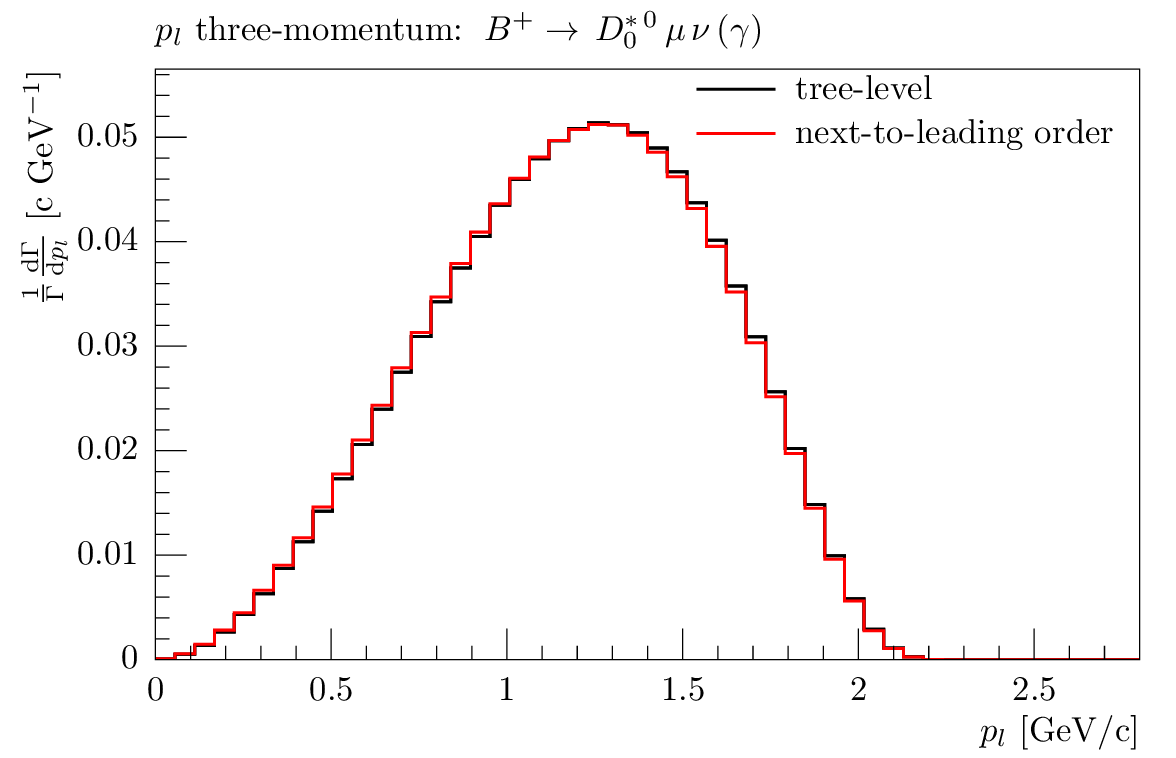} \\
 \includegraphics[width= 0.49\textwidth]{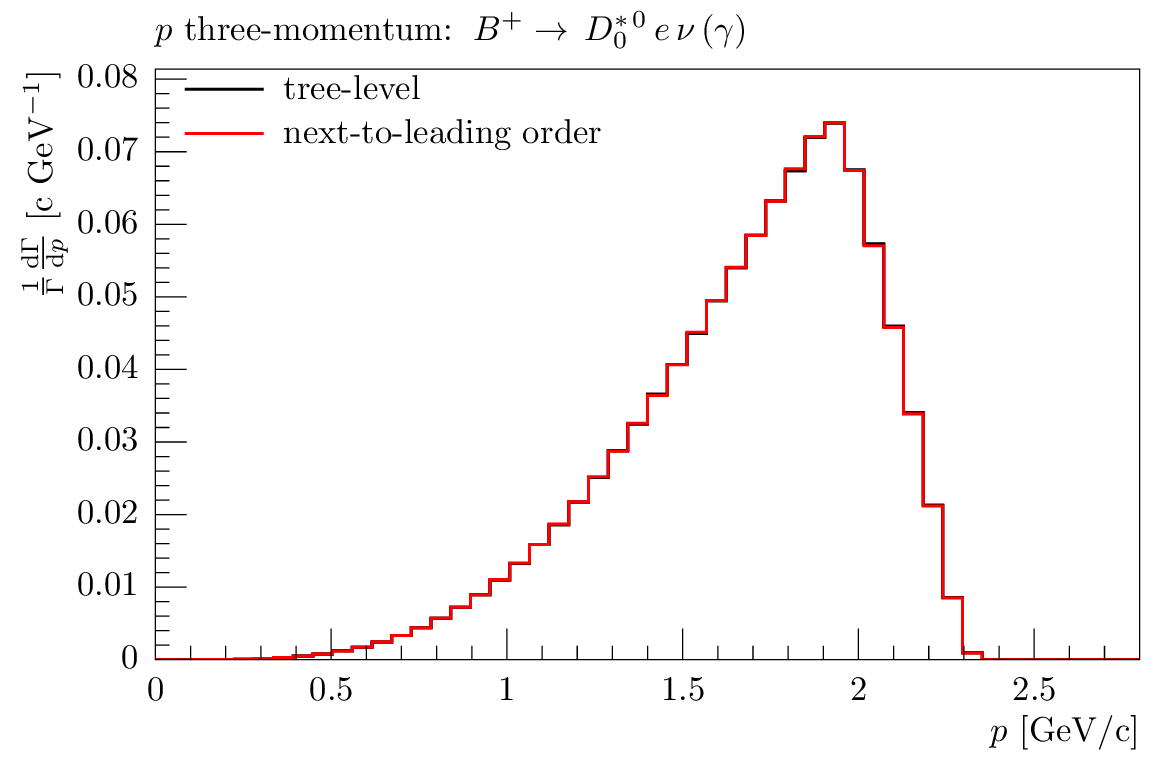} 
 \includegraphics[width= 0.49\textwidth]{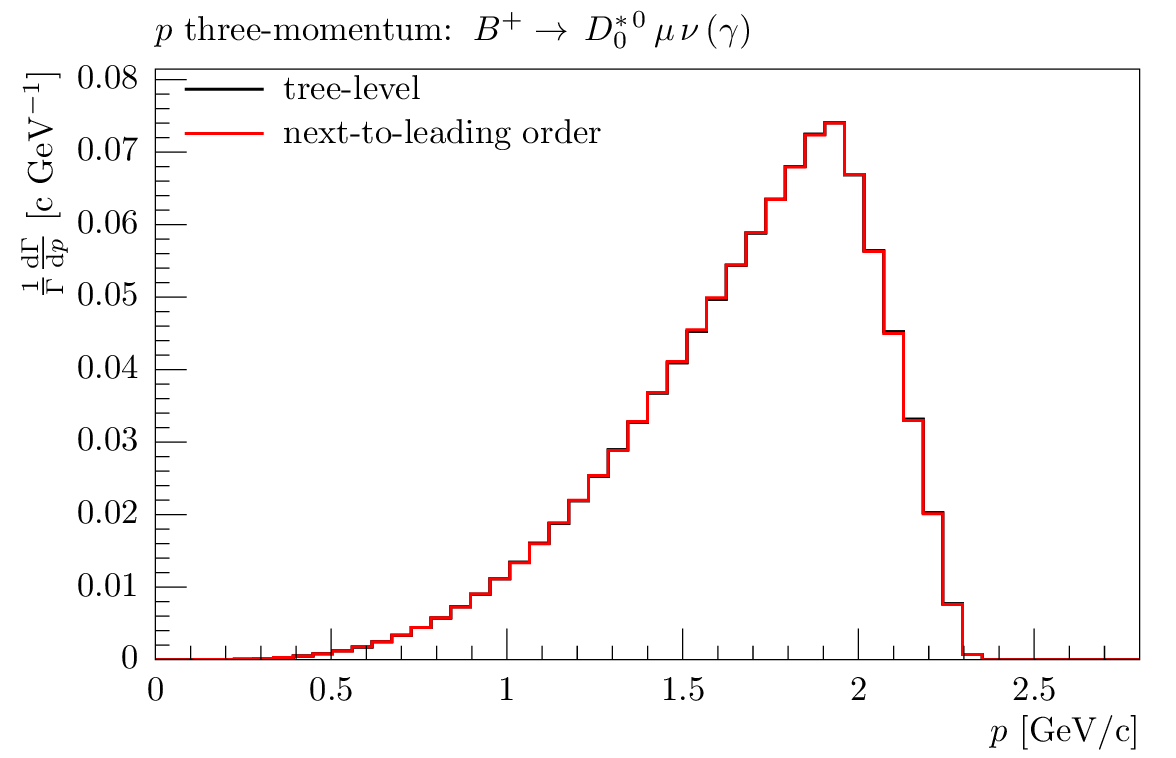} \\
 \includegraphics[width= 0.49\textwidth]{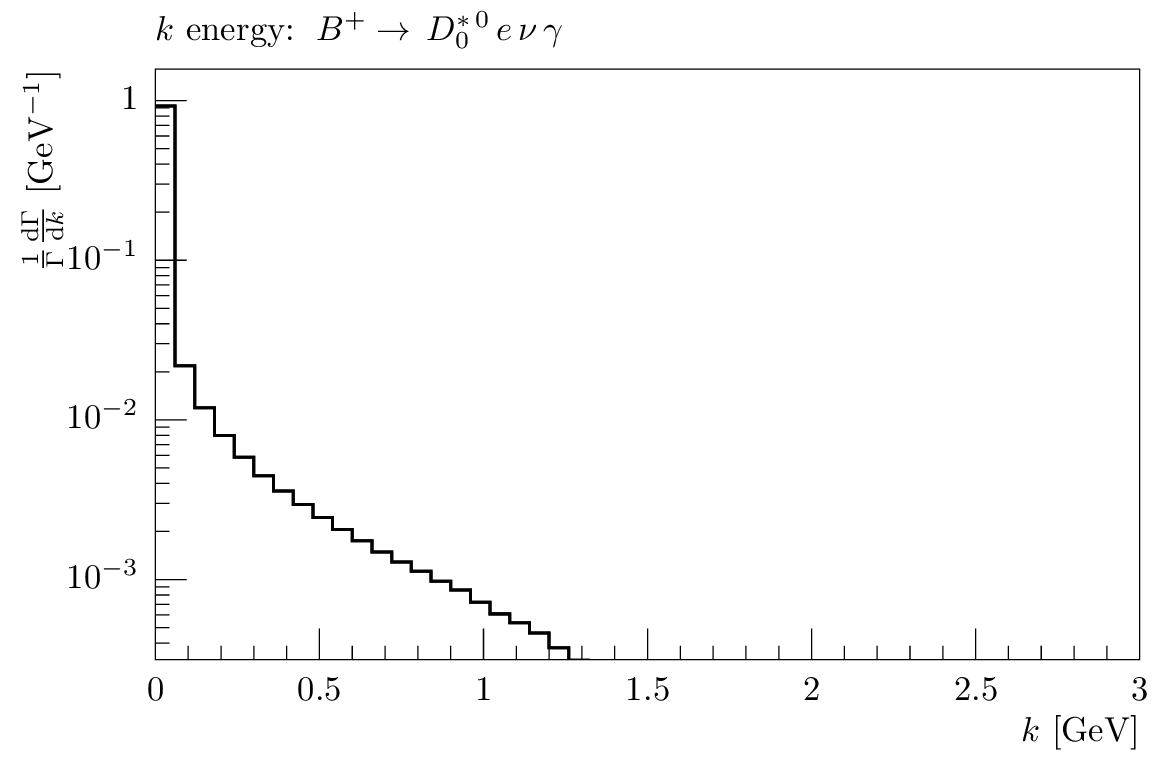} 
 \includegraphics[width= 0.49\textwidth]{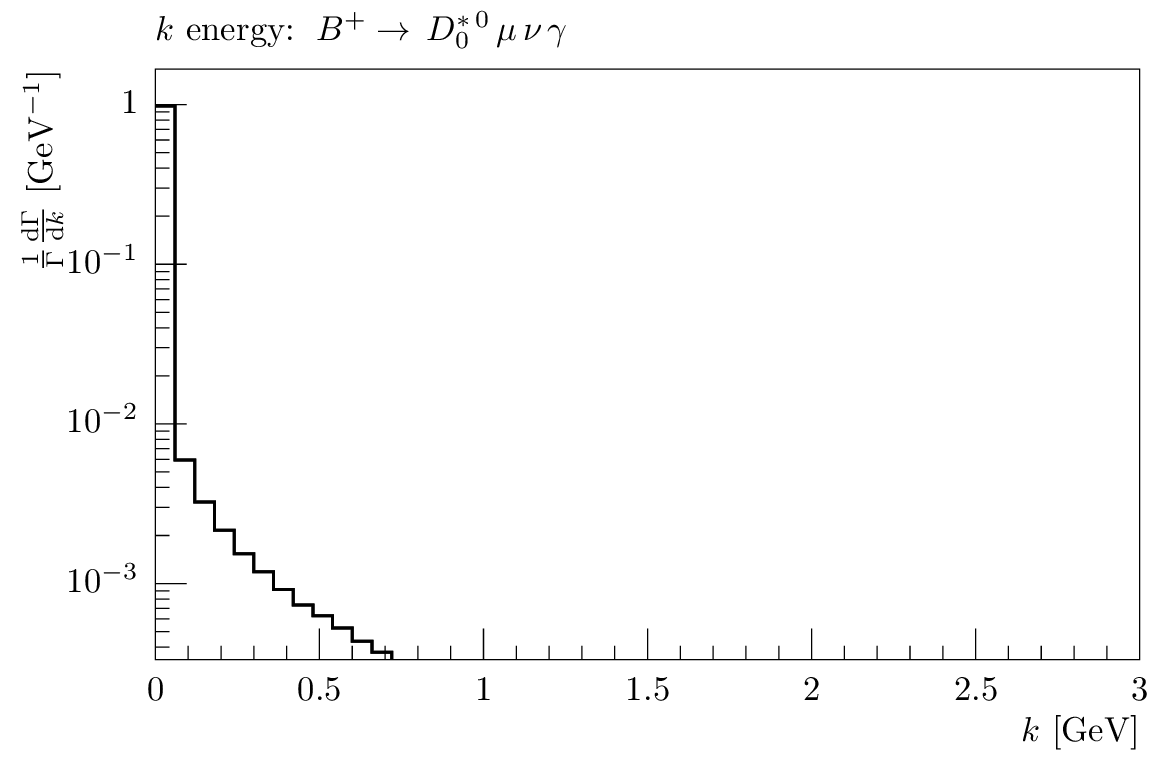} 
 \caption{ $B^+ \to \bar D^{*\, 0}_0 \, l \, \nu\, (\g)$: The predicted lepton and $D^*_0$ three-momentum magnitudes, $p_l$ and $p$, for tree-level and next-to-leading order and the logarithmic photon energy $k$ distribution are shown. }   \label{llswpartialratepred3}
 \end{center}
 \end{figure}

Figs. \ref{llswpartialratepred1} present the predictions for the $B^0 \to D_0^{*\,+} \, e \, \nu \, (\g)$ and  $B^0 \to D_0^{*\,+} \, \mu \, \nu \, (\g)$ differential decay rates. The three-momentum of the $D_0^{+\,*}$, similarly as the $D^+$ case, only receives small corrections, whereas the electron three-momentum distribution is shifted to lower momenta, and the muon distribution receives small corrections. Similarly, Fig. \ref{llswpartialratepred3} depict the predictions for the $B^+ \to \bar D_0^{*\,0} \, e \, \nu \, (\g)$ and  $B^+ \to \bar D_0^{*\,0} \, \mu \, \nu \, (\g)$ differential decay rates.
%
%
%

\section{Conclusion}\label{conclusion}
The next-to-leading order corrections for exclusive semileptonic $B$-meson decays into pseudoscalar and scalar final states were calculated in our model approach
 for $B\to D \, l \, \nu\, (\g)$,  $B\to \pi \, l \, \nu\, (\g)$, and $B\to D^*_0 \, l \, \nu\, (\g)$ decays. We predict the previously unknown total enhancement to the total decay rate $\delta_{\text{total}}$ for $B\to D \, l \, \nu\, (\g)$: averaged over lepton species the $B \to D \, l \, \nu\, (\g)$ total decay rate receives a global correction of $1.86(29)\%$ in comparison to the tree-level prediction. This results in the reduction of the extracted value of $|V_{\text{cb}}|$ of $-0.91(7)\%$ in comparison to the tree-level result of the total decay rate. For the exclusive $B^0 \to \pi^+ \, l \, \nu\, (\g)$ decays, we find a correction of $5.50(1.50)\%$ to the total rate due to next-to-leading order effects. This translates into a reduction of $|V_{\text{ub}}|$ by $-2.64(34)\%$. 
 
 The leading uncertainties in $B\to D \, l \, \nu\, (\g)$ and $B \to \pi \, l \, \nu\, (\g)$ results originate from neglecting short-distance contributions and due to the matching procedure. The large matching uncertainty for $B \to \pi \, l \, \nu\, (\g)$  indicates, that the intermediate region between long- and short-distance regime is not modeled adequately with our simplified approach. The correction for $|V_{\text{ub}}|$ should therefore be used with care. This seems not to be the case for $B\to D \, l \, \nu\, (\g)$, where swapping contributions between the long- and short-distance regime only provokes a modest scale dependence. Note that in addition to the estimated uncertainties both results might suffer from additional model dependencies, which cannot be quantified within the model itself. The result for $B\to D^*_0 \, l \, \nu\, (\g)$ has the same feature, but due to the lack of further knowledge due to SD related real corrections, the pure IB result should be used with care. 
 
The extracted correction factors for $B \to D \, l \, \nu \, (\g)$ can be combined with the strong corrections to determine the corrections of the semileptonic decay rates at $\mathcal{O}( \alpha \,  \GF +  \alpha_s \,  \GF)$. Corrections of mixing terms, which are at next-to-next-to-leading order $\mathcal{O}(  \alpha \, \alpha_s \, \GF)$ are negligible. By providing the experimental community with these results and a next-to-leading order Monte Carlo generator, we hope to make a first step towards closing this important knowledge gap and improve the future extractions of $|V_{\text{cb}}|$ and $|V_{\text{ub}}|$.

 \acknowledgments
  
  We thank Marek Sch\"onherr for many great discussions and his help on the matter. Additional thanks go to S\'ebastien Descotes-Genon, Dominik St\"ockinger, Gino Isidori, and Julie Michaud, who provided us with many useful comments, for which we are grateful.
   Last but not least we thank Troy Andre, Sasha Glazov, and Rick Kessler for providing the \texttt{KLOR} Monte Carlo generator code. 

\clearpage

 \begin{table}[ph!]\begin{center}
\begin{tabular}{l  llll} 
&  $\mu_0$ & $ \delta_{\text{sd}}$ &  $\delta_{\text{ld}}$ & $\delta_{\text{total}}$   \\ \hline 
 $B^0 \to D^{+} \, e \, \nu \, (\g)$         & $\half \mu_0$ &  $0.0213$  &   $0.0028(1)$ & $0.0241(1)$\\
                                                      & $\mu_0$    &   $0.0181$  & $0.0042(1)$     & $0.0222(1)$\\
                                                      & $2 \mu_0$    &  $0.0148$   & $0.0059(1)$        & $0.0207(1)$\\ \hline
 $B^0 \to  D^{+}  \, \mu \, \nu\, (\g)$        &  $\half \mu_0$ &    $0.0213 $   &   $0.0028(1)$ & $0.0241(1)$\\
                                                      & $\mu_0$    &   $0.0181 $  & $0.0042(1)$     & $0.0222(1)$\\
                                                      & $2 \mu_0$    &  $0.0148 $   & $0.0059(1)$        & $0.0207(1)$\\ \hline
 $B^+ \to \bar D^{0}  \, e \, \nu\, (\g)$         &  $\half \mu_0$ &  $0.0213$   &   $-0.0025(1)$ & $0.0188(1)$\\
                                                                 & $\mu_0$ &   $0.0181$  & $-0.0035(1)$        & $0.0146(1)$\\
                                                    	           & $2 \mu_0$  &  $0.0149$  & $-0.0040(1)$       & $0.0109(1)$\\ \hline
 $B^+ \to \bar D^{0}  \, \mu \, \nu\, (\g)$      &  $\half \mu_0$ &  $0.0213 $   &   $-0.0024(1)$ & $0.0189(1)$\\
                                                                  & $\mu_0$     &   $0.0181 $ & $-0.0034(1)$     & $0.0147(1)$\\
                                                                  & $2 \mu_0$   &  $0.0149 $ & $-0.0039(1)$   & $0.0110(1)$\\ \hline
\end{tabular}
 \caption{The running of  $1 + \delta_{\text{sd}}+ \delta_{\text{ld}}$ as a function of $\mu_0$ is shown for $B \to D \, l \, \nu \, (\g)$. The averaged difference between the predictions of $\delta_{\text{total}}$ at $ \half \mu_0 $ and $2 \mu_0 $ to the prediction at $\mu_0 = m_{D^+}$ or $m_{D^0}$ is taken as the estimator due to matching uncertainty $\sigma_{\text{matching}}$. The uncertainty in the parentheses is numerical.  } \label{matchD}
 \end{center}\end{table}
 
  \begin{table}[ph!]\begin{center}
\begin{tabular}{l  llll} 
&  $\mu_0$ & $ \delta_{\text{sd}}$ &  $\delta_{\text{ld}}$ & $\delta_{\text{total}}$   \\ \hline 
 $B^0 \to \pi^+ \, e \, \nu\, (\g)$         &  $\half \mu_0$ &  $0.0333$   &   $0.0388(1)$ & $0.0721(1)$\\
                                                      & $\mu_0$ & $0.0301$        &   $0.0254(1)$ & $0.0555(1)$\\
                                                      & $2 \mu_0$  & $0.0269$       &  $0.0157(1)$ & $0.0426(1)$\\ \hline
 $B^0 \to \pi^+ \, \mu \, \nu\, (\g)$      &  $\half \mu_0$ &  $0.0333 $   &   $0.0362(1)$ & $0.0696(1)$\\
                                                      & $\mu_0$ & $0.0301 $        &   $0.0244(1) $ & $0.0545(1)$\\
                                                      & $2 \mu_0$  & $0.0269 $       &  $0.0154(1)$ & $0.0423(1)$\\ \hline
 $B^+ \to \pi^0 \, e \, \nu\, (\g)$         &  $\half \mu_0$ &  $0.0335$   &   $0.0186(1)$ & $0.0521(1)$\\
                                                                 & $\mu_0$ &   $0.0303$  & $0.0108(1)$        & $0.0411(1)$\\
                                                    	         & $2 \mu_0$  &  $0.0271$  & $0.0050(1)$       & $0.0320(1)$\\ \hline
 $B^+ \to \pi^0 \, \mu \, \nu\, (\g)$     &  $\half \mu_0$ &  $0.0335 $   &   $0.0160(1)$ & $0.0495(1)$\\
                                                                 & $\mu_0$ &   $0.0303 $  & $0.0098(1)$        & $0.0401(1)$\\
                                                    	         & $2 \mu_0$  &  $0.0271 $  & $0.0047(1)$       & $0.0318(1)$\\ \hline
\end{tabular}
\caption{The running of  $1 + \delta_{\text{sd}}+ \delta_{\text{ld}}$ as a function of $\mu_0$ is shown  for $B \to \pi \, l \, \nu \, (\g)$. The averaged difference between the predictions of $\delta_{\text{total}}$ at $ \half \mu_0 $ and $2 \mu_0 $ to the prediction at $\mu_0 = m_{\pi^+}$ or $m_{\pi^0}$ is taken as the estimator due to matching uncertainty $\sigma_{\text{matching}}$. The uncertainty in the parentheses is numerical.  }
\label{matchpi}
 \end{center}\end{table}

  \begin{table}[ph!]\begin{center}
\begin{tabular}{l  llll} 
&  $\mu_0$ & $ \delta_{\text{sd}}$ &  $\delta_{\text{ld}}$ & $\delta_{\text{total}}$   \\ \hline 
 $B^0 \to D^{*\,+}_0 \, e \, \nu\, (\g)$     &  $\half \mu_0$ &  $0.0201$   &   $0.0031(1)$ & $0.0232(1)$\\
                                                      & $\mu_0$ & $0.0169$        &   $0.0055(1)$ & $0.0224(1)$\\
                                                      & $2 \mu_0$  & $0.0137$       &  $0.0076(1)$ & $0.0212(1)$\\ \hline
 $B^0 \to  D^{*\,+}_0  \, \mu \, \nu\, (\g)$      &  $\half \mu_0$ &  $0.0201 $   &   $0.0032(1)$ & $0.0234(1)$\\
                                                      & $\mu_0$ & $0.0169 $        &   $0.0057(1)$ & $0.0226(1)$\\
                                                      & $2 \mu_0$  & $0.0137 $       &  $0.0078(1)$ & $0.0215(1)$\\ \hline
 $B^+ \to  \bar D^{*\,0}_0  \, e \, \nu\, (\g)$     &  $\half \mu_0$ &  $0.0202$   &   $-0.0023(1)$ & $0.0179(1)$\\
                                                                 & $\mu_0$ &   $0.0170$  & $-0.0028(1)$        & $0.0142(1)$\\
                                                    	         & $2 \mu_0$  &  $0.0138$  & $-0.0028(1)$       & $0.0109(1)$\\ \hline
 $B^+ \to  \bar D^{*\,0}_0  \, \mu \, \nu\, (\g)$      &  $\half \mu_0$ &  $0.0202 $   &   $-0.0022(1)$ & $0.0180(1)$\\
                                                                 & $\mu_0$ &   $0.0170 $  & $-0.0026(1)$        & $0.0144(1)$\\
                                                    	         & $2 \mu_0$  &  $0.0138 $  & $-0.0027(1)$       & $0.0111(1)$\\ \hline
\end{tabular}
\caption{The running of  $1 + \delta_{\text{sd}}+ \delta_{\text{ld}}$ as a function of $\mu_0$ is shown for  for $B \to D^*_0 \, l \, \nu \, (\g)$. The averaged difference between the predictions of $\delta_{\text{total}}$ at $ \half \mu_0 $ and $2 \mu_0 $ to the prediction at $\mu_0 = m_{D^{*\, +}_0}$ or $m_{D^{*\, 0}_0}$ is taken as the estimator due to matching uncertainty $\sigma_{\text{matching}}$. The uncertainty in the parentheses is numerical.  }\label{matchDs}
 \end{center}\end{table}

 \clearpage


%
 
 
 
 
%% file: formfactors.tex
\section{Form factor models of exclusive semileptonic $B$-meson decays reviewed}\label{appa}

\subsection{Form factors for $B \to D \, l \, \nu$}\label{hqetformfactorsintroduction}

Considering the heavy quark sector of the Standard Model Lagrangian in the limit of infinitely massive quarks yields interesting simplifications: The arising effective field theory is known as heavy quark effective theory \cite{Caprini1998153} and predicts the expansion of the hadronic current for exclusive $B \to D \, l \, \nu$ decays as a function of the four-velocity transfer $w = v_B \cdot v_D$, where $v_B$ and $v_D$ are the four velocities of the $B$- or $D$-meson, respectively. The relevant vector current is 
\bea\label{heavyquarkcurrent}
 \bra D | V^\mu  | B \ket  & \eq & \sqrt{m_B m_D} \Big( h_{+}(w)\bve{v_B+v_D}^{\mu} + h_{-}(w)\bve{v_B - v_D}^{\mu}\Big) \,,
\eea
with the heavy quark form factors 
\bea
 h_{+}(w) & \eq &  \mathcal{G}(1) \times \Big[ 1 - 8 \rho_D^2 \, z + (51 \rho_D^2 - 10) \, z^2 - (252 \rho_D^2 - 84) \, z^3) \Big] \,, \\
 h_{-}(w) & \eq & 0 \,,
\eea
where $z =\frac{\sqrt{w+1} - \sqrt{2}}{ \sqrt{w+1}+\sqrt{2} } \, .$ The expansion depends on $\rho_D^2$ and $\mathcal{G}(1)$, which are the form factor slope and the normalization at $w \eq 1$, respectively. The numerical value of the slope used in this analysis is listed in Table \ref{theoryparshqet}.

 \begin{table}[h!]\begin{center}
\begin{tabular}{r l} 
 Parameter &  Value   \\ \hline 
 $\rho_D^2$  &$1.19$ \\
\end{tabular}
 \caption{Used slope taken from \cite{hfag08}.} \label{theoryparshqet}
 \end{center}\end{table}
 
\subsection{Form factors for $B \to D^* \, l \, \nu$}\label{hqetvectorformfactorsintroduction}

Similarly as for the pseudoscalar $D$-meson final state, HQET can predict the total and differential rates for $B \to D^* \, l \, \nu$ decays as a function of the four-velocity transfer $w = v_B \cdot v_{D^*}$. The vector and axial-vector currents are \cite{Caprini1998153}
\bea\label{hqetdsvecaxveccur}
  \bra {D^*} | V^\mu  | B \ket  & \eq & 2 \, i \, \varepsilon^{\mu\nu\alpha\beta} \epsilon^*_{\nu} \bve{p_B}_{\alpha} \bve{p_{D^*}}_{\beta} \,g(t) \,, \\
  \bra {D^*} | A^\mu  | B \ket  & \eq &  \epsilon^{*\, \mu} \, f(t) + \bve{p_B + p_{D^*}}^\mu \, \epsilon^* \cdot p_B \, a_{+}(t)  + \bve{p_B - p_{D^*}}^\mu \, \epsilon^* \cdot p_B \, a_{-}(t)   \,,
\eea
with
\bea
g(t) & \eq &  \frac{1}{m_B \, R \, (1+r)} \, R_1(w) \, h_{A1}(w) \, , \\
f(t) & \eq &  \frac{m_B \, R \, (1+r)}{2} \, (w+1) \, h_{A1}(w) \, , \\
a_{+}(t) & \eq & - \frac{1}{m_B \, R \, (1+r)} \, R_2(w) \, h_{A1}(w) \, , \\
a_{-}(t) & \eq & 0 \, . 
\eea
 and
\bea
 h_{A1}(w) & \eq & \mathcal{F}(1) \times \Big[1 - 8 \rho_{D^*}^2 \, z + (53  \rho_{D^*}^2 - 15) \, z^2 - (231  \rho_{D^*}^2 - 91) \, z^3 \Big] \\
 R_1(w) & \eq &  R_1(0) - 0.12 (w-1) + 0.05 (w-1)^2 \,, \\
 R_2(w) & \eq & R_2(0) + 0.11(w-1) - 0.06 (w-1)^2  \,, 
\eea
where $\rho_{D^*}^2$ and $\mathcal{F}(1)$ are, respectively, the form factor slope and normalization at $w \eq 1$. The used numerical values of the slope and ratios are stated in Table \ref{theoryparshqet2}.

  \begin{table}[h!]\begin{center}
\begin{tabular}{r l} 
 Parameter & Value   \\ \hline 
 $\rho_D^{*2}$&   $1.16$ \\
 $R_1(0)$ & $1.369$ \\
 $R_2(0)$ & $0.846$ \\
\end{tabular}
 \caption{Used slope from \cite{hfag08}.} \label{theoryparshqet2}
 \end{center}\end{table}

\subsection{Form factors for $B \to \pi \, l \, \nu$}\label{bzformfactorsintroduction}

Extrapolating results from lattice QCD calculations and light-cone sum rules, the form factors of exclusive $B \to \pi \, l \, \nu$ can be parametrized as a function of the four-momentum transfer squared $t$. The relevant vector current is \cite{PhysRevD.71.014015} 
\bea
 \bra \pi | V^\mu  | B \ket  & \eq &  \Big( \bve{p_B + p_{\pi}}^{\mu} - \frac{m_B^2 - m_{\pi}^2}{t} \bve{p_B - p_{\pi}}^\mu \Big) \, f_{+}(t) + \Big(\frac{m_B^2 - m_{\pi}^2}{t}  \bve{p_B - p_{\pi}}^\mu   \Big) \, f_{0}(t) \,, \nonumber \\
 \eea
with form factors parametrized as
\bea
 f_{+}(t) & \eq &  \frac{r_{f_{+1}} }{ 1 - \frac{t}{m_{f_{+1}}^2}}  +  \frac{r_{f_{+2}} }{ 1 - \frac{t}{m_{f_{+2}}^2}} \,, \\
 f_{0}(t) & \eq &  \frac{r_{f_{0}} }{ 1 - \frac{t}{m_{f_{0}}^2 }}  \,,
\eea
with the normalizations $r_{f_{+1}} $,$r_{f_{+2}}$ , and $r_{f_{0}}$,  and the pole masses $m_{f_{+1}}$, $m_{f_{+2}}$, and $m_{f_{0}}$. The values used in this paper are stated in Table \ref{bzpars}. The $q^\mu$ proportional form factor $f_-(t)$ can be written as
\bea
 f_{-}(t) & \eq & \frac{m_B^2 - m_{\pi}^2}{t} \Big( f_0(t) - f_{+}(t) \Big) \, \eq \, \frac{m_B^2 - m_{\pi}^2}{t} \Bigg(  \frac{r_{f_{0}} }{ 1 - \frac{t}{m_{f_{0}}^2}}  -  \frac{r_{f_{+1}} }{ 1 - \frac{t}{m_{f_{+1}}^2}}  -  \frac{r_{f_{+2}} }{ 1 - \frac{t}{m_{f_{+2}}^2}} \Bigg) \, . \nonumber \\
\eea

 \begin{table}[h!]\begin{center}
\begin{tabular}{r l} 
 Parameter & Value   \\ \hline 
 $ r_{f_{+1}} $  & 0.744 \\ 
  $ r_{f_{+2}}  $  & -0.486 \\
 $  m_{f_{+1}}^2  $  & 28.40 $\text{GeV}^2/c^4$ \\
 $  m_{f_{+2}}^2  $ &  40.73  $\text{GeV}^2/c^4$ \\
  $ r_{f_{0}} $ &   0.258 \\ 
 $  m_{f_{0}}^2  $   & 33.81$\text{GeV}^2/c^4$  \\
\end{tabular}
 \caption{Form factor slopes and pole masses taken from \cite{PhysRevD.71.014015}.} \label{bzpars}
 \end{center}\end{table}
 
 \subsection{Form factors for $B \to D^*_0 l \nu$ }\label{llswformfactorsintroduction}
 
 Heavy quark effective theory can be used to extract differential and total rates for other charmed resonances in an approximative manner, as done in \cite{Leibovich:1997tu,Leibovich:1997em} for $B \to D^*_0 l \nu$ decays. Since the $D^*_0$ is a scalar particle, the hadronic vector current vanishes and only axial contributions are non-zero. Note that the entire formalism discussed in Sec. \ref{nloform} can be adapted to derive the IB axial-vector contributions $A_{\mu\nu}^{\text{IB}}$ with SD related corrections. The relevant hadronic contribution is given by
\bea\label{llswheavyquarkcurrent}
  \bra {D^*_0} | A^\mu  | B \ket  & \eq &   \sqrt{m_B m_{D^*_0}} \Big(  g_{+}(w) \bve{v_B + v_{D^*_0}}^\mu + g_{-}(w) \bve{v_B - v_{D^*_0}}^\mu \Big)  \,  ,
\eea
with the form factors $g_{\pm}$. They are parametrized by 
\bea\label{llswformfactexp}
  g_{+}(w)  & \eq & \epsilon_c \Big[ 2 (w - 1) \zeta_1(w) - 3 \zeta(w) \frac{w \bar \Lambda^* - \bar\Lambda}{w+1} \Big] \nonumber \\
   & &\ph{+} - \epsilon_b \Big[  \frac{ \bar\Lambda^* (2w+1) - \bar\Lambda(w+2) }{w+1} \zeta(w) - 2 (w-1) \zeta_1(w) \Big] \,,  \nonumber \\
    g_{-}(w)  & \eq & \zeta(w)  \,,  \nonumber \\
\eea
where  
\bea
  \zeta(w)  & \eq & \frac{w+1}{\sqrt{3}} \, \tau(w) \,.
\eea
with $\tau(w)$ is the leading Isgur-wise function \cite{Isgur:1989vq,Isgur:1989ed}. This can be simplified to
 \bea
 \zeta(w)  & \eq & \zeta(1) \times \big[ 1 + \zeta'(w-1) \big] \,. ,
\eea
where $ \zeta'$ denotes the form factor slope. The value of $\zeta_1$ depends on the phase-space region, but its exact value only leads to small corrections. As a consequence, its contributions can either be neglected or approximatively taken into account. The latter is done by setting
\bea
 \zeta_1 = \bar\Lambda \zeta(w) \,. 
\eea
The numerical values of the remaining slopes and pole masses are stated in Table \ref{llswpars}.

 \begin{table}[h!]\begin{center}
\begin{tabular}{rl} 
 Parameter &  Value   \\ \hline 
 $  \epsilon_b  $ & 0.1042 $c$ GeV${}^{-1}$ \\ 
 $  \epsilon_c  $  & 0.3571 $c$ GeV${}^{-1}$ \\ 
  $\bar\Lambda$ &   0.4 GeV$/c^2$   \\
 $\bar\Lambda^*$ & 0.75  GeV$/c^2$ \\
  $\zeta' $ &   -1.0 \\
  \end{tabular}
 \caption{Form factor slopes and pole masses taken from \cite{Leibovich:1997tu}.} \label{llswpars}
 \end{center}\end{table}

%% file: loopintegrals.tex
\section{Loop integrals}\label{appc}

\subsection{Passarino Veltman reduction of higher order tensor integrals}

For one-loop tensor integrals, a systematic algorithm has been worked out by Passarino and Veltman \cite{Passarino:1978jh} to reduce any given higher order tensor integral as a sum of four-vectors multiplied by scalar integrals. For integrals with one, two, or three external legs, we use the notation 
\bea
A_{0,\mu\nu}(m^2) & \eq & \tilde\mu^{4-D} \, \int \frac{\ud^D k}{(2\pi)^{D/2}} \frac{1,k_\mu k_\nu}{\bve{-k^2 + m^2}} \, , \\ 
B_{0,\mu,\mu\nu} (p^2; m_0^2, m^2) & \eq &  \tilde\mu^{4-D} \, \int \frac{\ud^D k}{(2\pi)^{D/2}} \frac{1,k_\mu,k_\mu k_\nu}{\bve{-k^2 + m_0^2} \, \bve{(-p + k)^2 + m^2}} \, , \\
C_{0,\mu,\mu\nu} (p_1^2, s, p_2^2; m_0^2, m_1^2, m_2^2) & \eq &  \tilde\mu^{4-D} \, \int \frac{\ud^D k}{(2\pi)^{D/2}} \frac{1,k_\mu,k_\mu k_\nu}{\bve{-k^2 + m_0^2} \, \bve{(-p_1 + k)^2 + m_1^2} \, \bve{(-p_2 + k)^2 + m_2^2} } \, , \nonumber \\
\eea
respectively, with an apparent generalization towards more external legs and higher rank tensor integrals. Note that we redefined $\tilde\mu^{4-D} = - i \, (4\pi)^2 \, \mu^{4-D}$ and neglected $A_{\mu}$. The latter vanishes due to its odd transformation property when integrated. In addition, we defined $s = (p_1 - p_2)^2$. Due to Lorentz symmetry the integration result can only depend on tensor structures from external momenta $p^\mu_j$ and the metric tensor $g^{\mu\nu}$. This implies
\bea
 A^{\mu\nu} & \eq & g^{\mu\nu} \,  A_{00} \, , \nonumber \\
 B^\mu & \eq & p^\mu \, B_1\, , \nonumber \\
 B^{\mu\nu} & \eq & p^\mu \, p^\nu \, B_{11} + g^{\mu\nu} B_{00} \, , \nonumber \\
 C^{\mu} & \eq & p_1^\mu \, C_{1} + p_2^\mu \, C_{2} \, , \nonumber \\
 C^{\mu\nu} & \eq & p_1^\mu \, p_1^\nu \, C_{11}  + p_2^\mu \, p_2^\nu \, C_{22}  + p_1^{[\mu} \, p_2^{\nu]} \, C_{12}  + g^{\mu\nu} \, C_{00} \, . \nonumber 
\eea
Contracting both sides of these equations with external momenta and the metric tensor yields a set of equations that allow the determination of the loop-form factors $B_i, B_{ij}, C_i, C_{ij}$ with $i,j=0,1,2$. 

\subsubsection{Derivatives of two-point functions}

The derivative of a two-point function is defined as
\bea
 \dot B_{i} = \frac{\ud}{\ud p^2} B_{i}(p^2;m_1^2,m^2) \, ,
\eea
with $i = 0,1$.